\documentclass[times,preprint,authoryear]{elsarticle}
\usepackage[top=2.5cm,bottom=3cm,width=18cm]{geometry}

\usepackage{framed,multirow}

\usepackage{amsmath}
\usepackage{amssymb}
\usepackage{latexsym}

\usepackage{url}
\usepackage{xcolor}
\definecolor{newcolor}{rgb}{.8,.349,.1}

\usepackage{hyperref}

\usepackage{graphicx}
\usepackage{subfig}
\usepackage{float}

\usepackage{threeparttable}
\usepackage{array}
\usepackage{multirow}
\usepackage{multicol}
\usepackage[font=small]{caption}
\newenvironment{Table}
  {\par\bigskip\noindent\minipage{\columnwidth}\centering}
  {\endminipage\par\bigskip}
  
\graphicspath{{figures/}}

\journal{Advances in Space Research}

\makeatletter
\def\ps@pprintTitle{%
     \let\@oddhead\@empty
     \let\@evenhead\@empty
     \def\@oddfoot
       {\hbox to \textwidth%
        {\ifnopreprintline\relax\else
        \@myfooterfont%
         \ifx\@elsarticlemyfooteralign\@elsarticlemyfooteraligncenter%
           \hfil\@elsarticlemyfooter\hfil%
         \else%
         \ifx\@elsarticlemyfooteralign\@elsarticlemyfooteralignleft%
           \@elsarticlemyfooter\hfill{}%
         \else%
         \ifx\@elsarticlemyfooteralign\@elsarticlemyfooteralignright%
           {}\hfill\@elsarticlemyfooter%
         \else%
               Accepted manuscript for \ifx\@journal\@empty%
                 Elsevier%
            \else\@journal\fi\hfill\@date\fi%
         \fi%
         \fi%
         \fi%
         }
       }%
     \let\@evenfoot\@oddfoot}
\makeatother

\begin{document}

\date{}

\begin{frontmatter}

\title{The antenna phase center motion effect in high-accuracy spacecraft tracking experiments}%

\author[asc,bmstu]{D.~A.~Litvinov\corref{mycorrespondingauthor}}
\cortext[mycorrespondingauthor]{Corresponding author at: Astro Space Center, Lebedev Physical Institute, Profsoyuznaya 84/32, 117997 Moscow, Russia.}
\ead{litvirq@yandex.ru}

\author[yorku]{N.~V.~Nunes}

\author[sai]{A.~I.~Filetkin}

\author[yorku]{N.~Bartel}

\author[jive,tudelft]{L.~I.~Gurvits}

\author[hobart]{G.~Molera~Calves}

\author[sai]{V.~N.~Rudenko}

\author[kiam]{M.~V.~Zakhvatkin}

\address[asc]{Astro Space Center, Lebedev Physical Institute, Profsoyuznaya 84/32, 117997 Moscow, Russia}

\address[bmstu]{Bauman Moscow State Technical University, 2-ya Baumanskaya 5, 105005 Moscow, Russia}

\address[yorku]{York University, Toronto, Ontario M3J 1P3, Canada}

\address[sai]{Sternberg Astronomical Institute, Lomonosov Moscow State University,
Universitetsky pr.~13, 119991 Moscow, Russia}

\address[jive]{Joint Institute for VLBI ERIC, Oude Hoogeveensedijk 4, 7991 PD Dwingeloo, the Netherlands}

\address[tudelft]{Department of Astrodynamics and Space Missions, Delft University of Technology, 2629~HS~Delft, The Netherlands}

\address[hobart]{Department of Physics, University of Tasmania, Hobart 7001, Tasmania, Australia}

\address[kiam]{Keldysh Institute for Applied Mathematics, Russian Academy of Sciences, Miusskaya sq. 4, 125047 Moscow, Russia}


\begin{abstract}

We present an improved model for the antenna phase center motion effect for high-gain mechanically steerable ground-based and spacecraft-mounted antennas that takes into account non-perfect antenna pointing. Using tracking data of the RadioAstron spacecraft we show that our model can result in a correction of the computed value of the effect of up to $2\times10^{-14}$ in terms of the fractional frequency shift, which is significant for high-accuracy spacecraft tracking experiments. The total fractional frequency shift due to the phase center motion effect can exceed $1\times10^{-11}$ both for the ground and space antennas depending on the spacecraft orbit and antenna parameters. We also analyze the error in the computed value of the effect and find that it can be as large as $4\times10^{-14}$ due to uncertainties in the spacecraft antenna axis position, ground antenna axis offset and misalignment, and others. Finally, we present a way to reduce both the ground and space antenna phase center motion effects by several orders of magnitude, e.g. for RadioAstron to below $1\times10^{-16}$, by tracking the spacecraft simultaneously in the one-way downlink and two-way phase-locked loop modes, i.e. using the Gravity Probe A configuration of the communications links. 
\end{abstract}


\end{frontmatter}


\begin{multicols}{2}

\section{Introduction}
\label{sec:intro}

The phase center motion effect exhibited by high-gain mechanically steerable antennas is well-known in the fields of very long baseline interferometry (VLBI)
and orbit determination (OD) of deep space probes, and constitutes an essential part of the corresponding
data reduction models \citep{wade-1970-apj, moyer-1971-techreport,moyer-2005-book}.
We consider novel aspects of this effect that are relevant to tracking near-Earth spacecraft (SC). In this case both the ground-based
and spacecraft-mounted antenna usually do not point  at each other precisely
but there
are small pointing errors caused primarily by inaccuracy of the predicted orbit.
To take this into account the well-known equations for the antenna phase center motion (APCM) effect need to be modified. For small pointing errors
the required modification of the original equations consists simply in the replacement of the actual antenna pointing
angles with those that correspond to the true position of the signal source (or target, if the antenna is transmitting).

The magnitude of the ground APCM effect is proportional to the offset between
the rotation axes of the antenna and is formally zero for antennas with intersecting
axes. For SC-mounted antennas the APCM effect is proportional to the
distance between the SC center of mass and the intersection point of the antenna rotation axes and is usually non-zero. 

In this paper we are primarily concerned with the influence
of the APCM\ effect on the frequency of received and transmitted signals.
Stated in terms of the fractional frequency shift, $\Delta f/f$, our results are independent of the base frequency of the signal and apply
universally to tracking SC at S-, X-, Ka- or any other frequency
band.
Equations for the additional phase delay due to the effect, which might be
of interest to the problem of simultaneously tracking SC by several ground antennas in the VLBI regime \citep{duev-2012-aa}, are also provided. 

The problem of tracking SC of space very-long-baseline interferometry (space-VLBI or SVLBI) missions represents one of the primary applications of our results (for details on the techniques of VLBI and SVLBI see \cite{thompson-moran-swenson-2017-book} and \cite{gurvits-2020-asr}). Indeed, APCM effects for such spacecraft are usually large due to the use of large high-gain antennas that are needed to transmit large amounts of captured astronomical data at high data rates: 128 Mbit/s for the recent RadioAstron mission \citep{kardashev-2013-ar} and up to many Gbit/s for prospective successors \citep{gurvits-2020-asr}. Also, the usually high orbit ellipticity, which provides for broad coverage of lengths and orientations of the space-to-ground interferometer baseline vectors (the vectors between the ground and space antennas), additionally increases the effect near perigee passages by several orders of magnitude. Finally, VLBI requires highly stable atomic frequency standards to be used at each of the participating radio telescope, including those in space, to coherently time-tag the signals received from astronomical sources. Those frequency standards not only provide time and frequency reference signals for the on-board electronic equipment but also are usually used to synchronize downlink signal frequencies, which provides for performing high-accuracy radio science experiments with space-VLBI SC \citep{biriukov-2014-ar, litvinov-2018-pla, nunes-2020-asr, gurvits-2020-asr, litvinov-2021-cqg}. Therefore, to evaluate the significance of the changes we introduced to the APCM effect model we apply it to the case of the RadioAstron spacecraft \citep{kardashev-2013-ar} and the NRAO140 antenna of the Green Bank Earth Station \citep{ford-2014-spie} which served the RadioAstron SVLBI mission. 

We find that the correction we introduced is significant for high-accuracy Doppler tracking experiments. For the case of the NRAO140 antenna tracking the RadioAstron SC, it reaches $2\times10^{-14}$ in terms of the fractional frequency shift. The magnitude of the total fractional frequency shift due to the APCM effect exceeds $1\times10^{-11}$
at some parts of the orbit, both for the ground and spaceborne antenna. The effect is therefore large enough to be taken into account even for regular OD, which we demonstrate using data from Doppler tracking experiments performed with RadioAstron. 

Since the APCM effect is significant, the question arises if it can be taken into account accurately enough. In order to answer it we consider several error sources that may contribute to the APCM effect, such as the
uncertainties in the position of the intersection point of the SC antenna axes, ground antenna axis offset, and ground antenna axis alignment, among others.
Using RadioAstron we find that near perigees some of these errors can exceed $4\times10^{-14}$ in terms of the fractional frequency shift.

In order to ensure that these results are not specific to RadioAstron we consider a possible follow-up SVLBI mission with a
SC on a less eccentric orbit, similar to the one suggested in \citep{hong-2014-aa}. We find that in this case both the correction due to imperfect pointing of the ground and spaceborne antennas
and the error of estimating the APCM effect are of comparable magnitude to
those of RadioAstron. 

The significance
of the above-mentioned numbers becomes obvious if one considers the parameters of the frequency
stability and accuracy of the atomic frequency standards used in such experiments. For example, the instability
of the hydrogen maser of the RadioAstron spacecraft reached
$2\times10^{-15}$ in terms of the fractional frequency variations at the averaging time of one hour \citep{vremya-ch-vch-1010}. For the cesium fountain clock of the ACES experiment, which will be performed at the International Space Station, the instability is expected to reach $\sim10^{-16}$ at averaging times of
$\sim$~1~day \citep{aces-2011-acau}. Laboratory
 devices with the accuracy and stability parameters of $\sim10^{-18}$ at averaging times
of  $\sim$~1~hour have already been demonstrated
 \citep{bothwell-2019-metrologia}.

It is therefore of much interest to reduce the APCM
effect, or at least the error of estimating it, down to a level that is below the stability and accuracy of modern frequency
standards. One solution is to
avoid using mechanically steerable high-gain antennas in high-accuracy
SC tracking experiments. However, this may not be possible for SC on high Earth
orbits, other planet orbiters,  and deep space probes. 

An alternative approach, suggested in this paper, is to compensate for the APCM effect by using a specific configuration of the communication links, i.e. that of the simultaneously operating one-way downlink and two-way phase-locked loop. In the one-way
mode the SC's downlink signal is synchronized to
its on-board frequency standard, and in the two-way, or phase-locked loop mode, the phase of the spacecraft's downlink signal
is synchronized to that of the uplink signal transmitted by a ground tracking station (TS). Such configuration was first used in the Gravity Probe A mission \citep{vessot-levine-1979-grg} to compensate for the contributions of the non-relativistic Doppler effect and troposphere to the frequency shift of the signal transmitted by the SC. (The tropospheric frequency shift is compensated up to the fluctuations induced by the atmospheric refractive index variations, mostly due to water vapour, which occur on time scales shorter than the signal light travel time and length scales smaller than the separation between the up- and downlink signal paths.) We show that this scheme is also very effective in reducing the APCM effect, both for ground and spaceborne antennas. For example, for RadioAstron it
reduces the total fractional frequency shift due to the APCM effect by several orders of magnitude, down to below $1\times10^{-16}$. Thus, at least in gravity-related experiments, the Gravity Probe A compensation
scheme cancels, or significantly reduces, all the major unwanted frequency shifts except for the one due to the ionosphere. The ionospheric contribution can be cancelled by using multi-frequency links \citep{vessot-levine-1979-grg}. 


The outline of the paper is as follows. In Section~\ref{sec:theory} we present our generalized
equations for computing the APCM effect which take into account imperfect antenna pointing. We consider the cases of ground-based and SC-mounted antennas and also analyze
the errors in the estimated APCM effect. In Section~\ref{sec:radioastron} we apply these equations to the RadioAstron
spacecraft. In Section~\ref{sec:future-missions} we present the results
of a similar analysis
for a possible future follow-up SVLBI mission. In Section~\ref{sec:compensation} we develop equations for the compensation of the APCM effect using the Gravity Probe~A scheme and apply them to the case
of the RadioAstron spacecraft. We summarize our results in Section 6.

\section{Theory}
\label{sec:theory}

The antenna structure gives rise to several effects that influence the phase,
and thus the frequency,
of the received and transmitted signals. These effects include, among others, the variable geometric delay of the signal
propagation resulting from the antenna motion due to tracking the source/target, temperature-induced variations of the antenna
reference point position, gravity loading experienced by flexible antenna structures, and the differential feed rotation. Here we only consider the first of these effects and for the details on the rest refer the reader to \citep{wade-1970-apj, moyer-2005-book, sovers-1998-rmp,
sovers-1987-techreport} .

\subsection{Ground-based antennas}

\begin{figure*}[t]           
        \centering
        \subfloat[Alt-azimuth mount with a positive axis offset]{
                \label{fig:common-ground-antenna-mounts:a}
                \includegraphics[scale=0.15]{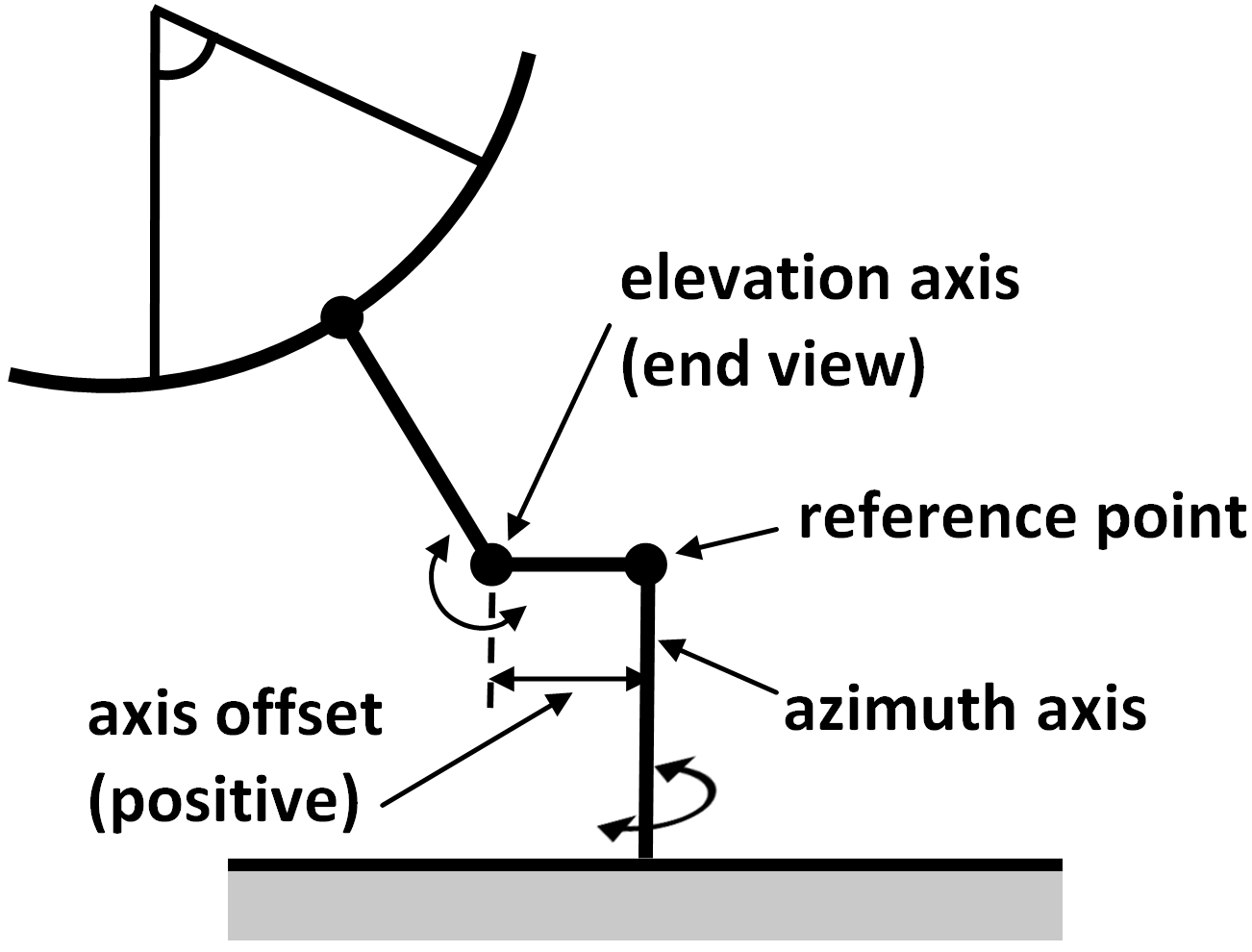}} \hspace{2cm}
        \subfloat[Alt-azimuth mount with a negative axis offset]{
                \label{fig:common-ground-antenna-mounts:b}
                \includegraphics[scale=0.15]{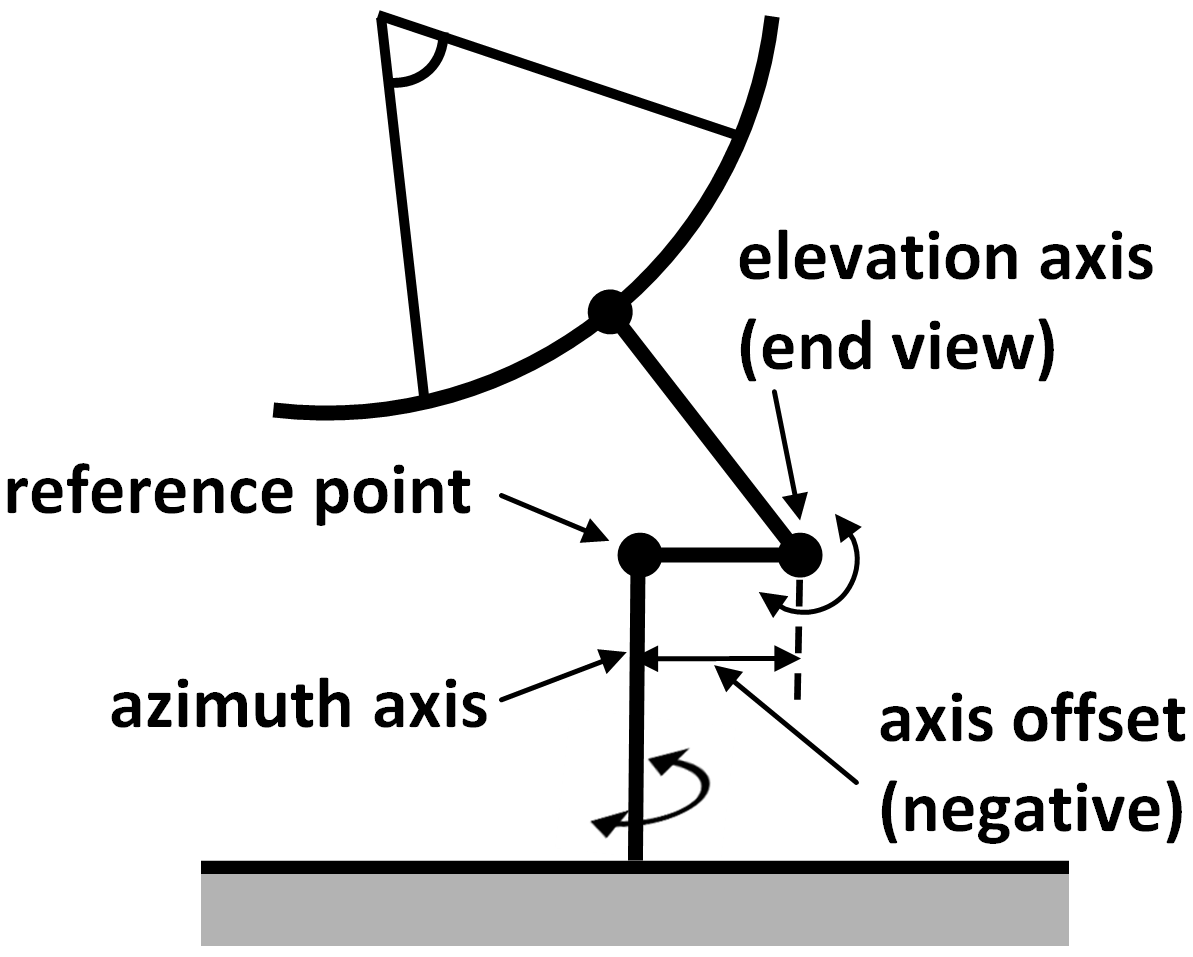}} \\
        \subfloat[Polar mount]{
                \label{fig:common-ground-antenna-mounts:c}
                \includegraphics[scale=0.15]{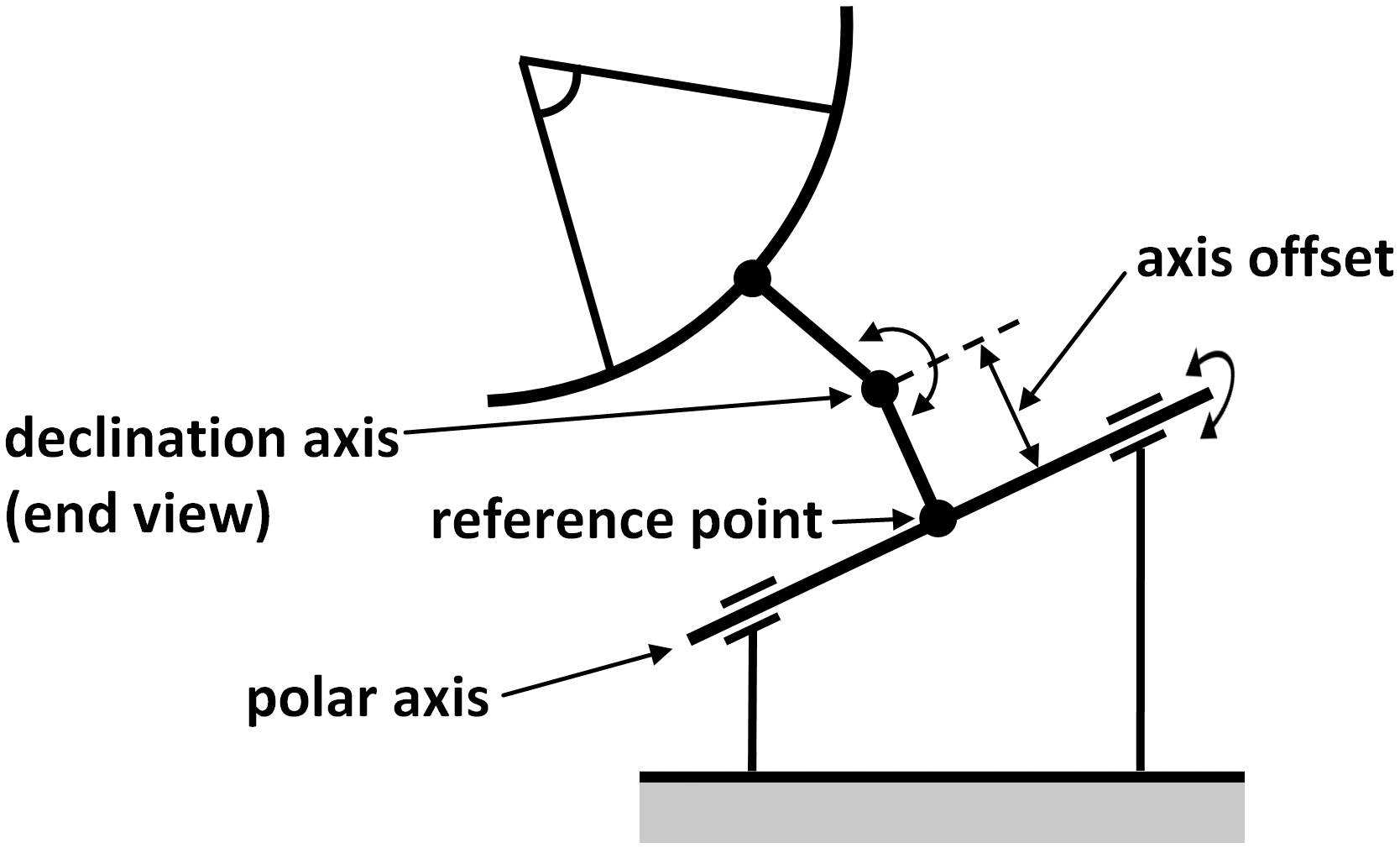}} \hspace{2cm}
        \subfloat[X-Y mount]{
                \label{fig:common-ground-antenna-mounts:d}
                \includegraphics[scale=0.15]{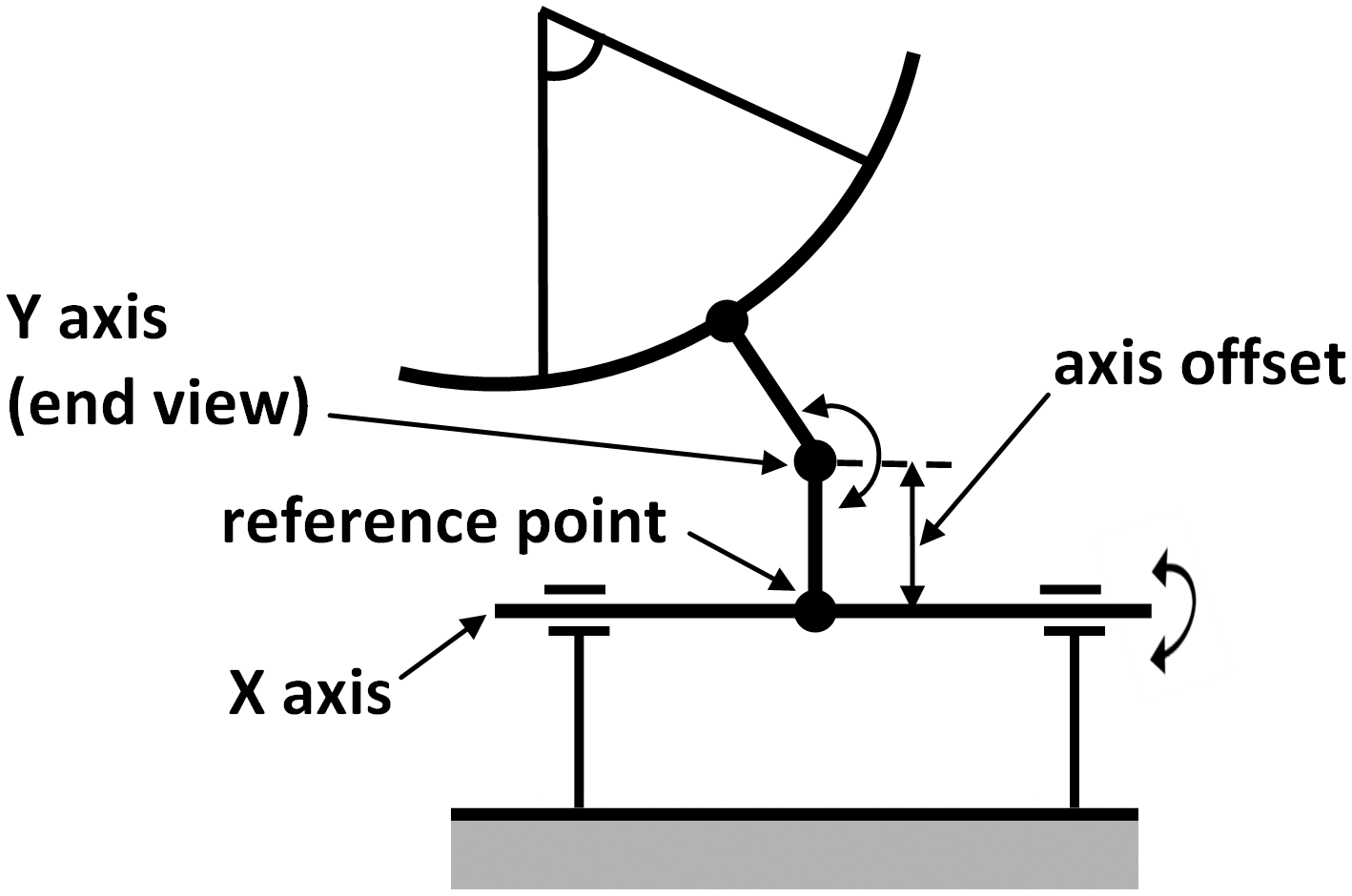}}
        \caption{Common mount types of ground steerable high-gain antennas.}
        \label{fig:common-ground-antenna-mounts}
\end{figure*}

High-gain steerable ground-based dish antennas usually use one of the following
three types of mounts: alt-azimuth (alt-az), polar (or equatorial) and X-Y (North-South or East-West) (Fig.~\ref{fig:common-ground-antenna-mounts}). For alt-az antennas the azimuth axis is oriented in the direction of the local zenith and is fixed relative to the Earth while the elevation axis lies in a plane
perpendicular to it and rotates around it. Alt-az antennas are often designed
such that their two axes nominally intersect and the axis offset is zero. However, designs with non-intersecting axes are also common. In the latter case the offset of the elevation axis from the azimuth axis can either be positive (the antenna dish is closer to the source compared to the zero-offset case, Fig.~\ref{fig:common-ground-antenna-mounts:a}) or negative (correspondingly, farther, Fig.~\ref{fig:common-ground-antenna-mounts:b}). Alt-az antennas with large axis offsets include the 70-meter Ussuriysk RT-70 antenna (4~m axis offset) and all antennas of the VLBA network (25~m diameter, 2.1~m axis offset) \citep{sked-antenna-cat}.

A polar (equatorial) mount antenna has its polar axis oriented along the Earth rotation axis and the declination axis lies in a plane that is perpendicular
to the polar axis. This antenna design makes it easy to track celestial sources since tracking requires rotating only the polar axis. Polar mounts usually have large axis offsets of order of a few meters,
with the largest one to date being that of the Green Bank NRAO140 43m antenna, for which it equals 14.9~m \citep{langston-2012-nrao-memo}.

An X-Y antenna has its Earth-fixed X axis in the horizontal plane, oriented
in the North-South or East-West direction, and its Y axis in the perpendicular
plane. X-Y antennas usually have axis offsets
of order of a few meters, with the 26m Hobart antenna
currently having the largest one of 8.2~m \citep{sked-antenna-cat}.

\end{multicols}

\begin{multicols}{2}

\begin{figure}[H]              
        \centering
        \includegraphics[scale=0.22]{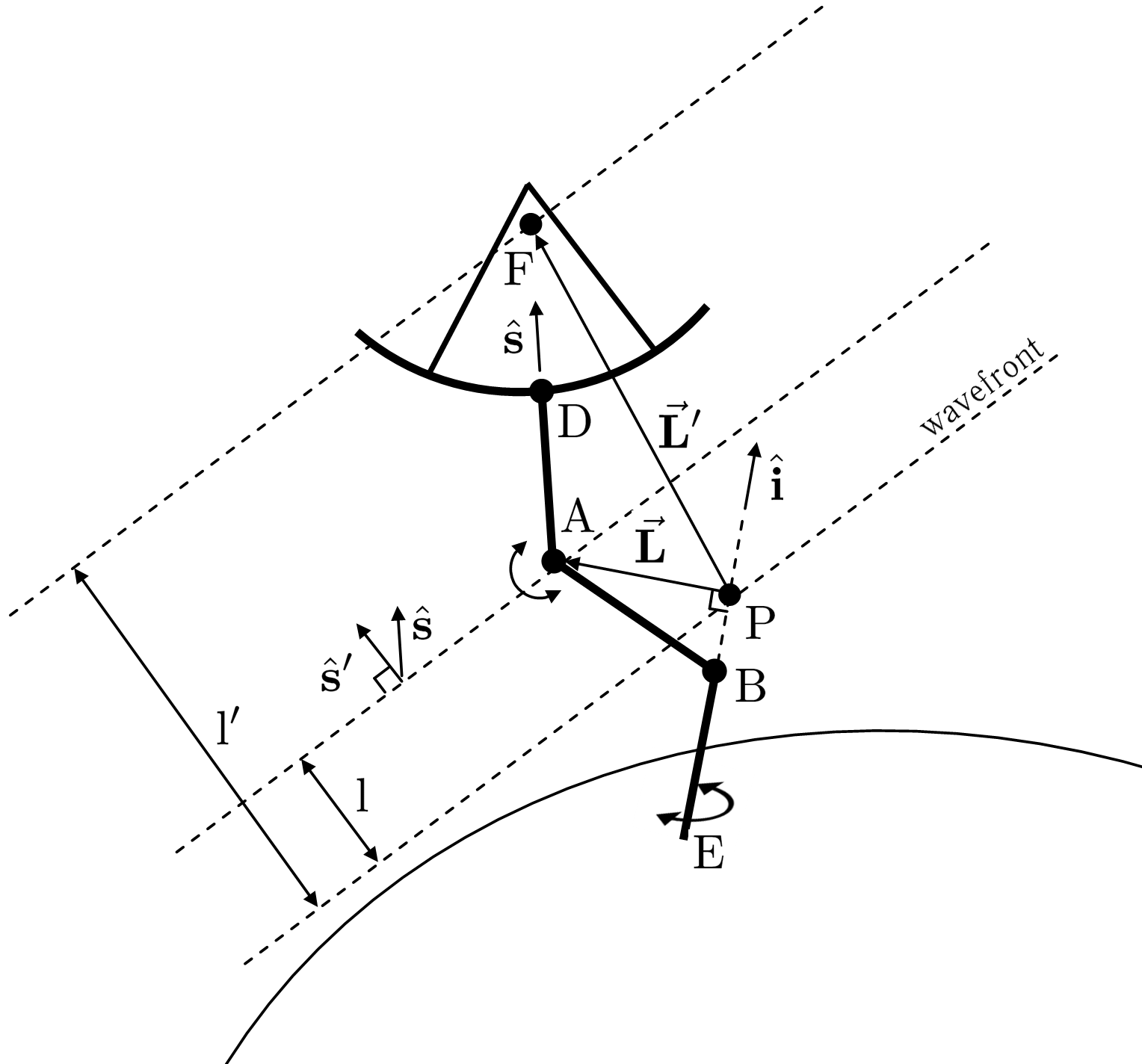}
        \caption{Generic mount of a ground steerable high-gain antenna. Dashed
lines mark the positions of the wavefront at the moments of its passing through
the antenna focus, $F$, antenna rotation axis $A$, and the antenna reference point, $P$.}
        \label{fig:generic-ground-antenna-mount}
\end{figure}

Let us now consider a generalized antenna structure that encompasses all the three antenna mounts described above (Fig.~\ref{fig:generic-ground-antenna-mount}). The antenna has two perpendicular rotation axes:
 $BE$, which is fixed relative to the Earth, and $A$ (end-view), which is
movable. The Earth-fixed antenna reference point, $P$, used to refer all the incoming/outgoing signals to, is usually chosen on axis $BE$ and defined
as its intersection with the plane that is perpendicular to $BE$ and contains axis $A$. If the
two axes intersect then $P$ is simply the intersection point of the axes. Let us further
introduce the unit vector $\hat{\mathbf{s}}$ along the antenna symmetry axis,
$AD$, the unit vector $\hat{\mathbf{i}}$ along axis $BE$, and the unit vector  $\hat{\mathbf{s}}'$ in the direction of propagation of the wavefront, which
we assume to be plane. (We ignore the wavefront curvature according to \citep{sovers-1998-rmp}.) Finally, we define the axis separation vector $\mathbf{L} = \overrightarrow{PA}$, denote the antenna focus by $F$, and introduce $\mathbf{L'} = \overrightarrow{PF}$.

The influence of the antenna motion on the parameters of received and
transmitted signals, such as phase, frequency, and time of arrival 
 can be easily understood from Fig.~\ref{fig:generic-ground-antenna-mount}. Without
loss of generality let us consider the case of the antenna receiving a signal from a SC and assume a prime focus antenna construction. While
the received signal characteristics are referred to the Earth-fixed reference point $P$, in 
reality the wavefront first reaches the movable antenna dish surface, then travels a fixed-length path to the focus, $F$, and then passes along  waveguides and cables, again of fixed length, to the data acqusition equipment. (As noted
above, we neglect the environmental effects.) The fixed-length parts of the signal
path in the antenna structure, waveguides, cabling, and equipment delays
can usually be neglected, both in VLBI and SC tracking, since they add constant offsets to the signal phase and delay and do not influence the frequency
\citep{sovers-1998-rmp}. It is the variation
of the distance between the positions of the wavefront at the moments when it passes through points $F$ and $P$, which we denote by $l'$, that gives a variable contribution to the measured signal characteristics. Obviously,
we have\begin{equation}
l' = \hat{\mathbf{s}}'\cdot \mathbf{L'},
\label{eq:new-expression-for-l}
\end{equation}
where $\mathbf{L'} = \overrightarrow{PF}$. Note that in general $\hat{\mathbf{s}}'$ does not lie in the plane of vectors $\hat{\mathbf{s}}$
and $\mathbf{L}$ since both of the antenna pointing angles, e.g. elevation and azimuth, may be in error.

Eq.~\eqref{eq:new-expression-for-l} gives a general expression for
the length of that part of the signal path which varies due to the antenna motion. The respective general expressions for the extra phase delay, $\tau'$, and the fractional frequency shift, $\Delta f / f$, are:
\begin{equation}
\tau' = \frac{l'}{c} = \frac{\hat{\mathbf{s}}'\cdot \mathbf{L'}}{c},
\label{eq:generic-expression-for-phase-delay}
\end{equation}
\begin{equation}
\frac{\Delta f}{f} = -\frac{1}{c} \frac{d}{dt} (\hat{\mathbf{s}}'\cdot \mathbf{L'}).
\label{eq:generic-expression-for-frac-freq-shift}
\end{equation}

Let us now consider the simplified case of negligible pointing errors:
\begin{equation}
\hat{\mathbf{s}} = \hat{\mathbf{s}}'.
\label{eq:s-equal-s-primed}
\end{equation}
In this case it is easy to see that Eqs.~\eqref{eq:generic-expression-for-phase-delay} and~\eqref{eq:generic-expression-for-frac-freq-shift} reduce to their unprimed analogues familiar from previous analyses~\citep{sovers-1998-rmp}:
\begin{equation}
\tau = \frac{l}{c} = \frac{\hat{\mathbf{s}}\cdot \mathbf{L}}{c},
\label{eq:prev-expression-for-phase-delay}
\end{equation}
\begin{equation}
\frac{\Delta f}{f} = -\frac{1}{c} \frac{d}{dt} (\hat{\mathbf{s}}\cdot \mathbf{L}),
\label{eq:prev-expression-for-frac-freq-shift}
\end{equation}
where $l$ is the distance between the wavefront positions at the
moments when it passes through axis $A$ and point $P$:
\begin{equation}
l = \hat{\mathbf{s}}\cdot \mathbf{L}.
\label{eq:old-expression-for-l}
\end{equation}
Indeed, if $\hat{\mathbf{s}} = \hat{\mathbf{s}}'$  we have:
\begin{equation}
\begin{split}
l' &= \hat{\mathbf{s}}\cdot \left( \mathbf{L} + \overrightarrow{AF}  \right)\\
   &= l + AF,
\label{eq:new-expression-for-l-reduced-to-old}
\end{split}
\end{equation}
so that $l$ and $l'$ differ only by a constant term of $AF$ which results
in a constant phase delay term in Eq.~\eqref{eq:generic-expression-for-phase-delay}
and does not contribute to the frequency shift of Eq.~\eqref{eq:generic-expression-for-frac-freq-shift}.

Simple geometric considerations lead to the following useful equation for the axis
separation vector, which is independent of the assumption of $\hat{\mathbf{s}} = \hat{\mathbf{s}}'$:
\begin{equation}
\mathbf{L} = \pm L\frac{\hat{\mathbf{i}}\times (\hat{\mathbf{s}}\times\hat{\mathbf{i}})}
{\big|\hat{\mathbf{i}}\times (\hat{\mathbf{s}}\times\hat{\mathbf{i}})\big|}.
\label{eq:axis-separation-vector}
\end{equation}
Here, $L=|\mathbf{L}|$ and the plus sign is used for ``positive'' offset
antennas (such that when $\hat{\mathbf{s}}$ and $\mathbf{L}$ are parallel or
antiparallel the antenna comes closer to the source as
$L$ increases, e.g. as in Fig.~\ref{fig:common-ground-antenna-mounts:a}) and minus for ``negative'' offset antennas (vice
versa, e.g. as in Fig.~\ref{fig:common-ground-antenna-mounts:b}). Let us return to the case of $\hat{\mathbf{s}} = \hat{\mathbf{s}}'$. Using Eqs.~\eqref{eq:old-expression-for-l} and \eqref{eq:axis-separation-vector}, as well as the geometric relation of
\begin{equation}
\hat{\mathbf{i}}\times (\hat{\mathbf{s}}\times\hat{\mathbf{i}})
= \hat{\mathbf{s}} - \hat{\mathbf{i}} (\hat{\mathbf{i}} \cdot \hat{\mathbf{s}}),
\label{eq:bac-cab}
\end{equation}
it is straightforward to obtain the following well-known equation for
\begin{equation}
l = \pm L\sqrt{1 - (\hat{\mathbf{s}}\cdot\hat{\mathbf{i}})^2} = \pm L \cos\theta,
\label{eq:old-expression-for-l-simplified}
\end{equation}
where we denoted
\begin{equation}
\theta = \frac{\pi}{2}-\angle(\hat{\mathbf{s}}, \hat{\mathbf{i}}).
\label{eq:theta-angle-def}
\end{equation} 
For the three antenna mounts discussed above, alt-az, polar, and X-Y, this angle is, respectively, elevation, declination, and auxiliary angle $Y$.

The assumption of $\hat{\mathbf{s}} = \hat{\mathbf{s}}'$ is usually valid to a very good degree for the case of tracking celestial objects with well known coordinates. However, it can easily fail when tracking SC. Indeed, the accuracy of predicted orbits used for tracking SVLBI spacecraft can be as low as 5~km for each component of the position vector, which corresponds to  $\angle(\hat{\mathbf{s}},
\hat{\mathbf{s}}')\sim17'$ for the distance to the SC of $\sim1,000$~km \citep{zakhvatkin-2020-asr}. When a ground TS is receiving signals sent by a SC, its antenna operators usually
  can partially compensate for the inaccuracy of the predicted orbit by applying constant pointing offsets that maximize
 the level of the received signal. However, this does not eliminate the pointing error completely (e.g. when it changes
 during the communication session) and such procedure cannot be performed when the TS is transmitting signals to the
SC. It seems reasonable to assume that the pointing error
is at least of the order that corresponds to the accuracy of the reconstructed
SC orbit. For RadioAstron the accuracy of the reconstructed
orbit near perigee is $\sim 100$~m for each
component of the position vector, which corresponds to $\angle(\hat{\mathbf{s}},
\hat{\mathbf{s}}') \sim 20''$ at the distance  to the SC of $\sim1,000$~km (see Section~\ref{sec:radioastron:description}). This pointing error is usually larger for the SC-mounted antenna since there is no operator on board
to apply pointing corrections and also because predicted
orbits  used by spacecraft are often uploaded to the on-board computer a few
days in advance of the communication sessions and thus are less up-to-date and less accurate than those used by ground antennas.
Moreover, at least for RadioAstron, pointing angles for the onboard antenna are computed on the fly by
the on-board computer based on simplified models of the spacecraft motion and signal propagation.

Now, let us consider the  practically important case when 
\begin{equation}
\hat{\mathbf{s}} \ne \hat{\mathbf{s}}'
\label{eq:s-not-equal-s-prime}
\end{equation}
but  the pointing errors are small. If we introduce the pointing error
vector of
\begin{equation}
\delta\mathbf{s} = \hat{\mathbf{s}}' - \hat{\mathbf{s}},
\end{equation}
the assumption of small pointing errors can be stated as:
\begin{equation}
|\delta\mathbf{s}| \ll 1.
\label{eq:assumption-of-small-pointing-errors}
\end{equation}
Note that, since $|\hat{\mathbf{s}}'| = |\hat{\mathbf{s}}| = 1$, we have:
\begin{equation}
\begin{split}
\delta\mathbf{s}\cdot \hat{\mathbf{s}}' = O(\delta\mathbf{s}^2), \\
\delta\mathbf{s}\cdot \hat{\mathbf{s}} = O(\delta\mathbf{s}^2).
\end{split}
\label{eq:dot-product-of-s-with-delta-s}
\end{equation}
Substituting $\hat{\mathbf{s}} = \hat{\mathbf{s}}' - \delta\mathbf{s}$ into Eq.~\eqref{eq:new-expression-for-l} and expanding it in powers of $\delta\mathbf{s}$, and also using Eq.~\eqref{eq:dot-product-of-s-with-delta-s} and $\mathbf{L'}
= \mathbf{L} + \overrightarrow{AF}$, it is straightforward to obtain:
\begin{equation}
\begin{split}
l' &= \pm L \cos\theta' + O(\delta\mathbf{s}^2) \\ 
  &= \pm L \cos\theta' + O(\delta\mathbf{\theta}^2),
\label{eq:new-expression-for-l-simplified}
\end{split}
\end{equation}
where 
\begin{equation}
\theta' = \frac{\pi}{2}-\angle(\hat{\mathbf{s}}', \hat{\mathbf{i}})
\label{eq:theta-prime-angle-def}
\end{equation} 
is the ``true'' elevation, declination, or auxiliary angle $Y$ of the SC
relative to the ground antenna (respectively, for the alt-az, polar, and X-Y mounts) and $\delta\theta$ is the error in this angle: 
\begin{equation}
\delta\theta = \theta' - \theta.
\label{eq:pointing-error-angle}
\end{equation}
Although Eqs.~\eqref{eq:old-expression-for-l-simplified}
and \eqref{eq:new-expression-for-l-simplified} formally look similar, their
meaning is different. In Eq.~\eqref{eq:old-expression-for-l-simplified}
$\theta$ is the angle determined by the antenna pointing
direction, $\hat{\mathbf{s}}$, while in Eq.~\eqref{eq:new-expression-for-l-simplified}
$\theta'$ is determined by the actual source/target position relative
to the antenna, $\hat{\mathbf{s}}'$. Also, while Eq.~\eqref{eq:old-expression-for-l-simplified}
is exact, Eq.~\eqref{eq:new-expression-for-l-simplified} is valid only for small pointing
errors.

The expression for the fractional frequency shift due to the APCM effect can be
obtained from
Eq.~\eqref{eq:new-expression-for-l-simplified} using
Eq.~\eqref{eq:generic-expression-for-frac-freq-shift} and assuming the rate of change of the $O(\delta\mathbf{\theta}^2)$ term is small, i.e. $\frac{d}{dt} O(\delta\mathbf{\theta}^2) = O(\delta\mathbf{\theta}^2)$.
Thus we have: 
\begin{equation}
\frac{\Delta f}{f} = \pm \frac{L}{c} \dot\theta'\sin\theta' + O(\delta\mathbf{\theta}^2),
\label{eq:generic-expression-for-ground-frac-freq-shift}
\end{equation}
where the plus sign is for ``positive'' offset antennas and minus for the ``negative'' (as above).

The equations for the extra phase delay and fractional frequency shift for the three common antenna mounts are summarized in Table~\ref{table:apcme-equations-for-three-antenna-mounts}.
\begin{table*}[t]              
\small
\renewcommand{\arraystretch}{2.0}
        \centering
        \begin{tabular}{|c|c|c|c|} \hline
                Antenna mount& Secondary angle & Delay correction, $\tau'$  & Frequency correction, $\Delta f/f$ \\ \hline
                alt-az & elevation $ \gamma' $ & $ \pm (L/c) \cos{\gamma'} $ & $\pm\;\;\;(L/c) \; \dot\gamma' \sin\gamma'$ \\
                polar & declination $ \delta' $ & $ \pm (L/c) \,  \cos{\delta'} $ & $\pm\;\;\;(L/c) \; \dot\delta' \sin\delta'$ \\
                X-Y & auxiliary angle $ Y' $ & $  \pm (L/c) \,  \cos{Y'} $ & $\pm\;\;\;(L/c) \; \dot Y' \sin Y'$ \\ \hline
        \end{tabular}
        \captionof{table}{Equations to compute the antenna phase center motion effect
for common antenna mounts. The upper sign is for ``positive'' axis offset antennas while the lower one, respectively, for the negative  (see the text for details). }
        \label{table:apcme-equations-for-three-antenna-mounts}
\end{table*}

\subsection{Spaceborne antennas}

\begin{figure}[H]              
        \centering
        \includegraphics[scale=0.4]{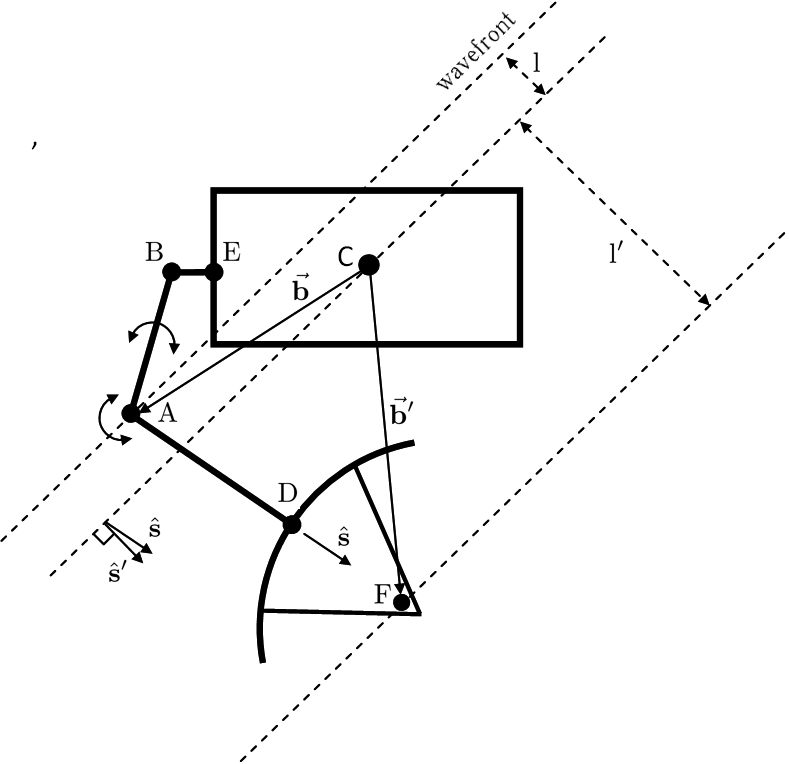}
        \caption{Generic spacecraft steerable high-gain antenna mount. Dashed
lines mark the positions of the wavefront at the moments of its passing through
the antenna focus, $F$, the antenna axis intersection point, $A$, and the spacecraft center of mass, $C$, which also serves as the antenna reference point.}
        \label{fig:generic-sc-antenna-mount}
\end{figure}

Let us now consider SC-mounted antennas. The equations for the APCM effect for spaceborne antennas are obtained relatively straightforwardly and in close analogy to the case of ground antennas. The description of the
SC motion is usually given in terms of the position of its center of
mass, point $C$ on Fig.~\ref{fig:generic-sc-antenna-mount}, and its attitude in a selected reference frame. For Earth-orbiting SC an Earth-centered
inertial reference frame is usually used, e.g. EME2000. The center of mass, $C$, also serves as the antenna reference point, i.e. the characteristics of the signals received and/or transmitted
by the antenna are referred to this point. An insignificant difference from
ground antenna mounts is that in this case it is $AB$ that usually serves
as one of the two antenna axes instead of $BE$. The other axis is oriented normal to the plane of Fig.~\ref{fig:generic-sc-antenna-mount} and located at the axis intersection point $A$.

Now, we define the unit vector $\hat{\mathbf{s}}'$
in the direction of the source/target, i.e. normal to the wavefront (which we again consider to be plane), the unit
vector $\hat{\mathbf{s}}$ in the direction of the antenna symmetry axis, $AD$,
the displacement vector of the antenna primary focus from the center of mass, $\overrightarrow{CF}=\mathbf{b'}$, and a similar displacement vector for the axes intersection point, $A$: $\overrightarrow{CA}=\mathbf{b}$.
In analogy to the ground case, the APCM effect is determined by the variable
distance
between the positions of the wavefront at the moments of its passing through the antenna focus, $F$, and the reference point, $C$:
\begin{equation}
l' = \mathbf{b'} \cdot \hat{\mathbf{s}}'.
\label{eq:space-apcme-for-path}
\end{equation}

Also similar to the ground case, when the pointing errors can be neglected, i.e. $\hat{\mathbf{s}}'=\hat{\mathbf{s}}$, we can use $\mathbf{b'} = \mathbf{b} + \overrightarrow{AF}$ and $\overrightarrow{AF} = AF\cdot\hat{\mathbf{s}}$ to obtain:
\begin{equation}
l' = l + \mathrm{const},
\label{eq:space-apcme-for-path-when-no-pointing-error}
\end{equation}
where $l$ is the distance travelled by the wavefront between its positions when passing through the axis intersection point, $A$, and the center of mass, $C$:
\begin{equation}
l = \mathbf{b}\cdot \hat{\mathbf{s}}.
\label{eq:sc-antenna-axis-offset-from-com}
\end{equation}

If there are small pointing errors:
\begin{equation}
|\delta\mathbf{s}| = |\hat{\mathbf{s}}' - \hat{\mathbf{s}}| \ll 1,
\end{equation}
it is straightforward to show that Eq.~\eqref{eq:space-apcme-for-path}, again in analogy with the ground case, simplifies to: 
\begin{equation}
l' = \mathbf{b} \cdot \hat{\mathbf{s}}'  + \mathrm{const} + O(\delta\mathbf{s}^2).
\label{eq:space-apcme-for-path-when-pointing-error-is-small}
\end{equation}
The advantage of this equation over Eq.~\eqref{eq:space-apcme-for-path}
is that vector $\mathbf{b}$ is fixed in the spacecraft reference frame while
vector $\mathbf{b'}$ moves according to the antenna motion.

The generic expression for the fractional frequency shift of the signal received or transmitted by the SC antenna can be obtained from Eq.~\eqref{eq:space-apcme-for-path} using Eq.~\eqref{eq:generic-expression-for-frac-freq-shift}:
\begin{equation}
\frac{\Delta f}{f} = -\frac{1}{c} \frac{d}{dt} (\mathbf{b'}\cdot \hat{\mathbf{s}}').
\label{eq:sc-apcme-frac-freq-shift-generic}
\end{equation}
For small pointing errors we have:
\begin{equation}
\frac{\Delta f}{f} = -\frac{1}{c} \frac{d}{dt} (\mathbf{b}\cdot \hat{\mathbf{s}}')
 + O(\delta\mathbf{s}^2),
\label{eq:sc-apcme-frac-freq-shift-for-small-pointing-errors}
\end{equation}
where we again assumed that the rate of change of $O(\delta\mathbf{s}^2)$ terms
in Eq.~\eqref{eq:space-apcme-for-path-when-pointing-error-is-small} is small.
In many cases, e.g. in SVLBI, the spacecraft maintains constant attitude in an inertial reference frame when
communicating with a TS. In this case vector
$\mathbf{b}$ is constant in an inertial reference frame and Eq.~\eqref{eq:sc-apcme-frac-freq-shift-for-small-pointing-errors}
further simplifies to:
\begin{equation}
\frac{\Delta f}{f} = -\frac{1}{c}  (\mathbf{b}\cdot \dot{\hat{\mathbf{s}}}')
 + O(\delta\mathbf{s}^2).
\label{eq:sc-apcme-frac-freq-shift-for-small-pointing-errors-and-constant-b}
\end{equation}

For numerical computations of the APCM effect, as well as its uncertainties, it is useful to specify vector $\mathbf{b}$ in the SC-fixed reference frame, which we denote by $\mathbf{b}_\mathrm{sc}$, while vector $\hat{\mathbf{s}}'$ is more naturally defined in the inertial reference frame. Denoting the transformation matrix from the SC-fixed to the inertial frame by $\mathrm{R}$, we obtain the following equation for the APCM effect due to the SC antenna (assuming a constant SC attitude in the inertial frame):
\begin{equation}
\frac{\Delta f}{f} = -\frac{1}{c}  (\mathrm{R} \mathbf{b}_\mathrm{sc} \cdot \dot{\hat{\mathbf{s}}}')
 + O(\delta\mathbf{s}^2).
\label{eq:sc-apcme-frac-freq-shift-for-small-pointing-errors-and-constant-b-with-transformation-matrix}
\end{equation}

\subsection{Error analysis}
\label{sec:error-analysis}
In high-accuracy experiments we are interested not only in the magnitude
of the APCM effect but also in estimating the errors in its computed values. The errors can be assessed using equations obtained above by varying them with respect to the particular parameters. For example, the error in the fractional frequency
shift of Eq.~\eqref{eq:sc-apcme-frac-freq-shift-for-small-pointing-errors-and-constant-b-with-transformation-matrix}
due to an error, $\delta\mathbf{b}_\mathrm{sc}$, in the position of the fixed SC antenna axis relative to the SC center of mass is:
\begin{equation}
\delta\left(\frac{\Delta f}{f}\right) = -\frac{1}{c}  ((\mathrm{R}\, \delta\mathbf{b}_\mathrm{sc}) \cdot \dot{\hat{\mathbf{s}}}')
 + O(\delta\mathbf{s}^2).
\label{eq:error-in-sc-apcme-frac-freq-shift}
\end{equation}
Then, for the uncertainty in the fractional frequency shift due to the APCM effect, which we will characterize by the standard deviation, we have:
\begin{equation}
\sigma_{\Delta f/f} = 
\left(\frac{1}{c^2}
(\dot{\hat{\mathbf{s}}}')^\mathrm{T}
\mathrm{R}\Sigma_b\mathrm{R}^\mathrm{T}
\dot{\hat{\mathbf{s}}}' + ...\right)^{1/2}
 + O(\delta\mathbf{s}^2),
\label{eq:uncertainty-in-sc-apcme-frac-freq-shift}
\end{equation}

\begin{table*}[t]              
\small
\centering
\begin{threeparttable}
\renewcommand{\arraystretch}{1.5}
        \centering
        \begin{tabular}{| p{8.1cm} |>{\centering\arraybackslash}m{2.0cm}|c| >{\centering\arraybackslash}m{2.8cm} |} \hline

                {Error source} & Affected parameter  & RadioAstron & Possible follow-up SVLBI mission \\  \hline
                
\renewcommand{\arraystretch}{2.0}
                Uncertainty in the offset between the ground antenna axes & $L$ & 0.002 m
                & 0.002 m\\
                
                
                Ground antenna axis misalignment  & $\hat{\mathbf{i}}$ & 5$'$ & 5$'$  \\

                Uncertainty in the position of the intersection point of the SC antenna axes relative to the SC center of mass & $\mathbf{b}_\mathrm{sc}$ & 0.005 m & 0.001 m \\
                
                Uncertainty in the SC attitude  & $\mathrm{R}$ & 10$''$ & 1$''$  \\
                
                Uncertainty in the TS-to-SC direction due to the SC position uncertainty
                 & $\hat{\mathbf{s}}$ & 100m $\rightarrow$ 20$''$ & 10m $\rightarrow$
                 2$''$  \\
                
 \hline
        \end{tabular}
        
        \captionof{table}{Uncertainties in the parameters that affect the computed phase and frequency shift due to the antenna phase center motion effect. Two sets of values are given: one is relevant for RadioAstron and the other is a set of tentative values
assumed for a possible follow-up SVLBI mission. For the unit vector along the ground antenna fixed axis, $\hat{\mathbf{i}}$,
and the unit vector in the TS-to-SC direction, $\hat{\mathbf{s}}$, the uncertainties in each of the two angles that define the orientation of these vectors are specified. The uncertainties in the direction of $\hat{\mathbf{s}}$ are computed from the
specified average uncertainties in the components of the SC position vector, assuming a distance to the TS of 1,000~km. For the on-board antenna axis position vector, $\mathbf{b}$, the uncertainty in each of its three components is specified. For the transformation matrix, $\mathrm{R}$, the uncertainty in each of the three angles that define its components are specified. In all cases the specified uncertainties in vector components and angles are assumed independent.}
        \label{table:errors}
\end{threeparttable}
\end{table*}

\noindent
where $\Sigma_b$ is the variance-covariance matrix of the components of vector $\mathbf{b}_\mathrm{sc}$ and the ellipsis denotes similar terms due to other parameters, which we assumed to be independent from $\mathbf{b}_\mathrm{sc}$.

The rate of change of the TS-to-SC direction vector, $\hat{\mathbf{s}}'$, significantly depends on the orbit parameters, the TS location, and the SC position on the orbit. Therefore, an error analysis of the APCM effect should be performed for each mission individually, taking into account its configuration and the particular values of uncertainties in the parameters. Moreover, it is necessary to characterize $\sigma_{\Delta f/f}$ not with a single, e.g. maximum, value but as a function of the SC position on its orbit.

In the next two sections we outline the results of two such error analyses based on Eq.~\eqref{eq:uncertainty-in-sc-apcme-frac-freq-shift} and a similar equation for
the uncertainty in the ground APCM effect (which can be straightforwardly obtained from Eq.~\eqref{eq:generic-expression-for-frac-freq-shift} or \eqref{eq:generic-expression-for-ground-frac-freq-shift}): one for the RadioAstron SC and another for a SC of a possible follow-up SVLBI mission. Section~\ref{sec:radioastron} can also serve as a rather detailed account of the part of the error budget of the RadioAstron gravitational redshift experiment \citep{litvinov-2018-pla,nunes-2020-asr} which is relevant to the APCM effect. For simplicity we consider only some of the possible error sources, see Table~\ref{table:errors}, and treat each of the listed parameter as independent. We also assume each component of the vector parameters to be independent. The particular values given in Table~\ref{table:errors} are commented on in the next Section.

\section{The APCM effect in the RadioAstron SVLBI mission}
\label{sec:radioastron}
\nopagebreak

\subsection{The RadioAstron spacecraft}
\begin{figure}[H]              
        \centering
        \includegraphics[scale=0.3]{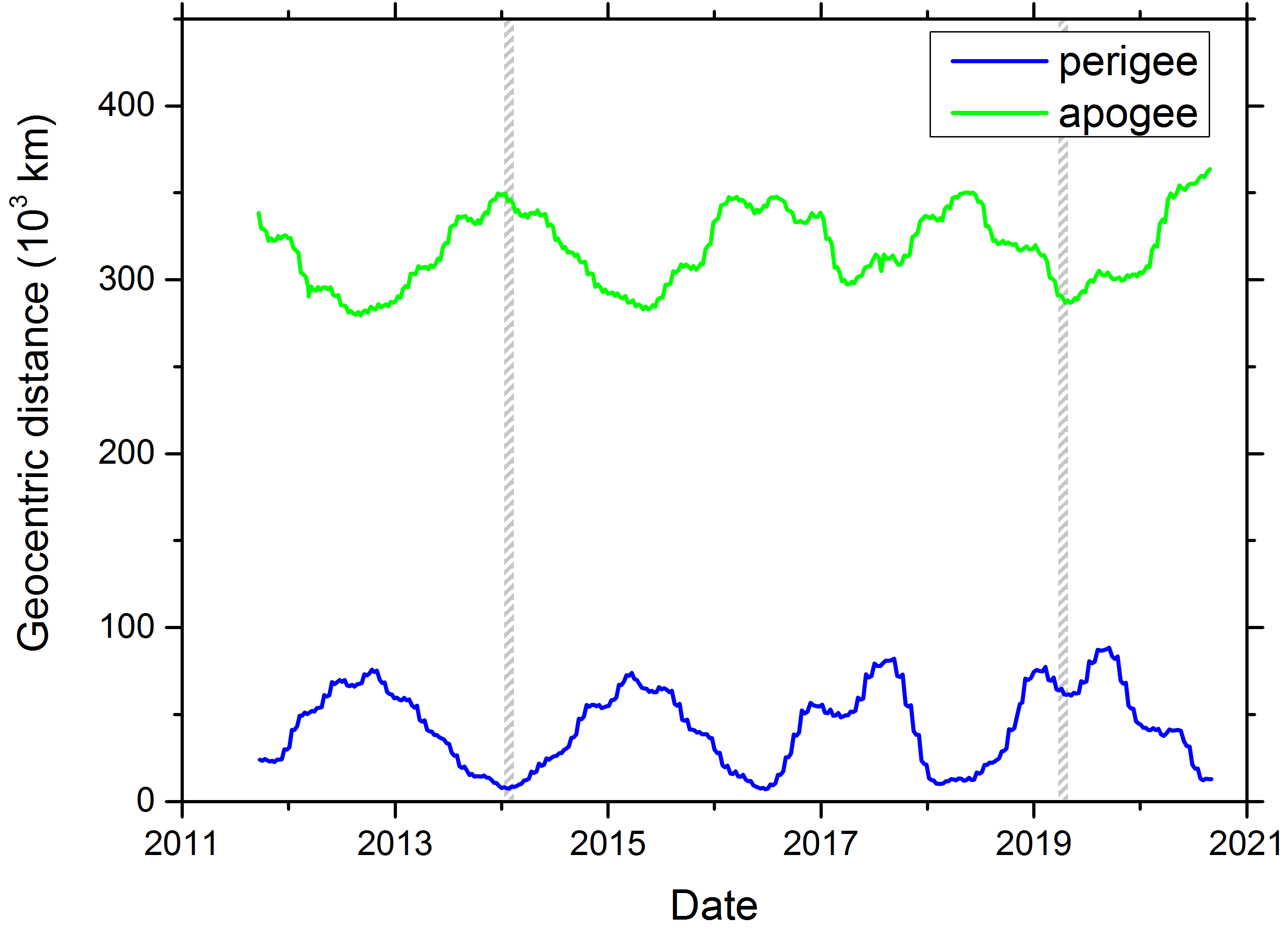}
        \caption{Evolution of the apogee and perigee of the RadioAstron spacecraft.
The evolution of the orbit is caused largely by the gravitational interaction with the Moon.
The two segments used in the analyses of Sections 3.3 and 5 are hatched.}
        \label{fig:radioastron-apogee-perigee-evolution}
\end{figure}

\begin{figure}[H]              
        \centering
        \includegraphics[scale=0.3]{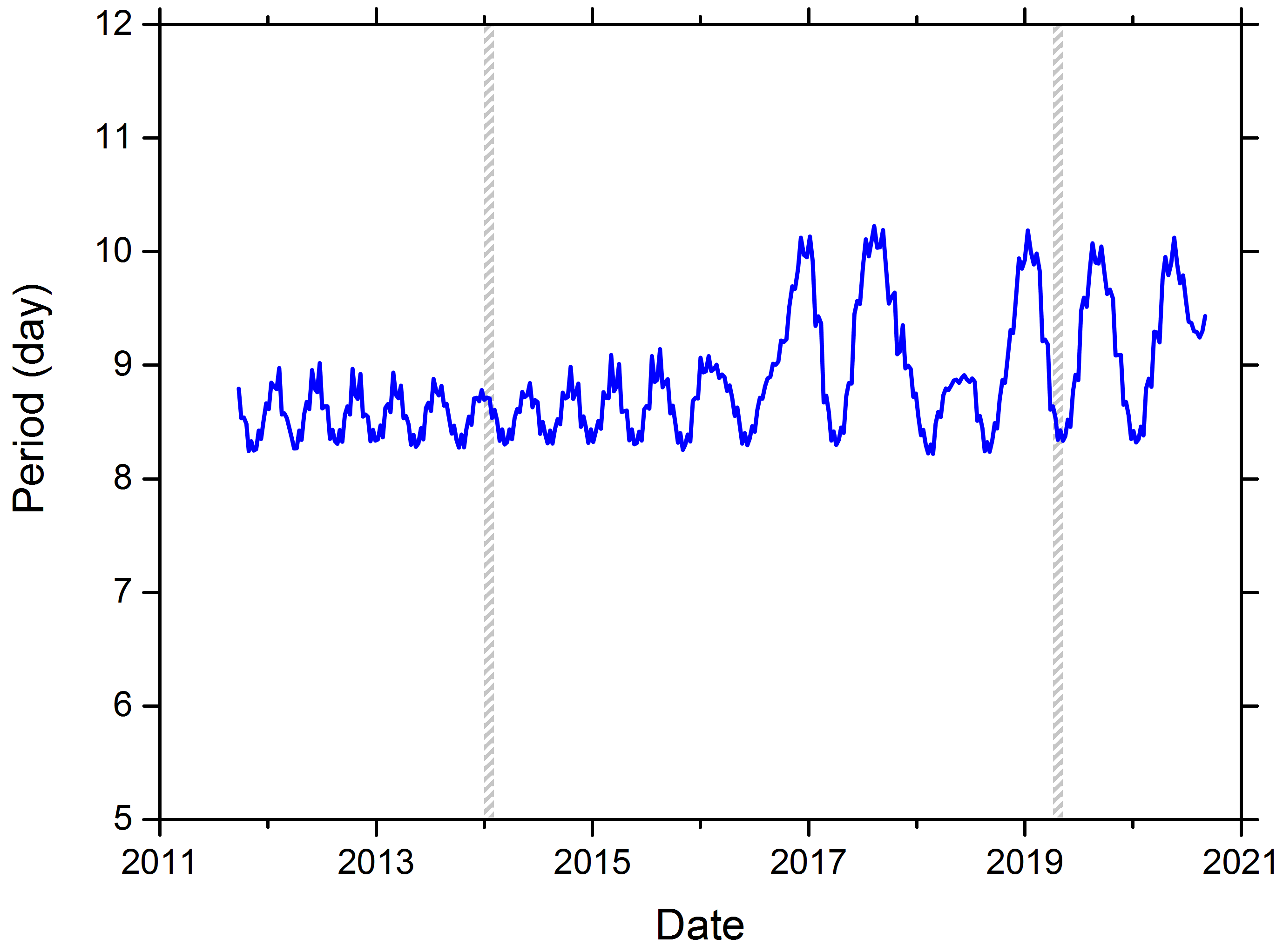}
        \caption{Evolution of the period of the RadioAstron spacecraft. The irregular low spike in 2018 is caused by an orbital maneuver applied to avoid colliding with the Earth. The two segments used in the analyses of Sections 3.3 and 5 are hatched.}
        \label{fig:radioastron-period-evolution}
\end{figure}

The RadioAstron spacecraft served the RadioAstron SVLBI mission  from 2011 to 2019 and helped astronomers observe various astrophysical objects at the highest angular resolution to date \citep{kardashev-2013-ar,johnson-2016-apj,giovannini-2018-nata,kravchenko-2020-apj}. In 2019 the spacecraft stopped responding to commands and the observational part of the mission ended. The satellite is on a highly eccentric orbit around the Earth which was designed to evolve significantly throughout
the mission under the gravitational influence of the Moon, as well as other factors, within a broad range of the orbital parameter space (perigee
altitude 1,000--82,000 km, apogee altitude 273,000--357,000~km, period 8.2--10.2~day, Figs.~\ref{fig:radioastron-apogee-perigee-evolution} and~\ref{fig:radioastron-period-evolution}).

The SC is equipped with five
antennas: 1) one 10-meter
high-gain parabolic spacecraft-fixed antenna used for observing celestial sources; 2) one 1.5m high-gain parabolic mechanically steerable antenna for a) transmitting
high-bit-rate observational data to the Earth, b) transmitting the highly stable signal of the on-board hydrogen maser frequency standard to the Earth for the purpose of Doppler tracking, and c) receiving a stable signal from an
Earth-based hydrogen maser for use as a reference on board (backup mode); 3) three omnidirectional
antennas for transmitting telemetry, receiving commands, Doppler and range tracking. The antenna
of interest to us is the 1.5~m high-gain steerable antenna (Fig.~\ref{fig:radioastron-spacecraft-drawing}).

This antenna communicated with the mission's two ground TS that were used
to collect science data, receive one- and two-way Doppler, and also to
provide uplink reference signal for the spacecraft. The mission's two tracking stations were the Pushchino TS
of the Pushchino Radio Astronomy Observatory (Moscow region, Russia) with
its 22m alt-az mount radio telescope RT-22 and the Green Bank Earth Station
of the National Radio Astronomy Observatory (West Virginia, US) with its 43-meter NRAO140 polar mount radio telescope.
The parameters of these two antennas are given in Table~\ref{table:ra-tracking-stations}.

\subsection{Description of the data, error sources, and data processing}
\label{sec:radioastron:description}

According to the equations derived in Section~\ref{sec:theory}, to compute the APCM effect both for the ground and space antennas the following data are needed: the SC orbit,  the position
of the SC-fixed antenna axis relative to its center of mass, the SC
attitude as a function of time,  the
coordinates of the ground an- tenna, \, and \, its \, axis \, offset. \, The information \, on \, the \, actual

\begin{figure}[H]              
        \centering
        \includegraphics[scale=0.2]{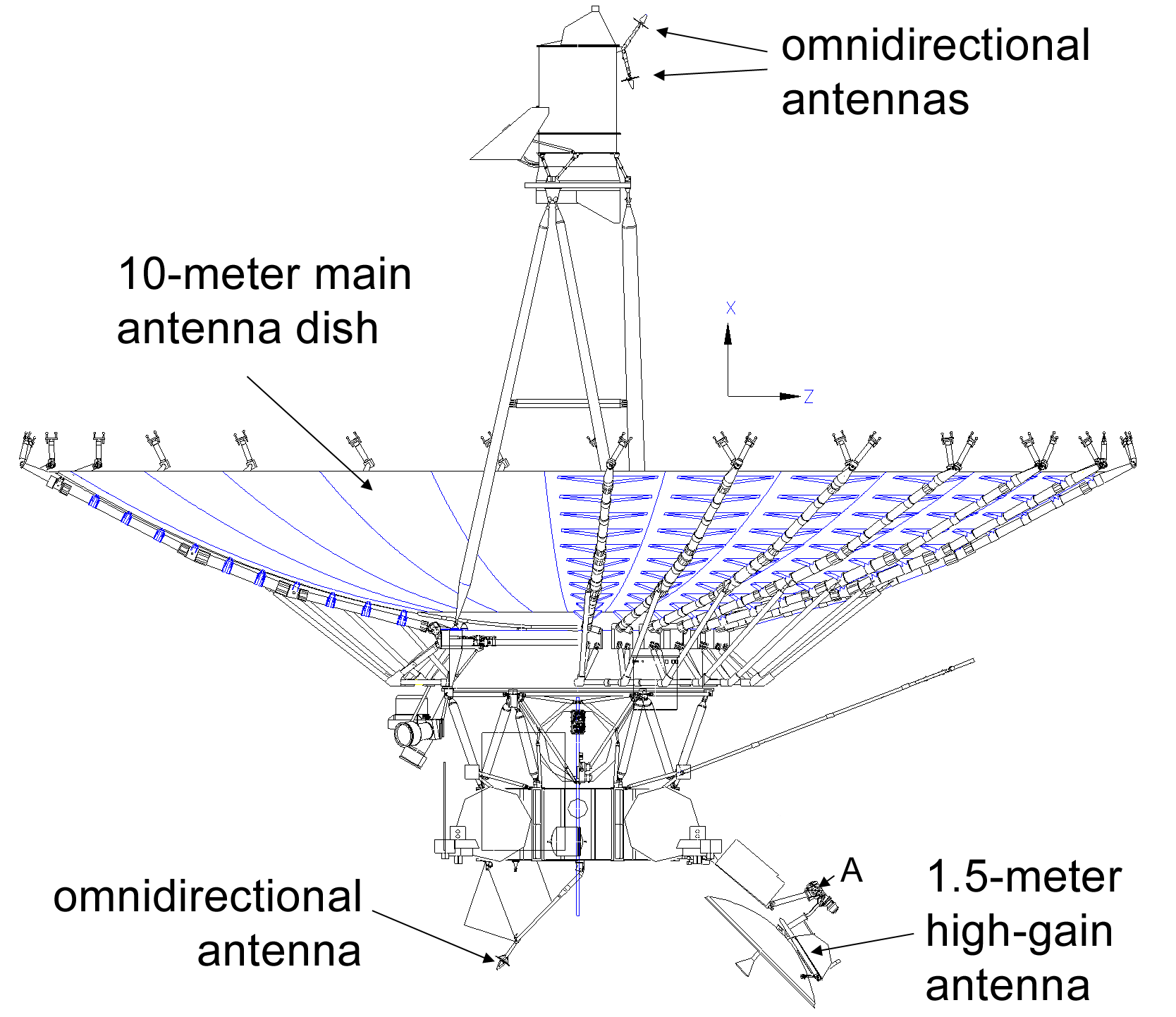}
        \caption{A technical drawing of the RadioAstron spacecraft \citep{fedorchuk-2014-cosres}.
The spacecraft is equipped with several antennas, the largest of which is
the 10-meter spacecraft-fixed parabolic dish used in astronomical observations.
The antenna of interest to us is the 1.5-meter high-gain mechanically
steerable antenna. Point $A$ marks the intersection of the rotation axes of this antenna.}
        \label{fig:radioastron-spacecraft-drawing}
\end{figure}

\noindent
pointing of the antennas during communication sessions is not needed as long as pointing errors are small.

In our analysis we use long-term predicted orbits provided for the RadioAstron mission by the ballistic
center of the Keldysh Institute for Applied Mathematics (KIAM) \citep{zakhvatkin-2020-asr}. Orbits of this type were produced primarily for the purpose of long-term planning of the mission operations. For a prediction time of about a month the average uncertainties in each component of the SC position and velocity vectors of such orbits can reach, correspondingly, 10 km and several cm/s. (These uncertainties are distributed unevenly between the vector components along and across the line of sight.) The KIAM also provides significantly more accurate aposteriori reconstructed orbits with average uncertainties in each component of the position and velocity vectors of $\sim$200~m and 2--3~mm/s, correspondingly. (These uncertainties are distributed relatively evenly between the vector components and also depend on the distance to the SC: near perigees the uncertainty in the SC position is usually several times lower than near apogees while that in the velocity is correspondingly larger.) In this paper we do not use reconstructed orbits to compute the APCM effect because orbits of this type are available only for the time segments when observations were performed and also because the accuracy of long-term predicted orbits is sufficient for our illustrative purposes. However, since in the data processing of experiments performed with RadioAstron one uses reconstructed orbits, we estimate the uncertainty in the TS-to-SC direction, $\hat{\mathbf{s}}'$, using the SC position uncertainty relevant to reconstructed orbits. For simplicity, we use a single value for the uncertainties in each of the two angles that determine the TS-to-SC direction and estimate it using the average uncertainty in the SC position that is relevant to perigees, i.e. 100~m (see Table~\ref{table:errors}).

\begin{Table}     
        \small
        \centering
\renewcommand{\arraystretch}{1.5}
        \begin{tabular}{|l|c|c|} \hline
          & Pushchino & Green Bank \\ \hline
        Location & Moscow region, Russia & West Virginia, USA \\ \hline
        Latitude & 54$^{\circ}$49.0$'$14.24 & +38$^{\circ}$26$'$16.166$'$$'$\\ \hline
        Longitude & 37$^{\circ}$37$'$41.84$'$$'$ & -79$^{\circ}$50$'$08.810$'$$'$\\ \hline
        Height & 239.09 m & 812.50 m\\ \hline
        Dish diameter & 22 m & 43 m \\ \hline
        Mount type & alt-az & polar\\ \hline
        Axis offset & 0.0 m & 14.94 m\\ \hline
        \end{tabular}
        \captionof{table}{Parameters of the antennas of the two tracking stations of the RadioAstron mission.}
        \label{table:ra-tracking-stations}
\end{Table}

In order to compute the components of the displacement of the intersection point of the SC rotation axes off the SC center of mass, $\mathbf{b}$, we determined the coordinates of that point from a high-resolution technical drawing version of Fig.~\ref{fig:radioastron-spacecraft-drawing}, while the position of the SC center of mass was obtained from internal technical documentation of the RadioAstron mission. The location of the center of mass depends on the amount of fuel in the tank, which for simplicity we assumed to be full. Thus we obtained:
\begin{equation}
\mathbf{b}_\mathrm{sc} = \left[ -2.299; \: 0; \: 2.546 \right]~\mathrm{m}.
\label{eq:radioastron-vector-b}
\end{equation}
The rigidity properties of the antenna mount construction and the accuracy of the position of the center of mass make it reasonable to attribute an uncertainty of $\sim5$~mm to each component of vector $\mathbf{b}_\mathrm{sc}$ (Table~\ref{table:errors}).

Eq.~\eqref{eq:radioastron-vector-b} gives the components of vector $\mathbf{b}$
in the SC-fixed reference frame. In order to transform them to an inertial
reference frame using the transformation matrix $\mathrm{R}$, see Eq.~\eqref{eq:sc-apcme-frac-freq-shift-for-small-pointing-errors-and-constant-b-with-transformation-matrix}, the knowledge of the SC attitude as a function of time is required. The SC attitude during a particular observation almost always was maintained constant but it differed from one observation to another (the duration of an observation was usually of order of an hour). For the actual data processing the information on the SC attitude is obtained from the telemetry data provided by the on-board attitude control system. The accuracy of these data, obtained mostly from RadioAstron's star trackers, corresponds to the uncertainty in each of the three angles that determine the SC orientation of $\sim 10''$ (Table~\ref{table:errors}). For our illustrative purposes we assume the SC always maintains a constant orientation in the inertial space, such that the transformation between the two reference frames is represented by the identity matrix, $\mathrm{R} = 1$. However, we allow for an error in the realization of this orientation by the attitude control system, $\delta\mathrm{R}$, and estimate it using the above uncertainties in the orientation angles (assumed independent from each other).

The geodetic (ITRS) coordinates
of the reference point of the Pushchino RT-22 antenna and its axis offset
were obtained from internal documentation of the RadioAstron mission, while those of the NRAO140 antenna from \citep{langston-2012-nrao-memo}. The uncertainties
in the antenna reference point coordinates are of  order of a cm and thus give negligible
contribution to the uncertainty in the TS-to-SC direction compared to that
of the SC position. The uncertainties in the axis offsets were
assumed to be 2~mm based on the difference between the value provided in
\citep{langston-2012-nrao-memo} and that obtained from data processing of global geodetic VLBI campaigns \citep{petrov-2009-url}.

Finally, we take into account the possibility for a small misalignment of the Earth-fixed axis
of the ground antenna from its intended direction. For example, for an alt-az
antenna its azimuth axis may be slightly offset from the true local zenith, for a polar mount antenna its polar axis may not point exactly along the Earth's rotation
axis. This error is reflected by an error in the direction of vector $\hat{\mathbf{i}}$ for which we assume a value of 5$'$. This value is very tentative and corresponds to the misalignment of the electrical and mechanical axes of the NRAO140 antenna \citep{mezger-1966-nrao-memo}. Antenna pointing calibrations can probably reduce it by at least half an order of magnitude.

The computation of the APCM effect for the ground and space antennas of the RadioAstron
mission involves direct application of equations of Section~\ref{sec:theory}.
The errors are estimated according to Section~\ref{sec:error-analysis}.

\subsection{Results}

Due to the highly evolving character of RadioAstron's orbit, we selected two distinct epochs for our analysis: a low-perigee epoch of January 2014 and a high-perigee epoch of April 2019. Some of the orbital parameters
relevant to these two epochs, which we denote, respectively, A and B, are given in Table~\ref{table:ra-orbital-params-for-two-epochs}.  

\begin{Table}
\small
\renewcommand{\arraystretch}{1.5}
        \centering
        \begin{tabular}{|l|r|r|} \hline
                 & Epoch A: Jan 2014 & Epoch B: Apr 2019 \\ \hline
                Perigee (km)& 7,361 & 64,745 \\ \hline
                Apogee (km)& 345,060 & 286,804 \\ \hline
                Period (day)& 8.6 & 8.5 \\ \hline
                Eccentricity & 0.96 & 0.63 \\ \hline
        \end{tabular}
        \captionof{table}{The orbital parameters of the RadioAstron spacecraft
        for the two selected epochs of low (A) and high (B) perigee. }
        \label{table:ra-orbital-params-for-two-epochs}
\end{Table}

\begin{figure*}[!h]              
        \centering
        \subfloat[]{
                \label{fig:ra-2014-jan-gb:a}
                \includegraphics[clip, scale=0.31]{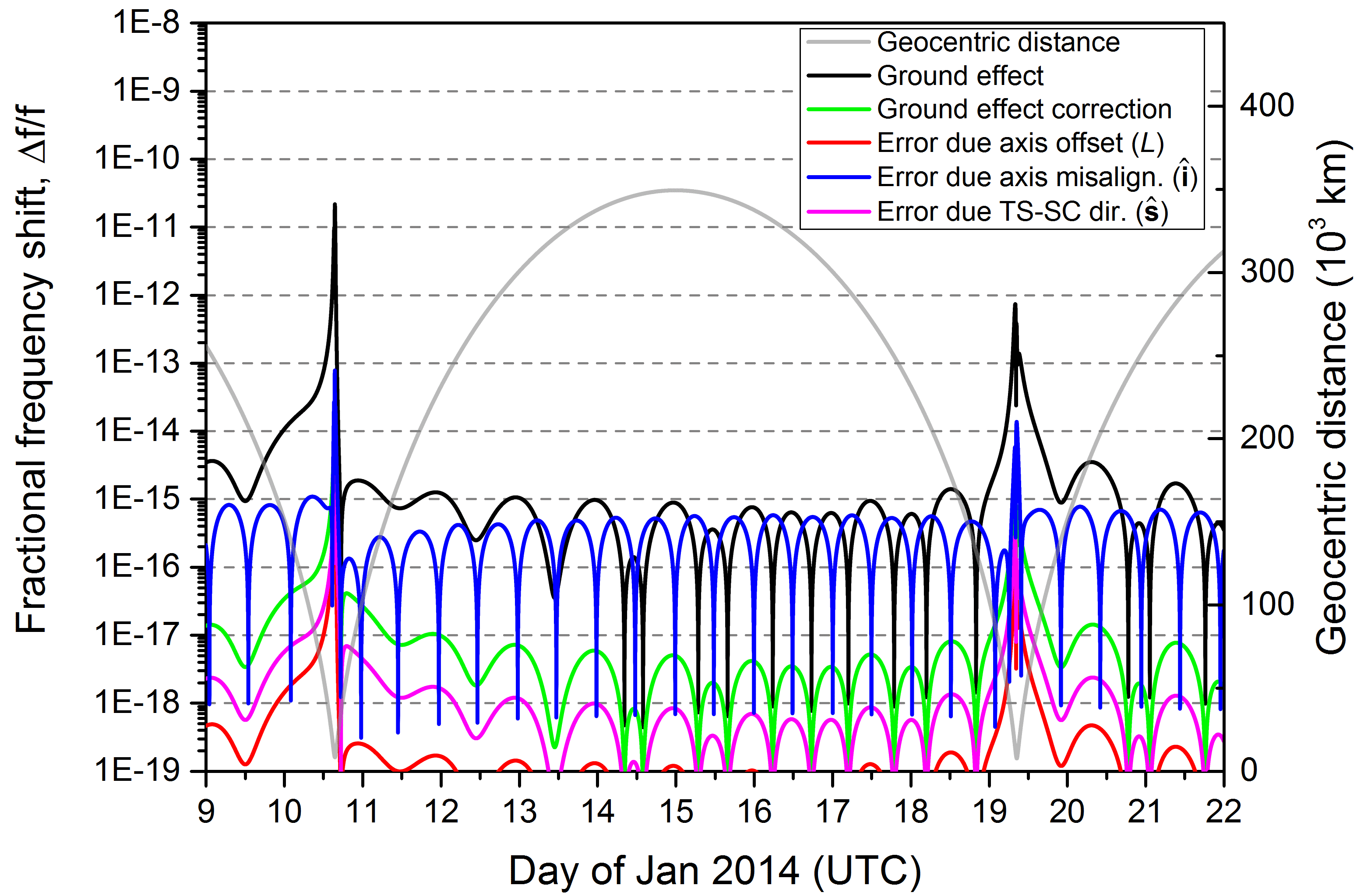}} \hspace{1cm}
        \subfloat[]{
                \label{fig:ra-2014-jan-gb:b}
                \includegraphics[clip, scale=0.31]{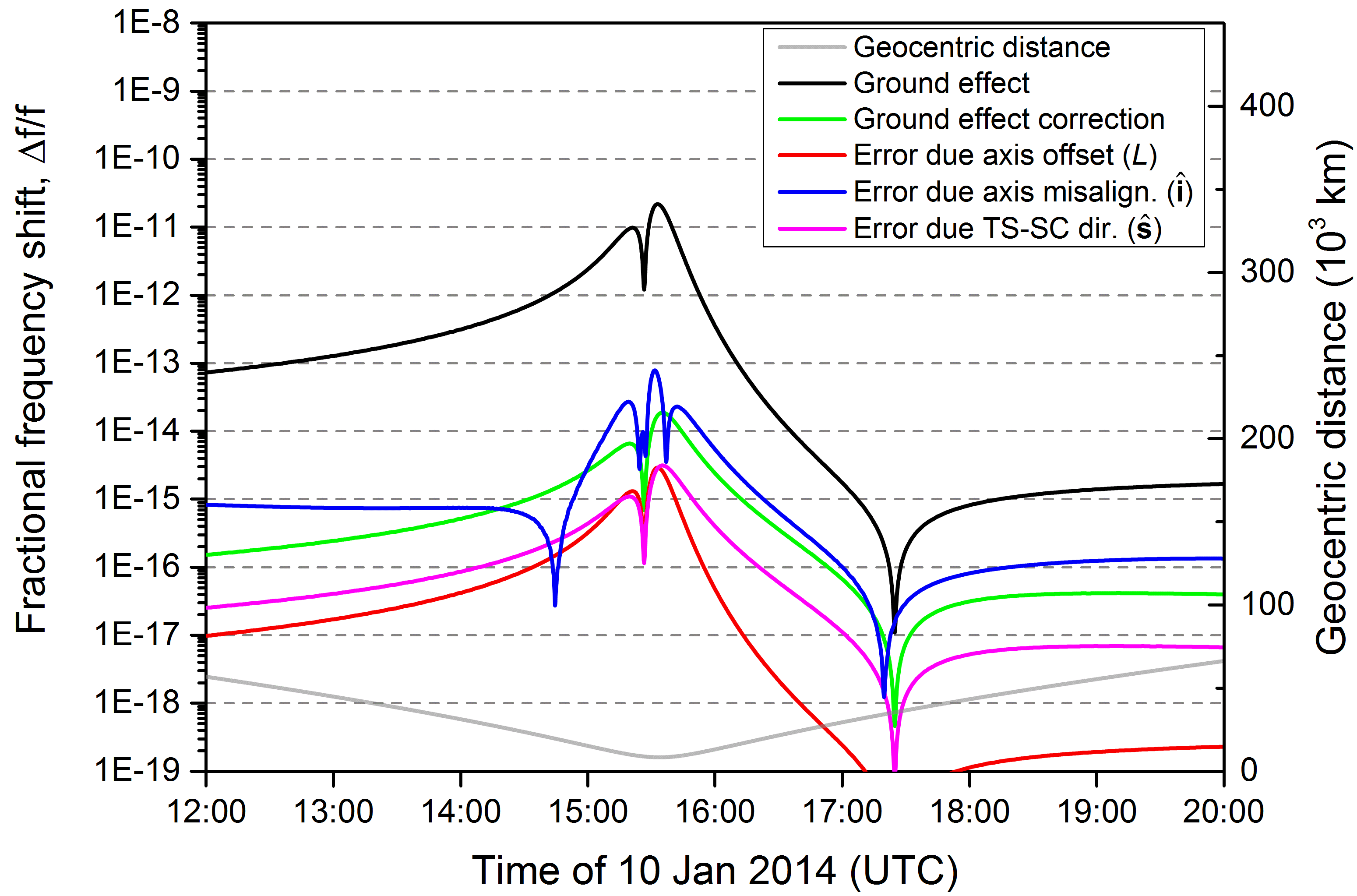}}
                \caption{The APCM effect and its estimation errors for the Green Bank ground antenna tracking the RadioAstron spacecraft during low-perigee epoch A: (a) 9--23 January 2014; (b) zoom-in into the perigee of 10 January 2014. The APCM effect, denoted ``ground effect'' in the legend, is computed using Eq.~\eqref{eq:generic-expression-for-ground-frac-freq-shift}. The ``ground effect correction'' is the error one makes in estimating the APCM effect if one uses the unprimed analogue of Eq.~\eqref{eq:generic-expression-for-ground-frac-freq-shift}, i.e. one uses the actual antenna pointing angle $\theta$ (declination for the Green Bank NRAO140 antenna) instead of the pointing angle $\theta'$ that corresponds to the true position of the SC. The three error curves depict the uncertainties in the computed values of the APCM effect due to the three error curves of Table~\ref{table:errors} that are relevant to the ground APCM effect: the uncertainty in the offset between the ground antenna axes ($L$), the ground antenna axis misalignment ($\hat{\mathbf{i}}$), and the uncertainty in the TS-to-SC direction due to the SC position uncertainty ($\hat{\mathbf{s}}$). The SC visibility and ground antenna tracking constraints are not taken into account for the sake of visual clarity.}
        \label{fig:ra-2014-jan-gb}
\end{figure*}

\begin{figure*}[t]              
        \centering
        \subfloat[]{
                \label{fig:ra-2014-jan-sc:a}
                \includegraphics[clip, scale=0.31]{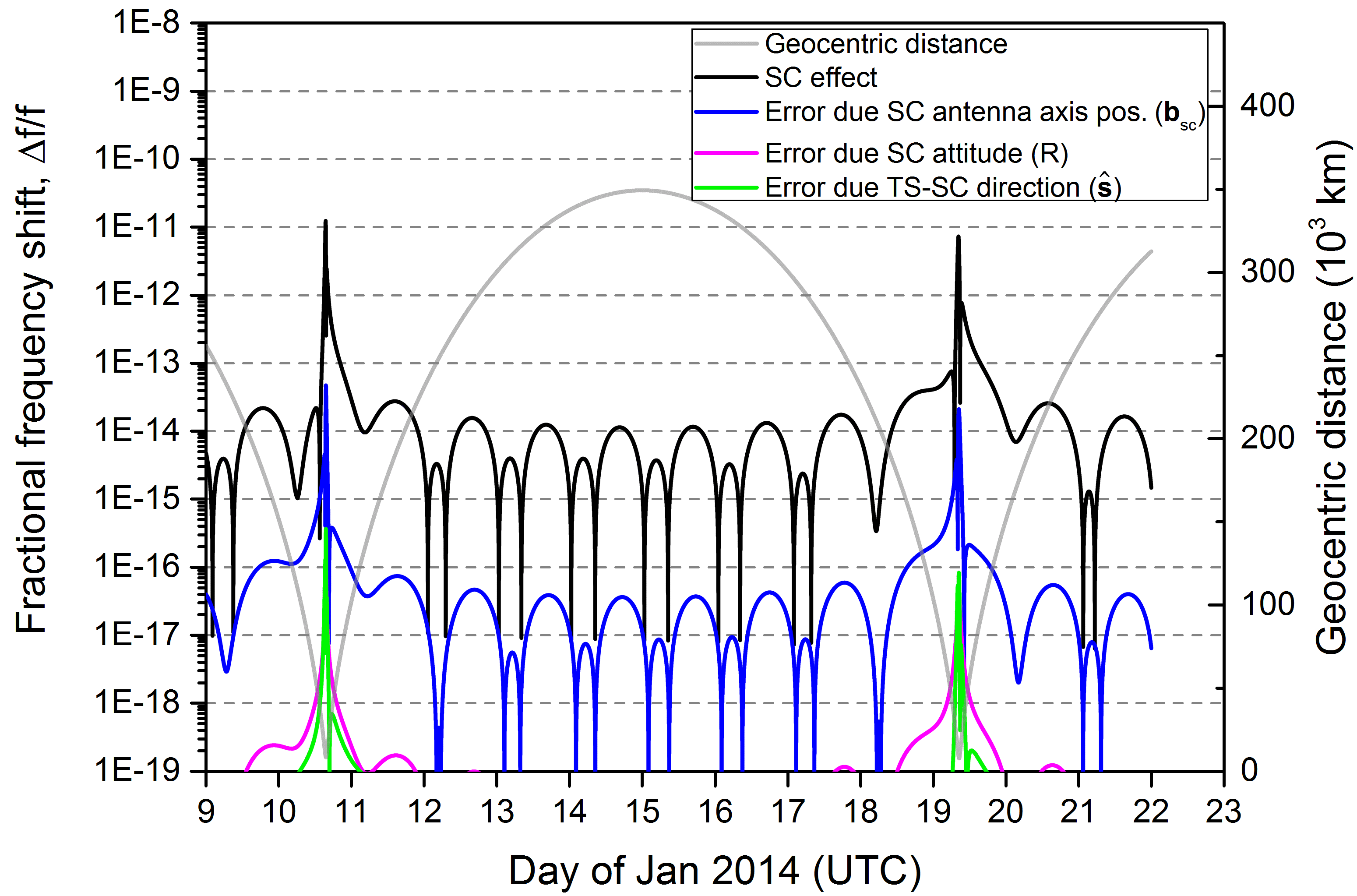}} \hspace{1cm}
        \subfloat[]{
                \label{fig:ra-2014-jan-sc:b}
                \includegraphics[clip, scale=0.31]{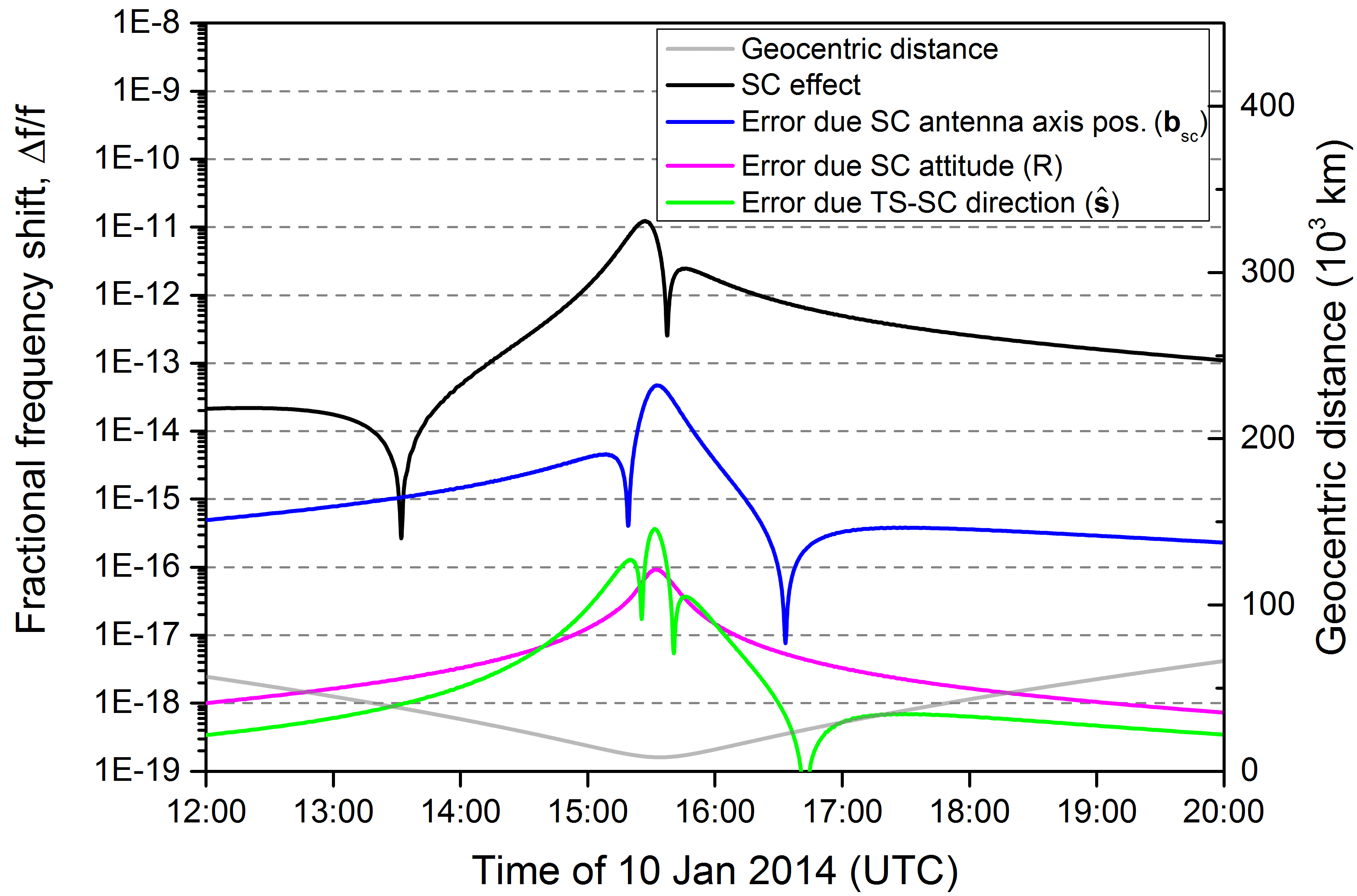}}
                \caption{The APCM effect and its estimation errors for the RadioAstron on-board antenna tracking the Green Bank station during low-perigee epoch A: (a) 9--23 January 2014; (b) zoom-in into the perigee of 10 January 2014. The ``SC effect'' is computed using Eq.~\eqref{eq:sc-apcme-frac-freq-shift-for-small-pointing-errors-and-constant-b-with-transformation-matrix} with $\mathrm{R}=1$. The three error curves depict the uncertainties in the computed values of the APCM effect due to the three error sources of Table~\ref{table:errors} that are relevant to the SC antenna: the uncertainty in the position of the intersection point of the SC antenna axes relative to the SC center of mass ($\mathbf{b}_\mathrm{sc}$), the uncertainty in the SC attitude ($\mathrm{R}$), and the uncertainty in the TS-to-SC direction due to the SC position uncertainty ($\hat{\mathbf{s}}$). Note that while here we assume that the SC maintains a specific constant orientation in the inertial reference frame, such that the rotation matrix $\mathrm{R}=1$, we allow for an error in the realization of this orientation by the attitude control system, $\delta\mathrm{R} \ne 0$. The SC visibility and ground antenna tracking constraints are not taken into account for the sake of visual clarity.}
        \label{fig:ra-2014-jan-sc}
\end{figure*}

\begin{figure*}[t]              
        \centering
        \subfloat[]{
                \label{fig:ra-2019-apr-gb:a}
                \includegraphics[clip, scale=0.31]{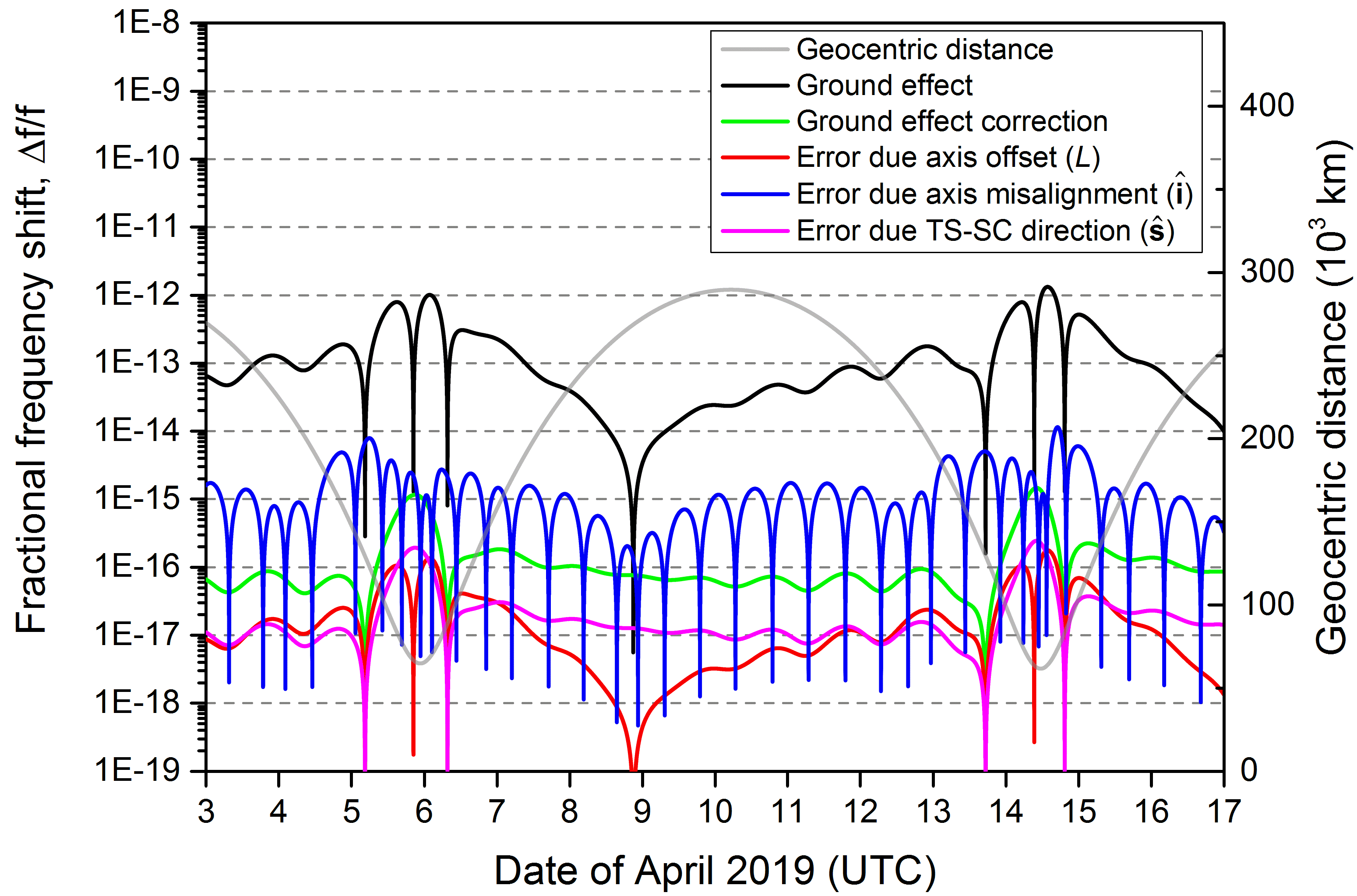}} \hspace{1cm}
        \subfloat[]{
                \label{fig:ra-2019-apr-gb:b}
                \includegraphics[clip, scale=0.31]{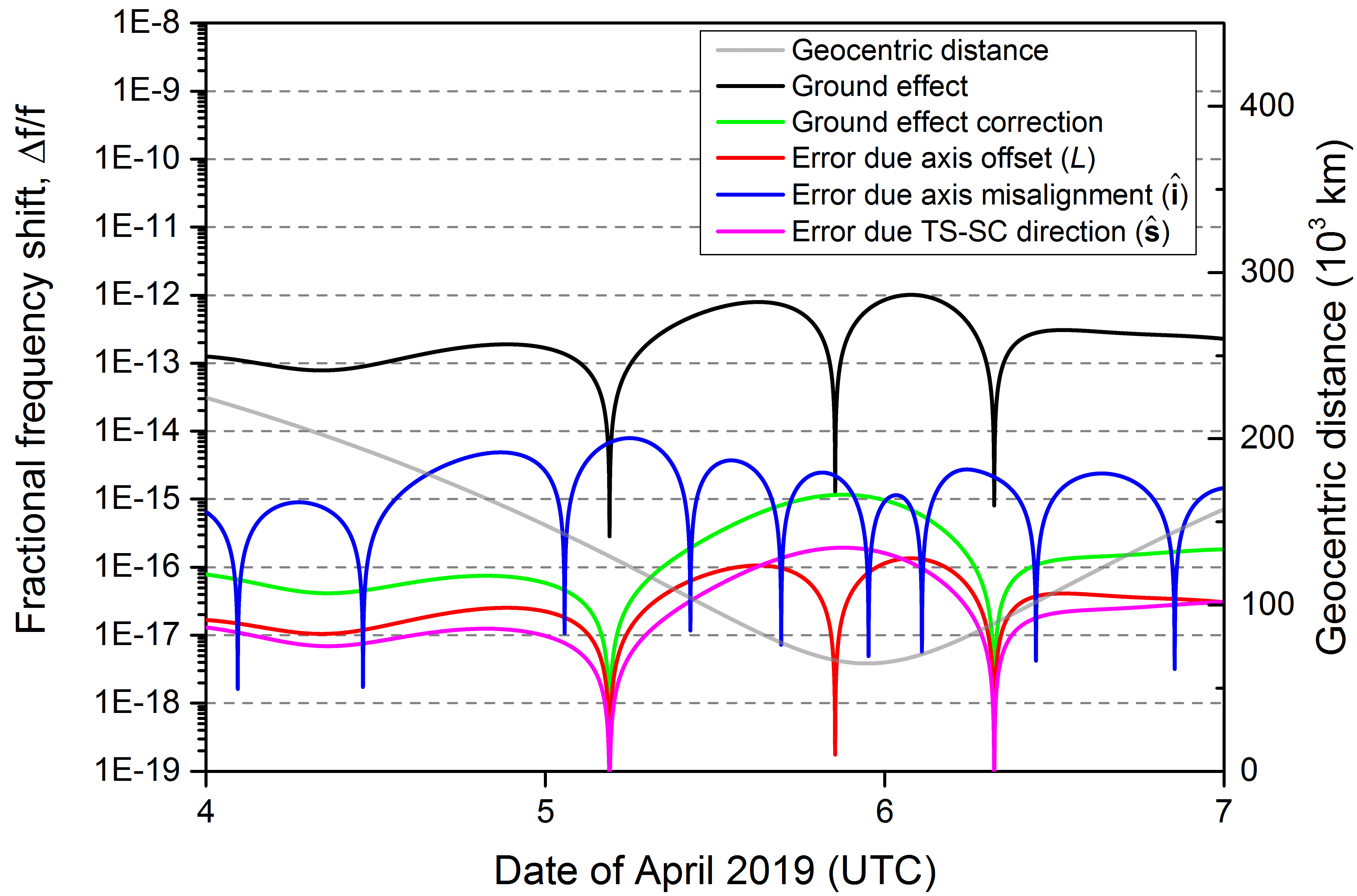}}                             
                \caption{The APCM effect and its estimation errors for the Green Bank ground antenna tracking the RadioAstron spacecraft during high-perigee epoch B: (a) 3--17 April 2019; (b) zoom-in into the perigee of 5 April 2019. See caption of Fig.~\ref{fig:ra-2014-jan-gb} for a description of the plots and other details.}
        \label{fig:ra-2019-apr-gb}
\end{figure*}

\begin{figure*}[t]              
        \centering
        \subfloat[]{
                \label{fig:ra-2019-apr-sc:a}
                \includegraphics[clip, scale=0.31]{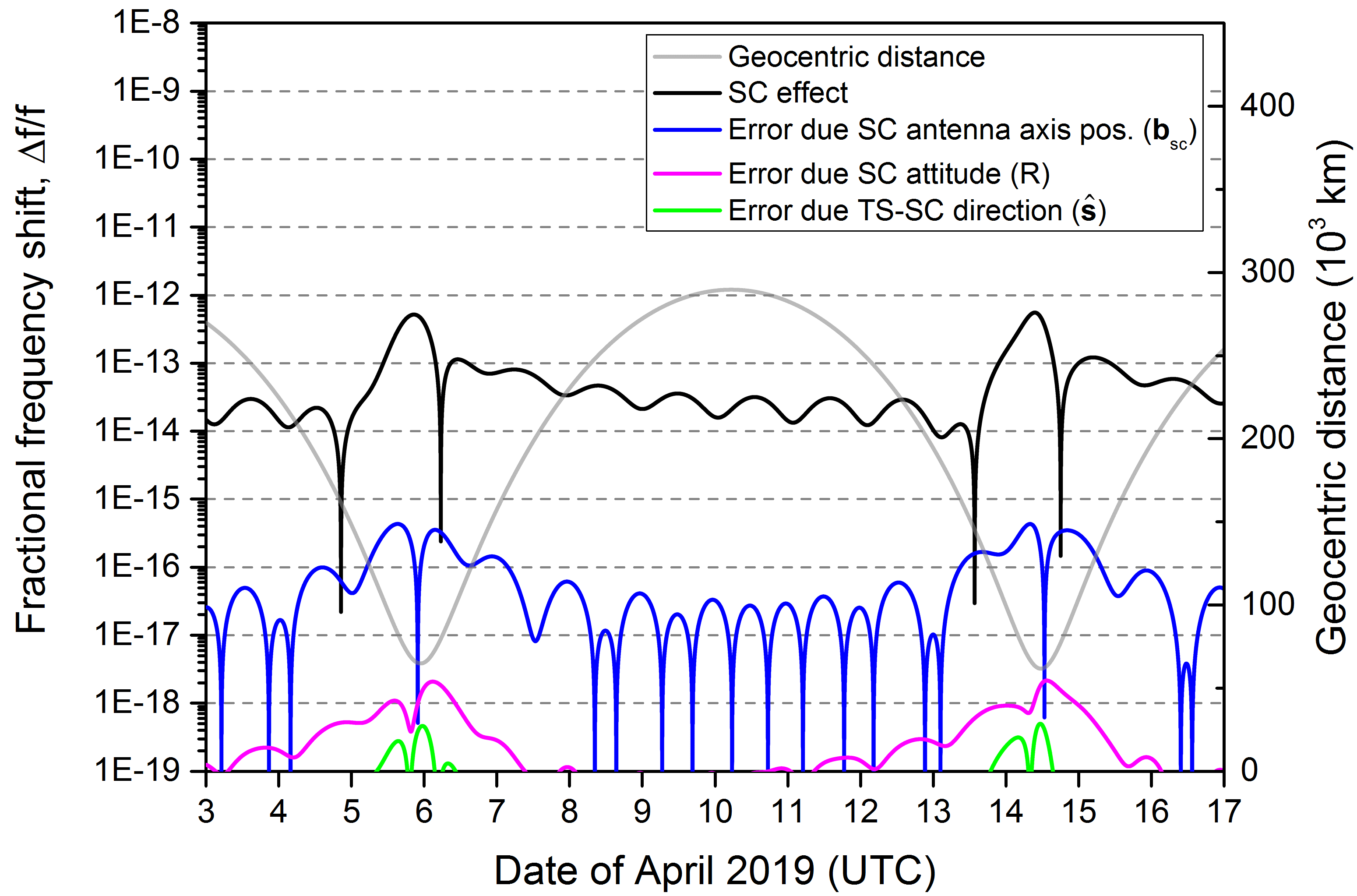}} \hspace{1cm}
        \subfloat[]{
                \label{fig:ra-2019-apr-sc:b}
                \includegraphics[clip, scale=0.31]{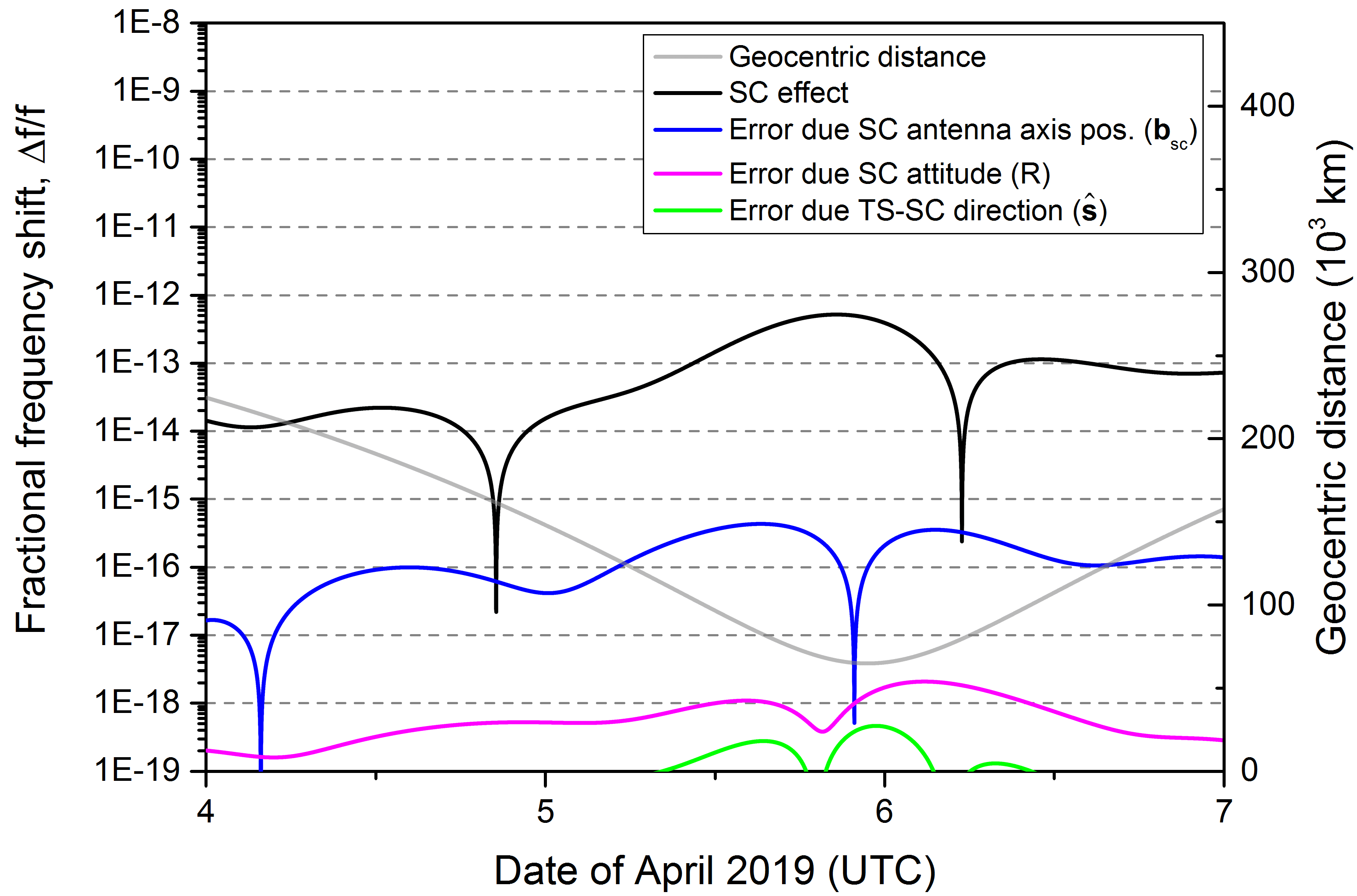}}
                \caption{The APCM effect and its estimation errors for the RadioAstron on-board antenna tracking the Green Bank station during high-perigee epoch B: (a) 3--17 April 2019; (b) zoom-in into the perigee of 5 April 2019. See caption of Fig.~\ref{fig:ra-2014-jan-sc} for a description of the plots and other details.}
        \label{fig:ra-2019-apr-sc}
\end{figure*}

We computed the ground APCM effect only for the Green Bank NRAO140 antenna
since the Pushchino antenna formally has a zero axis offset and thus does not exhibit it. For both selected epochs we show the evolution of the effect and its errors over a time span of approximately 1.5 orbital revolutions and a zoomed-in view into one of the perigees. The results are presented in Figs.~\ref{fig:ra-2014-jan-gb}--\ref{fig:ra-2019-apr-sc}.

Several aspects of these results are worth noting. First, since the effect and its errors vary over many orders of magnitude, we use
a logarithmic scale which allows us to plot only the absolute values of the effect. This
results in the many visible dips which are merely due to the effect changing its sign. Second, the curves are plotted even for the time segments when the SC was below the horizon for NRAO140 or the tracking constraints for the ground or space antenna were not met. This is done for the sake of visual clarity and because the particular SC visibility conditions and tracking constraints are affected in a semi-random way by the interplay of the particular values of the SC orbital parameters. It is worth noting, however, that the SC was visible to NRAO140 near and at the top of the spikes of the APCM effect in Fig.~\ref{fig:ra-2014-jan-gb}. An example of the data processing of a series of experiments performed near one of those spikes is presented below in Fig.~\ref{fig:residuals}. Another important observation is that the magnitudes of the space and ground effects are comparable. Further, as expected, near perigees the effect is larger in epoch~$A$ than in $B$, by an order of magnitude,
due to the more rapid motion of the SC across the sky. However, outside the near-perigee regions the situation is usually opposite, i.e. the effect is larger
in epoch $B$ than in $A$, also by an order of magnitude. For each epoch both the ground and SC effects are larger than $10^{-13}$ on significant parts of the orbit and almost never fall below $10^{-14}$ in the high-perigee epoch.

The correction to the ground antenna effect implied by Eq.~\eqref{eq:generic-expression-for-ground-frac-freq-shift},
that is, the necessity to use $\theta'$ computed from the true TS-to-SC direction instead of the respective antenna pointing angle, $\theta$, is larger
than $1\times10^{-14}$ on some parts of the orbit and thus is significant for high-accuracy
SC tracking experiments.

The errors of estimating the APCM effect are  more significant near perigees
as well. The largest are due to the SC antenna axis position
uncertainty, ground antenna axis misalignment and the ground antenna axis offset.

\end{multicols}


\begin{figure*}[t]              
        \centering
        \subfloat[]{
                \label{fig:residuals:a}
                \includegraphics[clip, scale=0.35]{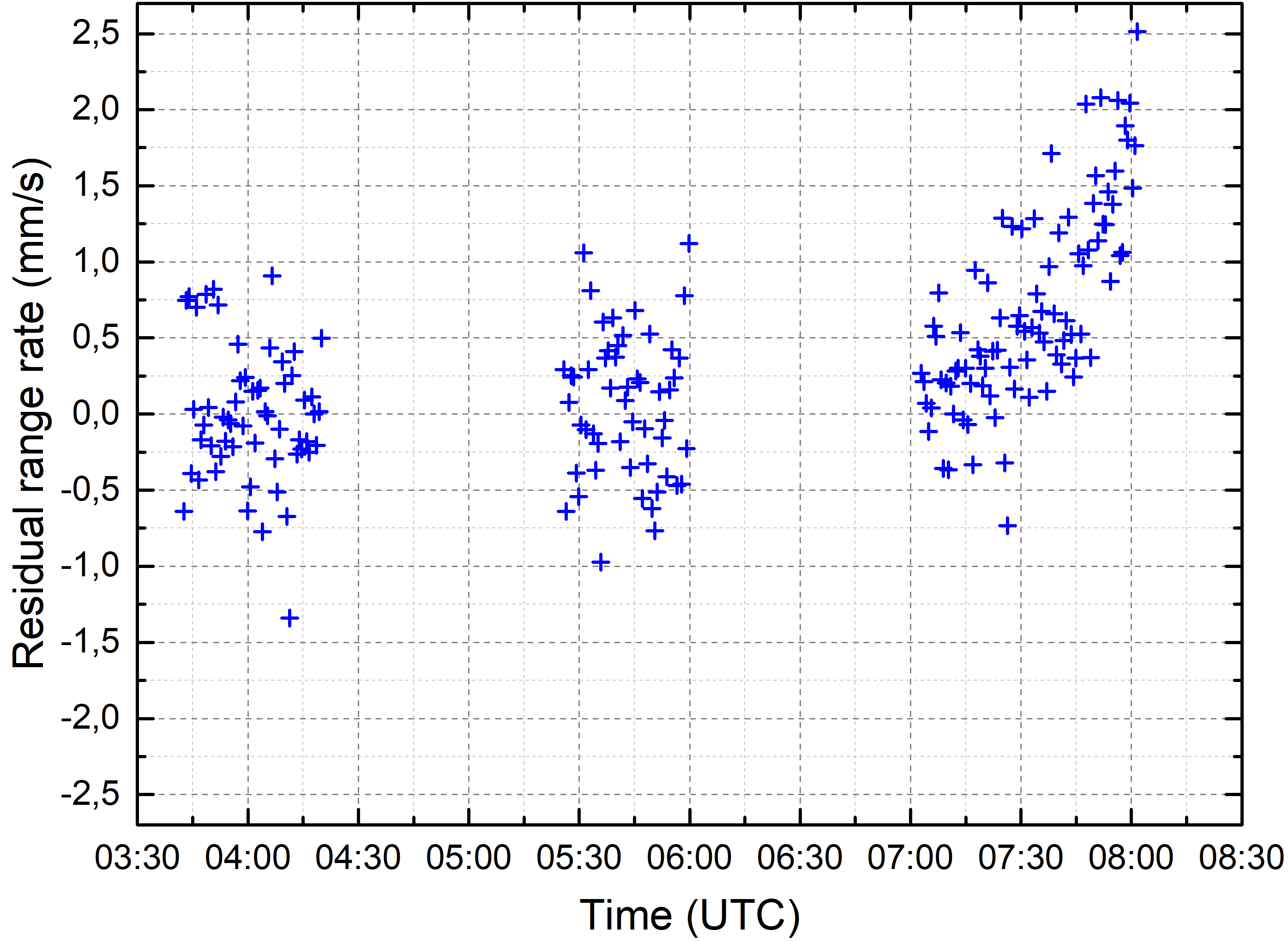}}
\hspace{1cm}
        \subfloat[]{
                \label{fig:residuals:b}
                \includegraphics[clip, scale=0.35]{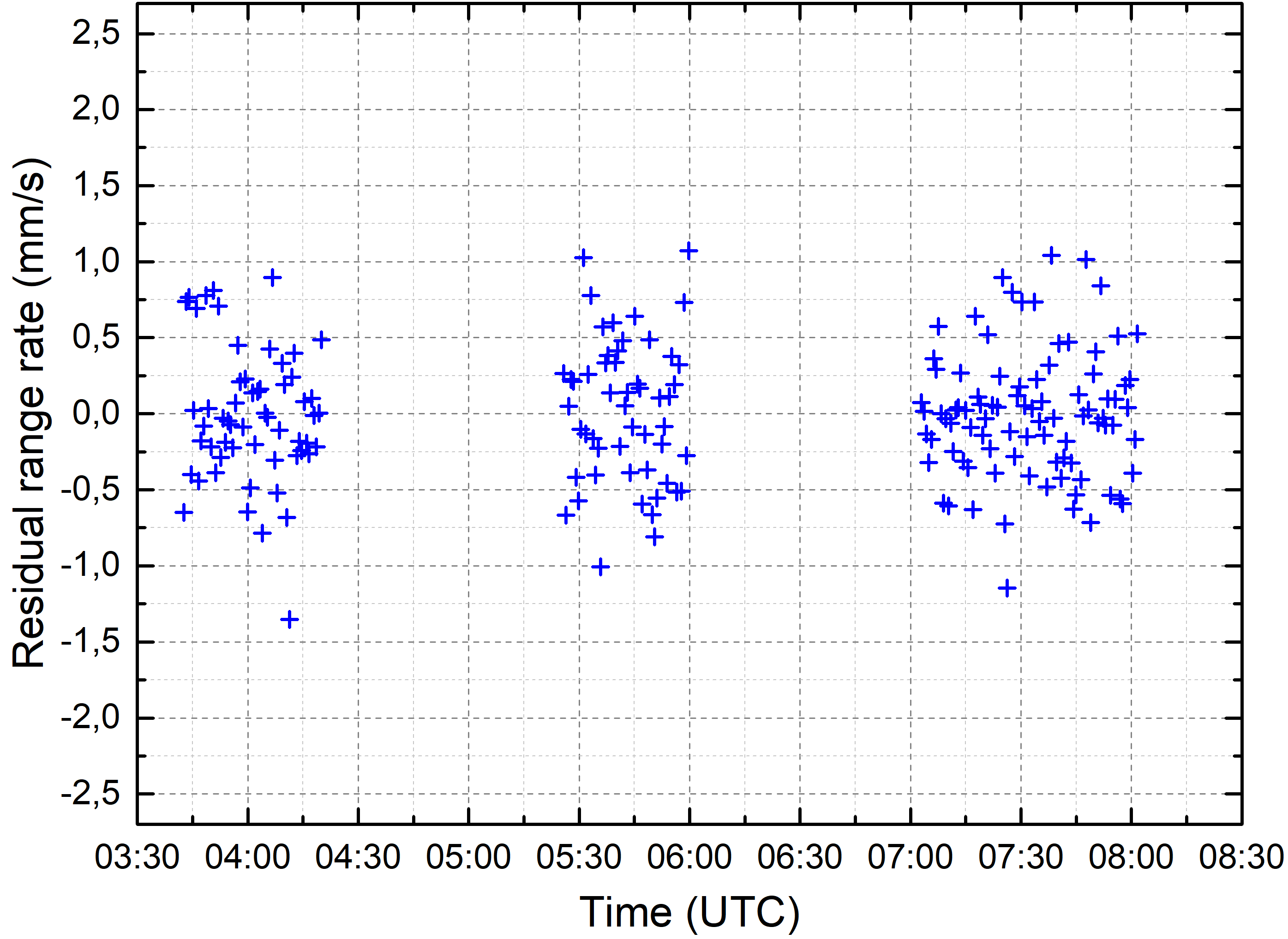}}
        \caption{One-way Doppler residuals for the Green Bank antenna tracking the RadioAstron spacecraft on 19 January 2014. The APCM effects for the ground and space antennas are: (a) not taken into account; (b) taken into account. In this case the dominant contribution to the APCM effect is due to the space antenna.}
        \label{fig:residuals}
\end{figure*}                                        

\clearpage

\begin{multicols}{2}

Near the perigees of epoch $A$ the magnitude of the fractional frequency shift due to the APCM effect reaches $\sim10^{-11}$ both for the ground and space
antennas, which is equivalent to 3~mm/s in terms of the required velocity correction. This is significant
not only for high-accuracy experiments but also for
the SC orbit determination. To demonstrate this we asked the ballistic center of the KIAM 
to generate two orbital solutions that include the data of near-perigee
observations of 19/01/2014 03:50--08:00~UTC (experiment code raks04c) performed
with the Green Bank TS: one with the APCM effects not included in the model and
the other taking them into account. The frequency residuals of the 8.4~GHz downlink signal, i.e. the differences between the observed frequency measurements and their computed values, that correspond to these two solutions are presented in Fig.~\ref{fig:residuals}. As expected, the inclusion of the APCM effects into the model results in a significantly better fit to the observational data.

\section{Future missions}
\label{sec:future-missions}
\nopagebreak
 
The technique of SVLBI, pioneered by VSOP \citep{hirabayashi-1998-sci}
and RadioAstron, proved to be a very useful tool for radio astronomy. Several possibilities for follow-up missions are now being discussed  \citep{gurvits-2020-asr}. Most of them share similar features with RadioAstron: a highly eccentric orbit, a highly stable on-board frequency standard, and a large mechanically steerable high-gain antenna (see Section~\ref{sec:intro}). This suggests that the results we obtained for RadioAstron will be useful in the preparation of the next generation of SVLBI missions. However,
since the orbits considered for future missions are usually more circular
than that of RadioAstron it seems reasonable to verify if the results presented in Section~\ref{sec:radioastron}
are general and not peculiar to RadioAstron.

In order to address this question we repeated the computations of the previous
Section for a SC with the orbital parameters specified in Table~\ref{table:future-svlbi-sc-orbit-parameters}.
These parameters are similar to those of the mission concept outlined in \citep{hong-2014-aa}. Compared to RadioAstron, the orbit of this SC is more circular, not evolving, and has a shorter
period. For our simulations we used the same ground antenna, the Green Bank Earth
Station's NRAO140 (Table~\ref{table:ra-tracking-stations}), with the assumption of the same values for the relevant uncertainties
(Table~\ref{table:errors}).
For the SC we used the same values for the components of vector $\mathbf{b}$ as in Eq.~\eqref{eq:radioastron-vector-b} but assumed a 5-fold decrease in their uncertainties. We also assumed a 10-fold improvement in the accuracies of the spacecraft attitude and the TS-to-SC direction. All these assumptions are tentative and based on our analysis of the current
state of the technology in the respective  areas. The 10-fold improvement of the OD accuracy, which determines the uncertainty in the TS-to-SC direction, might be considered too conservative in view of the availability of high-accuracy
satellite tracking means provided by satellite laser ranging and onboard
GNSS receivers. However, both of these technologies currently have limited capabilities to support satellites on high Earth orbits \citep{winternitz-2017-nav}.

\begin{Table}
\small
\renewcommand{\arraystretch}{1.5}
        \centering
        \begin{tabular}{|l|c|} \hline
                Perigee (km)& 10,000 km \\ \hline
                Apogee (km)& 57,131 km \\ \hline
                Period (hr)& 17.0 hr \\ \hline
                Eccentricity & 0.7 \\ \hline
                Inclination & 28.5$^{\circ}$\\ \hline
                RA of asc. node &  220.0$^{\circ}$ \\ \hline
                Mean anomaly at epoch & 0.0$^{\circ}$ \\ \hline
                Argument of perigee & 0.0$^{\circ}$ \\ \hline
                Epoch & 01/01/2030 00:00:00 UTC\\ \hline
        \end{tabular}
        \captionof{table}{Orbital parameters assumed for the spacecraft of a possible future SVLBI
mission. }
        \label{table:future-svlbi-sc-orbit-parameters}
\end{Table}

\begin{figure*}[h]
        \centering
        \subfloat[]{
                \label{fig:future-gb:a}
                \includegraphics[clip, scale=0.31]{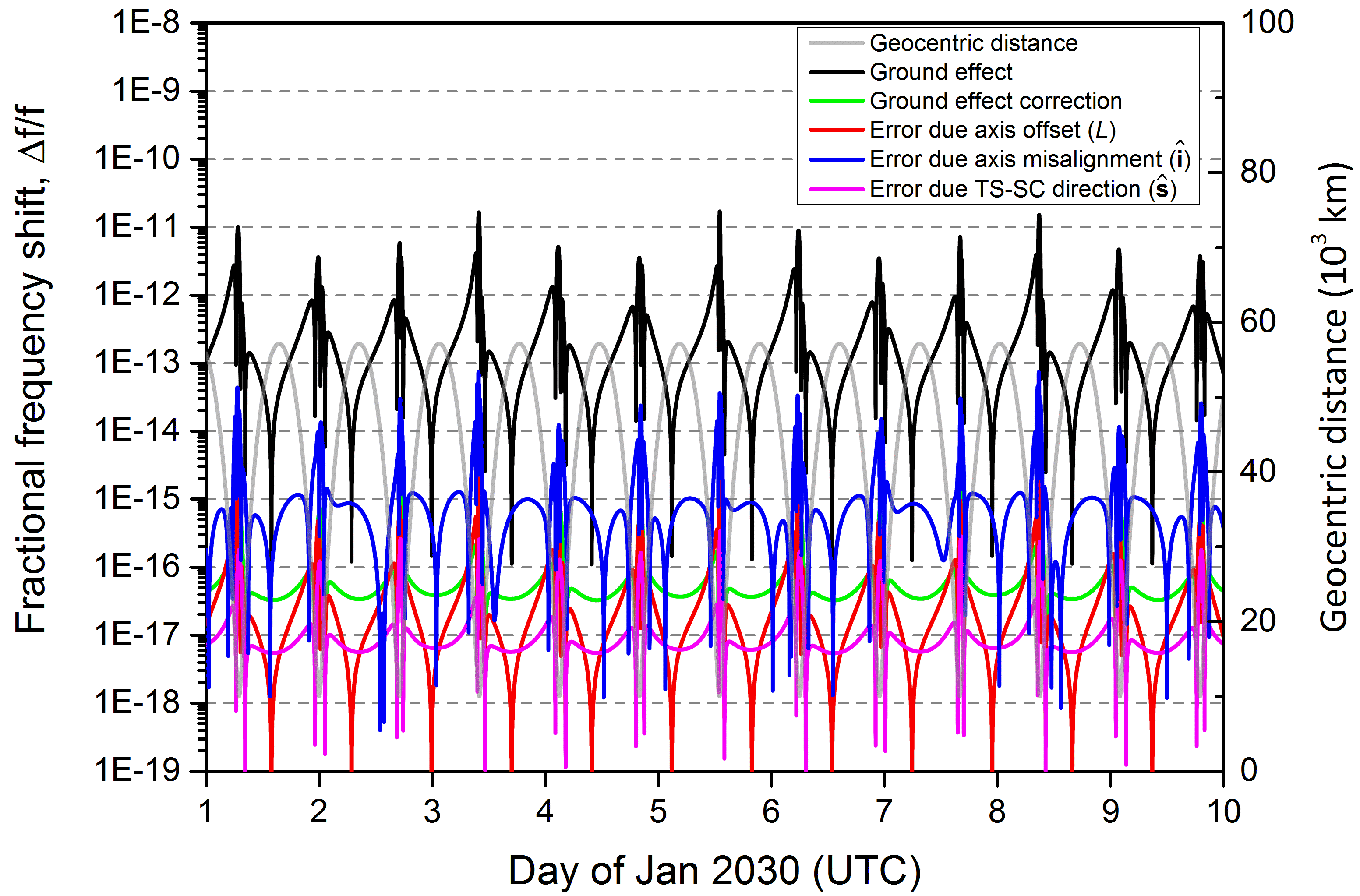}} \hspace{0.5cm}
        \subfloat[]{
                \label{fig:future-gb:b}
                \includegraphics[clip, scale=0.31]{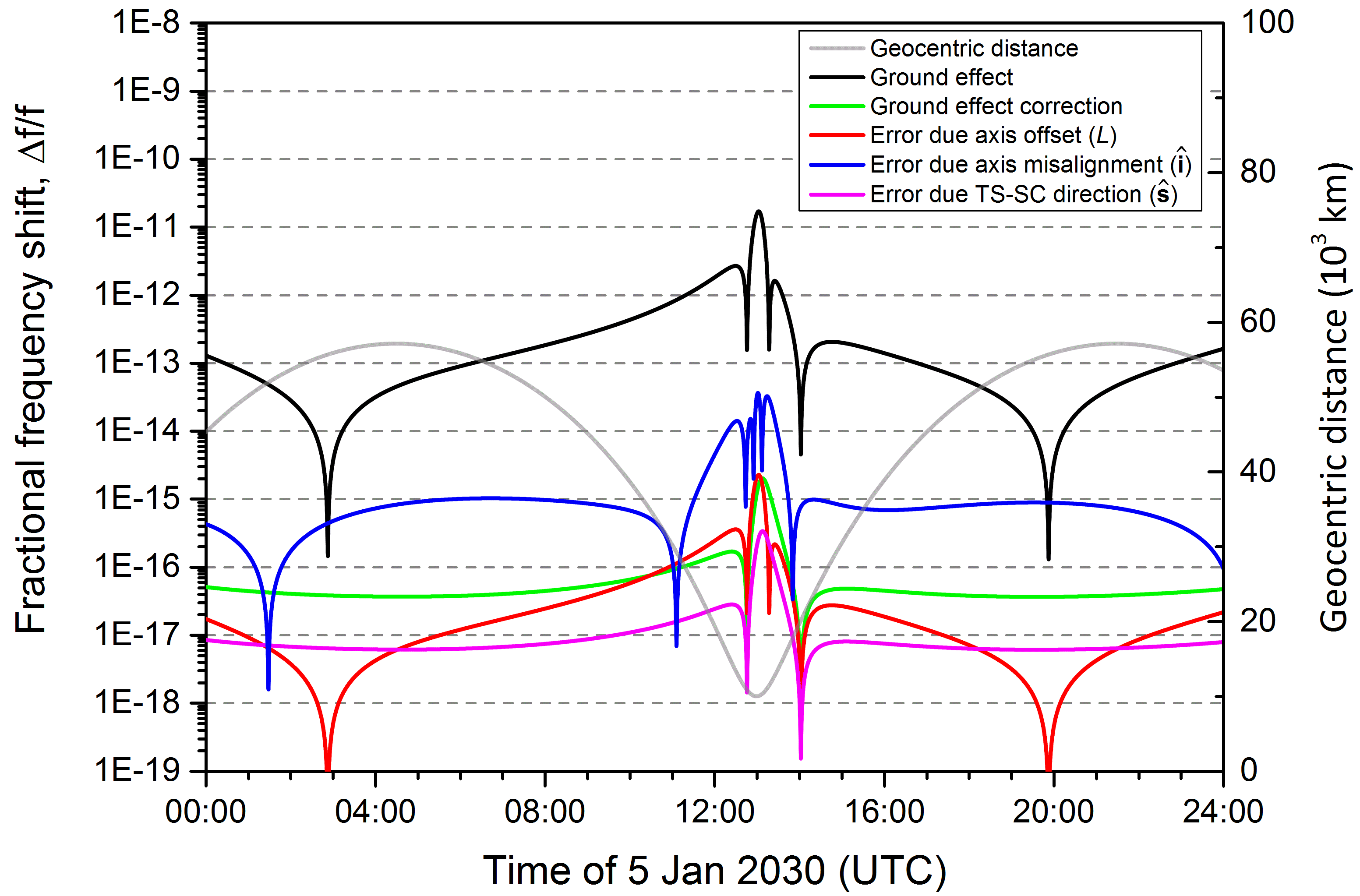}}
                \caption{The APCM effect and its estimation errors for the Green Bank ground antenna tracking the spacecraft of a possible future SVLBI mission: (a) 1--10 January 2030; (b) zoom-in into the perigee of 5 January 2030. See caption of Fig.~\ref{fig:ra-2014-jan-gb} for a description of the plots and other details.}
        \label{fig:future-gb}
\end{figure*}

\begin{figure*}[h]
        \centering
        \subfloat[]{
                \label{fig:future-sc:a}
                \includegraphics[clip, scale=0.31]{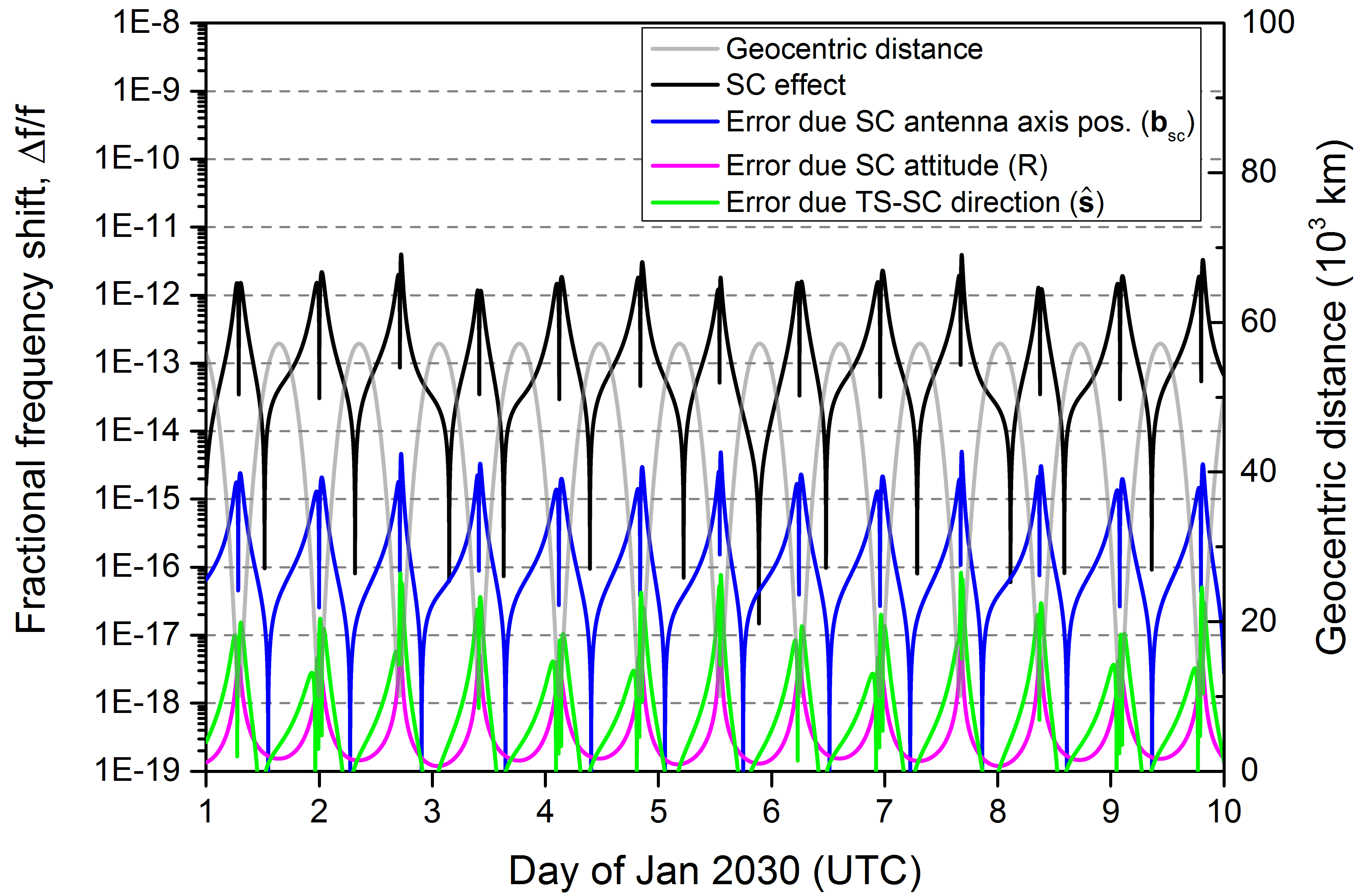}}\hspace{0.5cm}
        \subfloat[]{
                \label{fig:future-sc:b}
                \includegraphics[clip, scale=0.31]{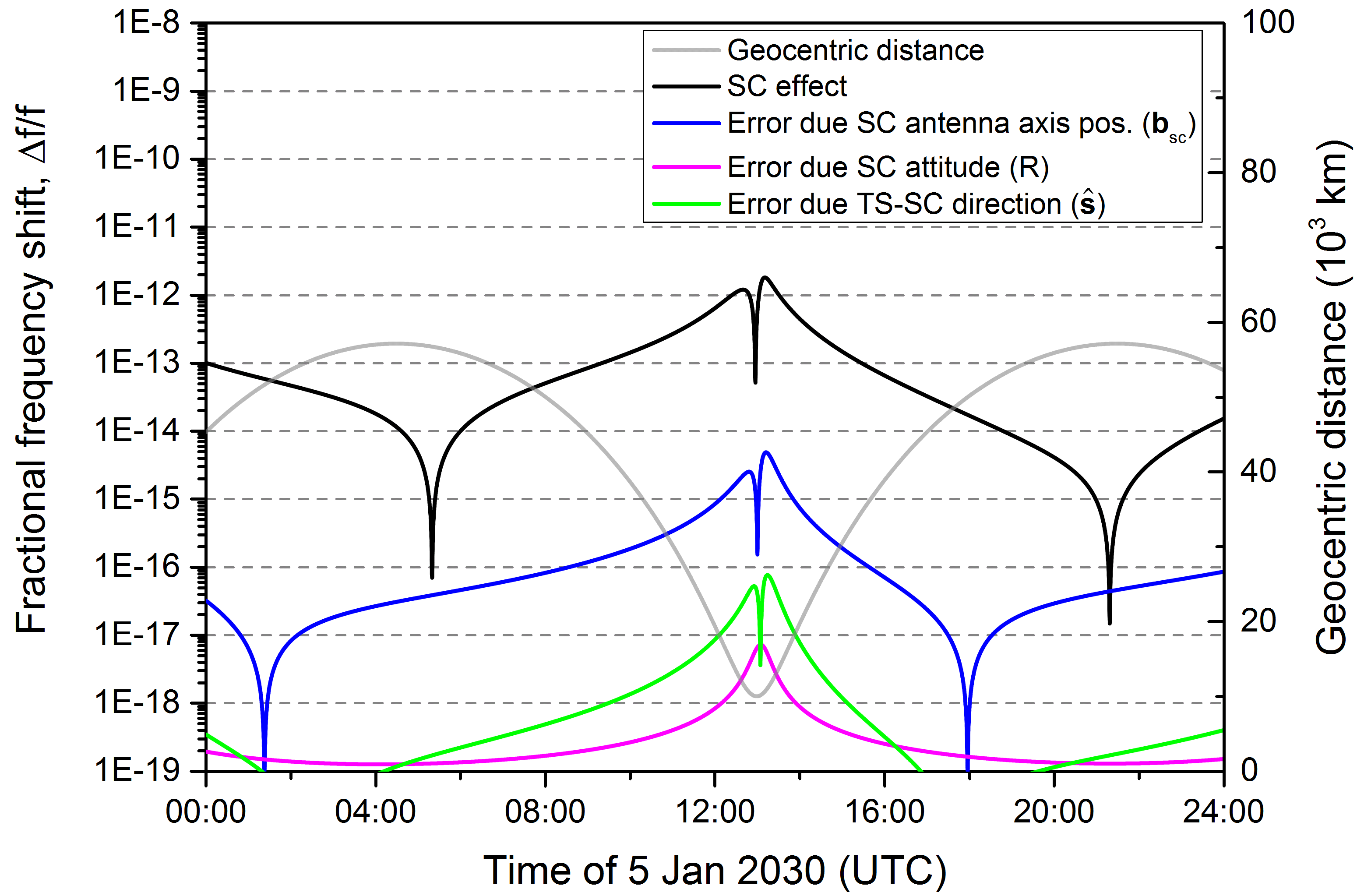}}

                \caption{The APCM effect and its estimation errors for the on-board antenna of the spacecraft of a possible future SVLBI mission tracking the Green Bank station: (a) 1--10 January 2030; (b) zoom-in into the perigee of 5 January 2030. See caption of Fig.~\ref{fig:ra-2014-jan-sc} for a description of the plots and other details.}
        \label{fig:future-sc}
\end{figure*}

The results of our simulations, spanning an arbitrarily  chosen 
decade of 1--10 January 2030, are shown
in Figs.~\ref{fig:future-gb} and~\ref{fig:future-sc}. The magnitude of the APCM effect for the ground and space
antennas is comparable to that of RadioAstron. The correction to the ground
effect implied by Eq.~\eqref{eq:generic-expression-for-ground-frac-freq-shift}
decreased according to our assumption of the better OD accuracy. Other
errors also decreased according to our assumptions on improvements of the uncertainties. However, the error
in the space APCM effect due to the uncertainty in vector $\mathbf{b}$ is still
unacceptably large for high-accuracy experiments, $\sim5\times10^{-15}$,
as well as the error in the ground APCM effect due to the ground antenna axis misalignment, $\sim3\times10^{-15}$.  

To summarize, in this case, as with RadioAstron, the APCM effect is large enough so that it needs to be taken into account even for regular OD. The errors in its
estimated values are negligible for regular OD but too large for the  tracking data obtained with such antennas to be usable in experiments with modern frequency standards with frequency
instability of $\sim10^{-15}$ or better.

\section{Compensation of the APCM effect}
\label{sec:compensation}

In Sections~\ref{sec:radioastron} and~\ref{sec:future-missions} we showed that
the APCM effect and the errors of estimating it can be significant in spacecraft
Doppler
tracking experiments that demand frequency measurements with stability of better than $\sim10^{-14}$.
Now, we consider an approach to considerably reduce the magnitude of the APCM effect that is
available when the SC communicates with the TS in one- and two-way modes simultaneously.
All the equations derived in Section~\ref{sec:theory} are relevant to the
one-way mode, i.e. when the SC transmits a signal synchronized to
its on-board frequency standard and the TS receives it. In the two-way, or phase-locked loop mode, the phase of the spacecraft's downlink signal
is synchronized to that of the uplink signal \, transmitted \, by \, the \, TS. \, In \, order \, to \, obtain \, the \, equation

\end{multicols}

\begin{figure}[H]             
        \centering
        \includegraphics[scale=0.6]{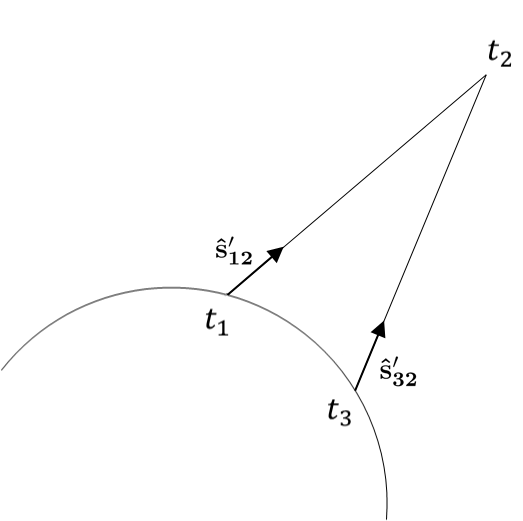}
        \caption{The schematic of the two-way signal propagation as viewed in an Earth-centered inertial reference frame. The signal is emitted by the tracking station at time $t_{1}$, received and coherently retransmitted by the spacecraft at time $t_{2}$ (neglecting the delays in the on-board equipment), and finally received by the tracking station at time $t_{3}$. During the signal travel time the tracking station slightly changes its position in the specified frame. Vector $\hat{\mathbf{s}}'_{12}(t_{1})$ points in the direction along the signal propagation in the up leg, vector $\hat{\mathbf{s}}'_{32}(t_{3})$ points in the direction opposite to that of the signal propagation in the down leg. The atmospheric refraction is ignored.}
        \label{fig:signal-propagation-in-two-way-mode}
\end{figure}

\begin{multicols}{2}

\noindent
for the APCM effect in this mode, we need to sum
the relevant expressions for the up and down legs, taking into account the finite signal propagation time and a slight change of its propagation direction in the down leg due to the TS motion:
\begin{equation}
\begin{split}        
        \frac{\Delta f_{\mathrm{2w}}^{\mathrm{(pcm)}}}{f}
         = \pm\frac{L}{c}\left(\sin{\theta'_{12}}(t_{1}) \cdot \dot\theta'_{12}(t_{1}) + \sin{\theta'_{32}}(t_{3}) \cdot \dot\theta'_{32}(t_{3})\right) \\
        +\:\frac{\mathbf{b'} \cdot \mathbf{\dot{\hat{s}}}'_{12}(t_{1})}{c}
        +\:\frac{\mathbf{b'} \cdot \mathbf{\dot{\hat{s}}}'_{32}(t_{3})}{c}
+ O(\delta\mathbf{\theta}^2) + O(\delta\mathbf{s}^2).
\end{split}
\label{eq:apcme-two-way}
\end{equation}
Here, $t_1$, $t_2$ and $t_3$ are the moments, respectively, when the signal is transmitted
by the TS, received and retransmitted by the SC, and received
by the TS (Fig.~\ref{fig:signal-propagation-in-two-way-mode}). Unit vector
$\mathbf{\hat{s}}'_{12}(t_{1})$ points in the direction of the signal propagation in the up leg,  $\mathbf{\hat{s}}'_{32}(t_{2})$ points in the direction
opposite to that of the signal propagation in the down leg. The two ground antenna pointing angles,  $\theta'$, are defined as:
\begin{equation}
\theta'_{12}(t_{1}) = \frac{\pi}{2}-\angle(\mathbf{\hat{s}}'_{12}(t_{1}), \mathbf{\hat{i}}(t_{1})),
\label{eq:theta12-prime-angle-def}
\end{equation} 
\begin{equation}
\theta'_{32}(t_{3}) = \frac{\pi}{2}-\angle(\mathbf{\hat{s}}'_{32}(t_{3}), \mathbf{\hat{i}}(t_{3})).
\label{eq:theta32-prime-angle-def}
\end{equation} 
The other designations were introduced in Section~\ref{sec:theory}. Note that in Eq.~\eqref{eq:apcme-two-way}
the SC terms enter with the positive sign since the direction
of the signal propagation in both legs is chosen op- posite to that assumed
in Eqs.~\eqref{eq:sc-apcme-frac-freq-shift-for-small-pointing-errors} and~\eqref{eq:sc-apcme-frac-freq-shift-for-small-pointing-errors-and-constant-b}. Also note that we

\noindent
cannot drop the subscripts at $\theta'$ since, for example, $\theta'_{12}(t_3)$ and $\theta'_{32}(t_3)$ are different antenna pointing angles that correspond to the SC positions, respectively, in the future and past of $t_3$.

Several assumptions were made in obtaining Eq.~\eqref{eq:apcme-two-way}. The first is that the pointing errors are small, which is the basis of our analysis. Second, the signal delay in the on-board hardware was treated as negligible (actually it can be of order of a $\mu$s). We also assumed that the base frequencies of the signals in the up and down legs, $f$, are the same. In reality these frequencies are usually chosen to be different to avoid self-excitation of the antenna (for RadioAstron they are 7.2 GHz for the uplink and 8.4 and 15 GHz for the downlinks). Both of the latter aspects can be taken into account and do not change our final conclusions. Finally, we ignored the signal propagation path bending due to atmospheric refraction. This bending can be significant for low elevation angles of the ground antenna, e.g. $\delta\theta \sim 0.1^\circ$ for an elevation angle of 6$^\circ$ \citep{sovers-1998-rmp}. However, since the up- and downlink signals are bent by almost exactly the same amount, $\delta\theta_{12}=\delta\theta_{32}$, the refractive corrections to the ground antenna pointing angles, $\theta'_{12}$ and $\theta'_{32}$, which could be introduced in Eq.~\eqref{eq:apcme-two-way} and Eq.~\eqref{eq:apcme-one-way} below, would cancel out in the final result of this Section given by Eq.~\eqref{eq:apcme-compensation-scheme}.
 

If, in addition to retransmitting the signal received from the ground station, the SC also transmits a one-way signal synchronized to its on-board frequency standard, then for the frequency shift of this signal received at the TS at a time $t_3$ we have:
\begin{equation}   
\begin{split}
        \frac{\Delta f_{\mathrm{1w}}^{\mathrm{(pcm)}}}{f}
        = \pm \frac{L}{c}\sin{\theta'_{32}}(t_{3}) \cdot \dot\theta'_{32}(t_{3})
        +\frac{\mathbf{b'} \cdot \dot{\hat{\mathbf{s}}}'_{32}(t_{3})}{c} \\
        + O(\delta\mathbf{\theta}^2) + O(\delta\mathbf{s}^2).
\end{split}
\label{eq:apcme-one-way}
\end{equation}

Again, this signal is usually transmitted at a frequency different from those of the up and down legs of the two-way signal, but we ignore it here.

The total frequency shift experienced by the one-way signal is a sum of contributions of various factors:
\begin{equation}       
        \Delta f_{\mathrm{1w}} = \Delta f_{\mathrm{kin}} + \Delta f_{\mathrm{grav}} + \Delta f_{\mathrm{media}} + \Delta f_{\mathrm{1w}}^{\mathrm{(pcm)}} + \Delta f_{\mathrm{1w}}^{\mathrm{(other)}},
\label{eq:generic-one-way-freq-shift}
\end{equation}          
where $\Delta f_{\mathrm{kin}}$ is the kinematic frequency shift of the first order in $v/c$, $\Delta f_{\mathrm{grav}}$ is the gravitational redshift,  $\Delta f_{\mathrm{media}}$ is the correction due to the propagation media, $\Delta f_{\mathrm{1w}}^{\mathrm{(pcm)}}$ is the correction due to the APCM effect given by Eq.~\eqref{eq:apcme-one-way}, and $\Delta f_{\mathrm{1w}}^{\mathrm{(other)}}$ is
the contribution of other factors such as higher-order kinematic terms, instrumental
effects, etc.
Similarly, for the two-way signal we have:
\begin{equation}   
\begin{split}
        \Delta f_{\mathrm{2w}} =
        2\Delta f_{\mathrm{kin}} + 2\Delta f_{\mathrm{media}} + \Delta f_{\mathrm{2w}}^{\mathrm{(pcm)}} + \Delta f_{\mathrm{2w}}^{\mathrm{(other)}}
+ \Delta f^{\mathrm{(res)}},
\end{split}
\label{eq:generic-two-way-freq-shift}
\end{equation}
where $\Delta f_{\mathrm{2w}}^{\mathrm{(pcm)}}$ is the correction due to the APCM effect given by Eq.~\eqref{eq:apcme-two-way} and $\Delta f_{\mathrm{2w}}^{\mathrm{(other)}}$
labels the contribution of the ``other'' terms (see above) relevant to the two-way
mode. Note that the two-way signal does not experience the gravitational frequency shift and, up to residual terms grouped into $\Delta f^{\mathrm{(res)}}$,
has twice the kinematic and media contributions compared to the one-way signal.
These residual terms are due to variations of the media properties on the time
scale of signal propagation and the spatial scale of non-reciprocity of the signal paths of the up and down legs. Usually these terms are smaller than the respective contributions of $\Delta f_{\mathrm{kin}}$ and $\Delta f_{\mathrm{media}}$ by several orders of magnitude \citep{vessot-levine-1979-grg, litvinov-2018-pla}. 

Now, let us consider the following combination of the one- and two-way measurements: 
$\Delta f_{\mathrm{1w}} - \dfrac{\Delta f_{\mathrm{2w}}}{2}$. Using Eqs.~\eqref{eq:generic-one-way-freq-shift} and~\eqref{eq:generic-two-way-freq-shift} we obtain:
\begin{equation}      
        \Delta f_{\mathrm{1w}} - \frac{\Delta f_{\mathrm{2w}}}{2} 
        = \Delta f_{\mathrm{grav}} 
        + \Delta f_{\mathrm{kin}}^{\mathrm{(res)}} 
        + \Delta f_{\mathrm{media}}^{\mathrm{(res)}}
        + \Delta f_{\mathrm{pcm}}^{\mathrm{(res)}}
        + \Delta f_{\mathrm{other}}^{\mathrm{(res)}}.
\label{eq:generic-compensation-scheme}
\end{equation}
Here we split the residual terms of Eq.~\eqref{eq:generic-two-way-freq-shift}
into its kinematic, media, and ``other''\ components and denoted 
\begin{equation}      
\Delta f_{\mathrm{pcm}}^{\mathrm{(res)}} = 
        \Delta f_{\mathrm{1w}}^{\mathrm{(pcm)}} - \frac{\Delta f_{\mathrm{2w}}^{\mathrm{(pcm)}}}{2}.
\label{eq:residual-pcm-def}
\end{equation}
A notable feature of Eq.~\eqref{eq:generic-compensation-scheme} is that its right-hand side fully retains the contribution of the gravitational frequency shift of the one-way downlink signal, $\Delta f_{\mathrm{grav}}$, but the
dominant contributions of the nonrelativistic Doppler shift and the troposphere

\end{multicols}

\begin{figure*}
        \centering
        \subfloat[]{
                \label{fig:radioastron-apcme-compensated:a}
                \includegraphics[clip, scale=0.31]{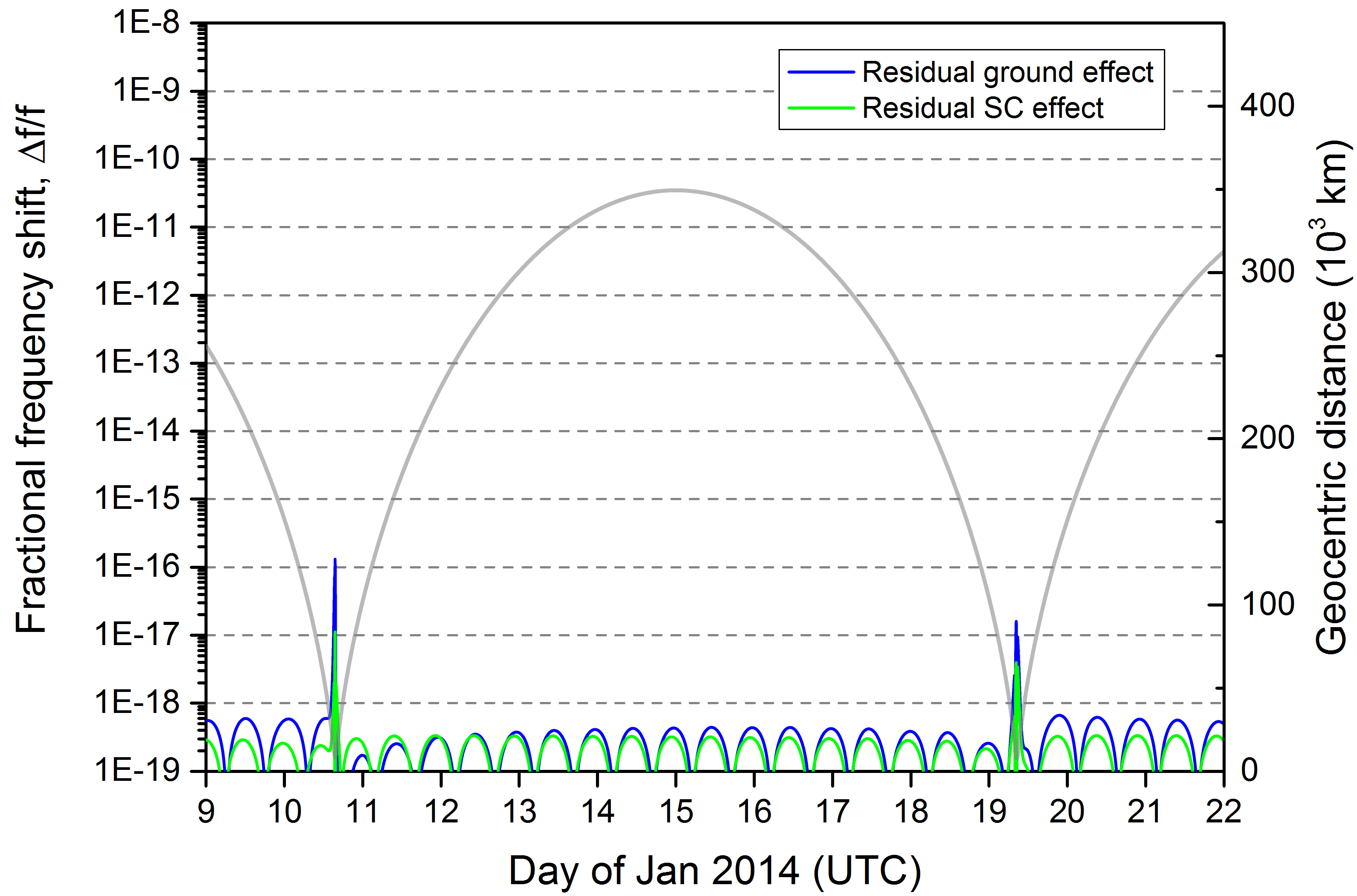}} \hspace{1cm}
        \subfloat[]{
                \label{fig:radioastron-apcme-compensated:b}
                \includegraphics[clip, scale=0.31]{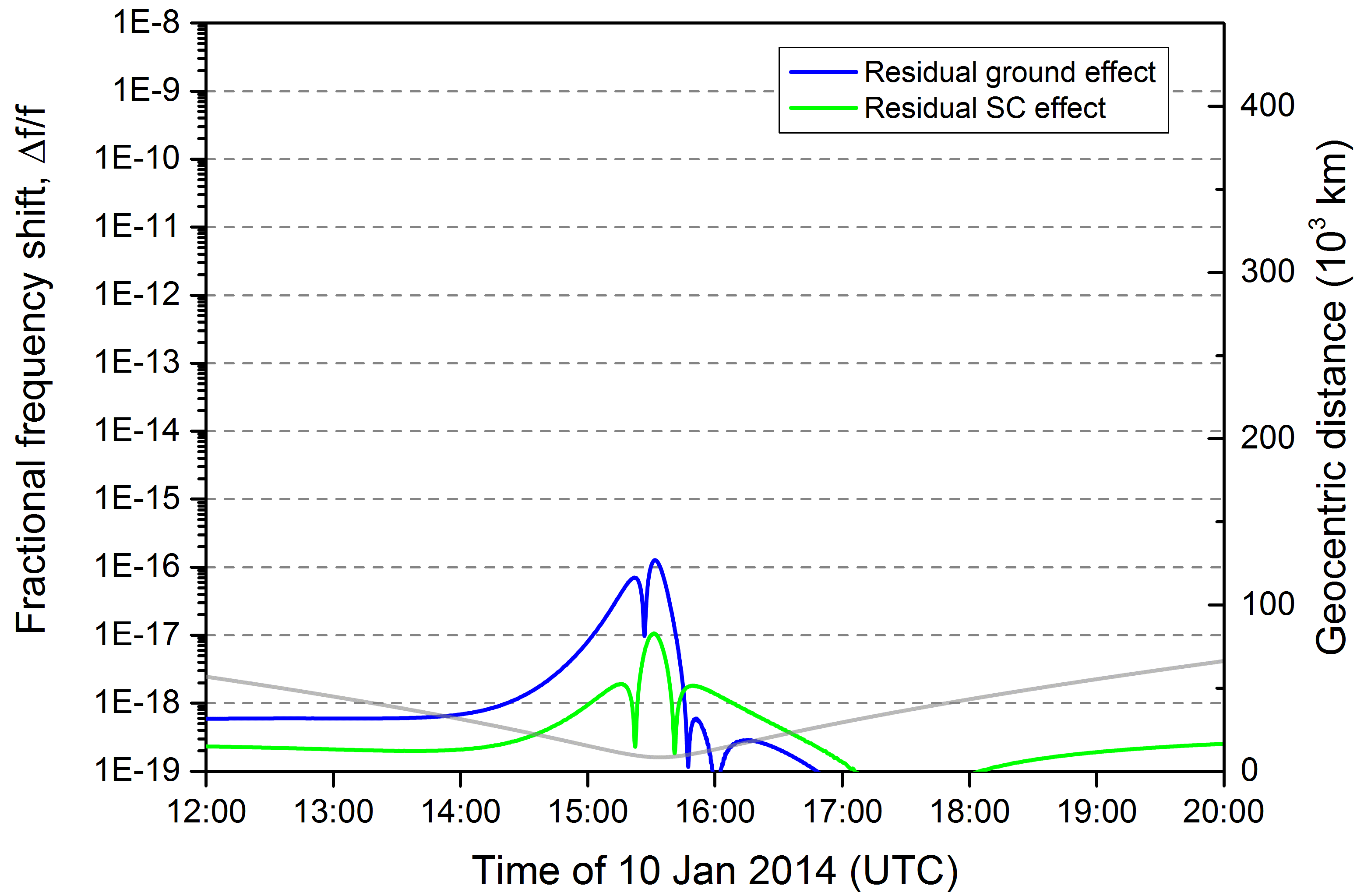}}
        \caption{The residual APCM effect for the on-board and ground antennas after application of the compensation scheme of Eq.~\eqref{eq:generic-compensation-scheme}. The Green Bank station antenna tracking the RadioAstron spacecraft: (a) 9--23 January 2014; (b) zoom-in into the perigee of 10 January 2014.}
        \label{fig:radioastron-apcme-compensated}
\end{figure*}                                        

\begin{multicols}{2}

\noindent
are cancelled or, at minimum, reduced by many orders of magnitude, down to their residual terms labelled $\Delta f_{\mathrm{kin}}^{\mathrm{(res)}}$ and $\Delta f_{\mathrm{media}}^{\mathrm{(res)}}$.

The importance of Eq.~\eqref{eq:generic-compensation-scheme}
for high-accuracy gravity-related experiments was first recognized in the Gravity Probe A 
mission that measured
the gravitational redshift, and the relevant violation parameter of the Einstein
Equivalence Principle, with an accuracy of about 0.01\% \citep{vessot-levine-1980-prl}. The space probe of that mission, however, used a low-gain non-tracking antenna
and the APCM effect was not considered. An approach
based on Eq.~\eqref{eq:generic-compensation-scheme} is also used in the gravitational redshift experiment with RadioAstron \citep{litvinov-2018-pla,nunes-2020-asr}
and now makes an integral part of almost every prospective mission to measure
the gravitational redshift \citep{altschul-2015-asr}.

Our goal is to demonstrate that the residual APCM effect term in Eq.~\eqref{eq:generic-compensation-scheme}, $\Delta f_{\mathrm{pcm}}^{(\mathrm{res})}$, is significantly reduced
compared to $\Delta f_{\mathrm{1w}}^{\mathrm{(pcm)}}$ and $\Delta f_{\mathrm{2w}}^{\mathrm{(pcm)}}$, similar to the reduction of the nonrelativistic Doppler and tropospheric terms. For simplicity let us consider
the case of a polar mount ground antenna, so that the vector $\hat{\mathbf{i}}$ along its primary axis does not depend on time in an Earth-centered inertial reference frame (the change of $\hat{\mathbf{i}}$ due to the Earth's nutation and polar motion can be neglected on the time scale of signal propagation). Then, using Eqs.~\eqref{eq:apcme-one-way}
and ~\eqref{eq:apcme-two-way}, as well as the expansion of
\begin{equation}
\begin{split}
\mathbf{\hat{s}}'_{12}(t_{1}) = \;&
\mathbf{\hat{s}}'_{32}(t_{3}) 
 + 2 \frac{\mathbf{v}_\mathrm{e}(t_3)}{c} \\ 
 & - 2 \left( \frac{\mathbf{v}_\mathrm{e}(t_3)}{c} \cdot \mathbf{\hat{s}}'_{32}(t_{3})
\right) \mathbf{\hat{s}}'_{32}(t_{3}) 
+ O\left(\frac{v_{\mathrm{e}}}{c}\right)^2,
\end{split}
\label{eq:unit-vector-expansion}
\end{equation}
which straightforwardly follows from the relation between successive positions of the TS and its velocity, $\mathbf{v}_{\mathrm{e}}$, in a chosen Earth-centered inertial reference frame, we obtain:
\begin{equation}
\begin{split}
        \frac{\Delta f_{\mathrm{pcm}}^{\mathrm{(res)}}}{f} 
        = \mp\frac{L}{c^2} \bigg\{
        \frac{\dot\theta'}{\cos^2\theta'}
         \left( \left( \mathbf{v} - 
         \left( \mathbf{v} \cdot \mathbf{\hat{s}}'
         \right)  \mathbf{\hat{s}}' \right) \cdot \hat{\mathbf{i}} \right)          \\
         + \tan\theta' \left[ 
         \left( \left( \mathbf{a} 
         -  \left( \mathbf{a} \cdot \mathbf{\hat{s}}' \right)  \mathbf{\hat{s}}'
         -  \left( \mathbf{v} \cdot \dot{\hat{\mathbf{s}}}' \right) \mathbf{\hat{s}}'
         -  \left( \mathbf{v} \cdot \mathbf{\hat{s}}' \right)  \dot{\hat{\mathbf{s}}}'
          \right) \cdot \hat{\mathbf{i}} \right)
         \right] \bigg\} \\
         - \,\frac{1}{c^2} \bigg( \mathbf{b'} \cdot \left( \mathbf{a}
         -  \left( \mathbf{a} \cdot \mathbf{\hat{s}}' \right)  \mathbf{\hat{s}}'
         -  \left( \mathbf{v} \cdot \dot{\hat{\mathbf{s}}}' \right) \mathbf{\hat{s}}'
         -  \left( \mathbf{v} \cdot \mathbf{\hat{s}}' \right)  \dot{\hat{\mathbf{s}}}'
          \right) \bigg)
\\
        + O(\delta\mathbf{\theta}^2) + O(\delta\mathbf{s}^2) + O(v/c)^3,
\end{split}
\label{eq:apcme-compensation-scheme}
\end{equation}
where, for brevity, we labeled:
$\theta' = \theta'_{32}(t_{3})$,
$\mathbf{\hat{s}}' = \mathbf{\hat{s}}'_{32}(t_{3})$,
$\mathbf{v} = \mathbf{v}_\mathrm{e}(t_3)$, and
$\mathbf{a} = \mathbf{\dot v}_\mathrm{e}(t_3)$.
Although Eq.~\eqref{eq:apcme-compensation-scheme} looks cumbersome, it is
clearly of $O(1/c)^2$. Thus we expect
the residual APCM effect, $\Delta f_\mathrm{pcm}^{\mathrm{(res)}}$, to be
much smaller than those in the one-way and two-way
modes since the latter are of $O(1/c)$.

To demonstrate this we use the RadioAstron SC again. RadioAstron implemented the
compensation scheme described above not fully in that the one- and two-way modes could be
operated only intermittently but not simultaneously. This adds complication
to the actual data processing but is not
significant for our illustrative purposes. The residual
APCM effect for RadioAstron during the low perigee epoch of January 2014 is shown in Fig.~\ref{fig:radioastron-apcme-compensated} assuming simultaneous tracking in the one- and two-way modes.
Comparing it to Figs.~\ref{fig:ra-2014-jan-gb} and~\ref{fig:ra-2014-jan-sc} we note that the peak values of the ground APCM effect are reduced by 5 orders of magnitude and those of the space APCM effect, respectively, by 6. None of the two residual
effects exceeds $1.3\times10^{-16}$. In particular, such a significant reduction makes them negligible for the RadioAstron gravitational redshift experiment. Although such residual effect may not be negligible for future high-accuracy
SC tracking experiments using the next generation of frequency standards with accuracy and stability of $\sim10^{-16}$, it will clearly be possible to take it into account accurately enough using Eq.~\eqref{eq:apcme-compensation-scheme}.

\section{Conclusions}

We have improved the model for the antenna phase center motion (APCM) effect for high-gain mechanically steerable antennas by
taking into account pointing errors made by the antenna while tracking the source/target. This improvement is particularly
relevant for high-accuracy SC tracking experiments. Using the data from radio tracking experiments performed
with the RadioAstron spacecraft we showed that the magnitude of the APCM
effect can be very large for spacecraft on highly elliptic orbits, i.e. of order $10^{-11}$ in terms of the fractional frequency shift both for ground and spaceborne antennas. The fractional frequency shift due to our correction to the APCM effect model can reach $2\times10^{-14}$ for RadioAstron.

We also found that the error of taking the APCM effect into account can be as large $4\times10^{-14}$. This is significant for many kinds of high-accuracy experiments, e.g. those concerned with studies of gravity. The largest contribution to the
error is due to the uncertainty in the position of the intersection point of the SC
antenna rotation axes relative to the SC center of mass, the ground antenna axis offset and the ground antenna axis misalignment. We found that the error due to the latter can reach even higher values, up to $7\times10^{-14}$. However, we consider this value as preliminary according to the tentative value of the NRAO140 antenna axis misalignment we used.

We also considered a possible future
SVLBI mission with a SC on a less eccentric orbit compared to RadioAstron
and with an improved error budget. We found that in this case  both the APCM effect and its errors are still unacceptably large for high-accuracy Doppler tracking experiments.

Finally, we showed that the APCM effect can be significantly reduced by using a specific configuration of the satellite communications links,
i.e. by combining the data of simultaneous one- and two-way frequency measurements. For
the case of RadioAstron this reduces both the ground and space
APCM effects down to below $1\times10^{-16}$, which provides a way for using high-gain
mechanically steerable antennas in high-accuracy SC tracking experiments in the near future.

\section*{Acknowledgements}

The authors wish to thank Y.~Y.~Kovalev for constant support in preparation of this paper and S.~Fedorchuk, N.~Babakin, Ya.~Podobedov, and R.~Cherny for helpful discussions on the on-board antenna of the RadioAstron spacecraft. We also wish to thank J.~Baars for reading the manuscript and making valuable remarks. Finally, we thank the anonymous referees for their valuable comments and suggestions. The RadioAstron project is led by the Astro Space Center of the Lebedev Physical Institute of the Russian Academy of Sciences and the Lavochkin Scientific and Production Association under a contract with the Russian Federal Space Agency, in collaboration with partner organizations in Russia and other countries.

\bibliographystyle{model5-names}
\biboptions{authoryear}
\bibliography{antenna_effect_v10}

\begin{thebibliography}{32}
\expandafter\ifx\csname natexlab\endcsname\relax\def\natexlab#1{#1}\fi
\providecommand{\url}[1]{\texttt{#1}}
\providecommand{\href}[2]{#2}
\providecommand{\path}[1]{#1}
\providecommand{\DOIprefix}{doi:}
\providecommand{\ArXivprefix}{arXiv:}
\providecommand{\URLprefix}{URL: }
\providecommand{\Pubmedprefix}{pmid:}
\providecommand{\doi}[1]{\href{http://dx.doi.org/#1}{\path{#1}}}
\providecommand{\Pubmed}[1]{\href{pmid:#1}{\path{#1}}}
\providecommand{\bibinfo}[2]{#2}
\ifx\xfnm\relax \def\xfnm[#1]{\unskip,\space#1}\fi
\bibitem[{{Altschul} et~al.(2015){Altschul}, {Bailey}, {Blanchet}, {Bongs},
  {Bouyer}, {Cacciapuoti}, {Capozziello}, {Gaaloul}, {Giulini}, {Hartwig},
  {Iess}, {Jetzer}, {Landragin}, {Rasel}, {Reynaud}, {Schiller}, {Schubert},
  {Sorrentino}, {Sterr}, {Tasson}, {Tino}, {Tuckey} \&
  {Wolf}}]{altschul-2015-asr}
\bibinfo{author}{{Altschul}, B.}, \bibinfo{author}{{Bailey}, Q.~G.},
  \bibinfo{author}{{Blanchet}, L.}, \bibinfo{author}{{Bongs}, K.},
  \bibinfo{author}{{Bouyer}, P.}, \bibinfo{author}{{Cacciapuoti}, L.},
  \bibinfo{author}{{Capozziello}, S.}, \bibinfo{author}{{Gaaloul}, N.},
  \bibinfo{author}{{Giulini}, D.}, \bibinfo{author}{{Hartwig}, J.},
  \bibinfo{author}{{Iess}, L.}, \bibinfo{author}{{Jetzer}, P.},
  \bibinfo{author}{{Landragin}, A.}, \bibinfo{author}{{Rasel}, E.},
  \bibinfo{author}{{Reynaud}, S.}, \bibinfo{author}{{Schiller}, S.},
  \bibinfo{author}{{Schubert}, C.}, \bibinfo{author}{{Sorrentino}, F.},
  \bibinfo{author}{{Sterr}, U.}, \bibinfo{author}{{Tasson}, J.~D.},
  \bibinfo{author}{{Tino}, G.~M.}, \bibinfo{author}{{Tuckey}, P.}, \&
  \bibinfo{author}{{Wolf}, P.} (\bibinfo{year}{2015}).
\newblock \bibinfo{title}{{Quantum tests of the Einstein Equivalence Principle
  with the STE-QUEST space mission}}.
\newblock {\it \bibinfo{journal}{Advances in Space Research}\/},  {\it
  \bibinfo{volume}{55}\/}, \bibinfo{pages}{501--524}.
  \DOIprefix\doi{10.1016/j.asr.2014.07.014}.
  \href{http://arxiv.org/abs/1404.4307}{\tt arXiv:1404.4307}.
\bibitem[{Biriukov et~al.(2014)Biriukov, Kauts, Kulagin, Litvinov \&
  Rudenko}]{biriukov-2014-ar}
\bibinfo{author}{Biriukov, A.~V.}, \bibinfo{author}{Kauts, V.~L.},
  \bibinfo{author}{Kulagin, V.~V.}, \bibinfo{author}{Litvinov, D.~A.}, \&
  \bibinfo{author}{Rudenko, V.~N.} (\bibinfo{year}{2014}).
\newblock \bibinfo{title}{{Gravitational redshift test with the space radio
  telescope ``RadioAstron''}}.
\newblock {\it \bibinfo{journal}{Astronomy Reports}\/},  {\it
  \bibinfo{volume}{58}\/}\bibinfo{issue}{(11)}, \bibinfo{pages}{783--795}.
  \DOIprefix\doi{10.1134/S1063772914110018}.
\bibitem[{{Bothwell} et~al.(2019){Bothwell}, {Kedar}, {Oelker}, {Robinson},
  {Bromley}, {Tew}, {Ye} \& {Kennedy}}]{bothwell-2019-metrologia}
\bibinfo{author}{{Bothwell}, T.}, \bibinfo{author}{{Kedar}, D.},
  \bibinfo{author}{{Oelker}, E.}, \bibinfo{author}{{Robinson}, J.~M.},
  \bibinfo{author}{{Bromley}, S.~L.}, \bibinfo{author}{{Tew}, W.~L.},
  \bibinfo{author}{{Ye}, J.}, \& \bibinfo{author}{{Kennedy}, C.~J.}
  (\bibinfo{year}{2019}).
\newblock \bibinfo{title}{{JILA SrI optical lattice clock with uncertainty of
  $2.0\times10^{-18}$}}.
\newblock {\it \bibinfo{journal}{Metrologia}\/},  {\it
  \bibinfo{volume}{56}\/}\bibinfo{issue}{(6)}, \bibinfo{pages}{065004}.
  \DOIprefix\doi{10.1088/1681-7575/ab4089}.
  \href{http://arxiv.org/abs/1906.06004}{\tt arXiv:1906.06004}.
\bibitem[{{Duev} et~al.(2012){Duev}, {Molera Calv{\'e}s}, {Pogrebenko},
  {Gurvits}, {Cim{\'o}} \& {Bocanegra Bahamon}}]{duev-2012-aa}
\bibinfo{author}{{Duev}, D.~A.}, \bibinfo{author}{{Molera Calv{\'e}s}, G.},
  \bibinfo{author}{{Pogrebenko}, S.~V.}, \bibinfo{author}{{Gurvits}, L.~I.},
  \bibinfo{author}{{Cim{\'o}}, G.}, \& \bibinfo{author}{{Bocanegra Bahamon},
  T.} (\bibinfo{year}{2012}).
\newblock \bibinfo{title}{{Spacecraft VLBI and Doppler tracking: algorithms and
  implementation}}.
\newblock {\it \bibinfo{journal}{Astronomy \& Astrophysics}\/},  {\it
  \bibinfo{volume}{541}\/}, \bibinfo{pages}{A43}.
  \DOIprefix\doi{10.1051/0004-6361/201218885}.
  \href{http://arxiv.org/abs/1203.4408}{\tt arXiv:1203.4408}.
\bibitem[{{Fedorchuk} \& {Arkhipov}(2014)}]{fedorchuk-2014-cosres}
\bibinfo{author}{{Fedorchuk}, S.~D.}, \& \bibinfo{author}{{Arkhipov}, M.~Y.}
  (\bibinfo{year}{2014}).
\newblock \bibinfo{title}{{On the assurance of the design accuracy of the space
  radio telescope RadioAstron}}.
\newblock {\it \bibinfo{journal}{Cosmic Research}\/},  {\it
  \bibinfo{volume}{52}\/}, \bibinfo{pages}{379--381}.
  \DOIprefix\doi{10.1134/S0010952514050049}.
\bibitem[{{Ford} et~al.(2014){Ford}, {Anderson}, {Belousov}, {Brandt}, {Ford},
  {Kanevsky}, {Kovalenko}, {Kovalev}, {Maddalena}, {Sergeev}, {Smirnov},
  {Watts} \& {Weadon}}]{ford-2014-spie}
\bibinfo{author}{{Ford}, H.~A.}, \bibinfo{author}{{Anderson}, R.},
  \bibinfo{author}{{Belousov}, K.}, \bibinfo{author}{{Brandt}, J.~J.},
  \bibinfo{author}{{Ford}, J.~M.}, \bibinfo{author}{{Kanevsky}, B.},
  \bibinfo{author}{{Kovalenko}, A.}, \bibinfo{author}{{Kovalev}, Y.~Y.},
  \bibinfo{author}{{Maddalena}, R.~J.}, \bibinfo{author}{{Sergeev}, S.},
  \bibinfo{author}{{Smirnov}, A.}, \bibinfo{author}{{Watts}, G.}, \&
  \bibinfo{author}{{Weadon}, T.~L.} (\bibinfo{year}{2014}).
\newblock \bibinfo{title}{{The RadioAstron Green Bank Earth Station}}.
\newblock In \bibinfo{editor}{L.~M. {Stepp}}, \bibinfo{editor}{R.~{Gilmozzi}},
  \& \bibinfo{editor}{H.~J. {Hall}} (Eds.), {\it
  \bibinfo{booktitle}{Ground-based and Airborne Telescopes V}\/} Proc. SPIE
  9145 (p. \bibinfo{pages}{91450B}).
\newblock \DOIprefix\doi{10.1117/12.2056761}.
\bibitem[{{Giovannini} et~al.(2018){Giovannini}, {Savolainen}, {Orienti},
  {Nakamura}, {Nagai}, {Kino}, {Giroletti}, {Hada}, {Bruni}, {Kovalev},
  {Anderson}, {D'Ammando}, {Hodgson}, {Honma}, {Krichbaum}, {Lee}, {Lico},
  {Lisakov}, {Lobanov}, {Petrov}, {Sohn}, {Sokolovsky}, {Voitsik}, {Zensus} \&
  {Tingay}}]{giovannini-2018-nata}
\bibinfo{author}{{Giovannini}, G.}, \bibinfo{author}{{Savolainen}, T.},
  \bibinfo{author}{{Orienti}, M.}, \bibinfo{author}{{Nakamura}, M.},
  \bibinfo{author}{{Nagai}, H.}, \bibinfo{author}{{Kino}, M.},
  \bibinfo{author}{{Giroletti}, M.}, \bibinfo{author}{{Hada}, K.},
  \bibinfo{author}{{Bruni}, G.}, \bibinfo{author}{{Kovalev}, Y.~Y.},
  \bibinfo{author}{{Anderson}, J.~M.}, \bibinfo{author}{{D'Ammando}, F.},
  \bibinfo{author}{{Hodgson}, J.}, \bibinfo{author}{{Honma}, M.},
  \bibinfo{author}{{Krichbaum}, T.~P.}, \bibinfo{author}{{Lee}, S.~S.},
  \bibinfo{author}{{Lico}, R.}, \bibinfo{author}{{Lisakov}, M.~M.},
  \bibinfo{author}{{Lobanov}, A.~P.}, \bibinfo{author}{{Petrov}, L.},
  \bibinfo{author}{{Sohn}, B.~W.}, \bibinfo{author}{{Sokolovsky}, K.~V.},
  \bibinfo{author}{{Voitsik}, P.~A.}, \bibinfo{author}{{Zensus}, J.~A.}, \&
  \bibinfo{author}{{Tingay}, S.} (\bibinfo{year}{2018}).
\newblock \bibinfo{title}{{A wide and collimated radio jet in 3C84 on the scale
  of a few hundred gravitational radii}}.
\newblock {\it \bibinfo{journal}{Nature Astronomy}\/},  {\it
  \bibinfo{volume}{2}\/}, \bibinfo{pages}{472--477}.
  \DOIprefix\doi{10.1038/s41550-018-0431-2}.
  \href{http://arxiv.org/abs/1804.02198}{\tt arXiv:1804.02198}.
\bibitem[{{Gurvits}(2020)}]{gurvits-2020-asr}
\bibinfo{author}{{Gurvits}, L.~I.} (\bibinfo{year}{2020}).
\newblock \bibinfo{title}{{Space VLBI: from first ideas to operational
  missions}}.
\newblock {\it \bibinfo{journal}{Advances in Space Research}\/},  {\it
  \bibinfo{volume}{65}\/}\bibinfo{issue}{(2)}, \bibinfo{pages}{868--876}.
  \DOIprefix\doi{10.1016/j.asr.2019.05.042}.
  \href{http://arxiv.org/abs/1905.11175}{\tt arXiv:1905.11175}.
\bibitem[{{He{\ss}} et~al.(2011){He{\ss}}, {Stringhetti}, {Hummelsberger},
  {Hausner}, {Stalford}, {Nasca}, {Cacciapuoti}, {Much}, {Feltham}, {Vudali},
  {L{\'e}ger}, {Picard}, {Massonnet}, {Rochat}, {Goujon}, {Sch{\"a}fer},
  {Laurent}, {Lemonde}, {Clairon}, {Wolf}, {Salomon}, {Proch{\'a}zka},
  {Schreiber} \& {Montenbruck}}]{aces-2011-acau}
\bibinfo{author}{{He{\ss}}, M.~P.}, \bibinfo{author}{{Stringhetti}, L.},
  \bibinfo{author}{{Hummelsberger}, B.}, \bibinfo{author}{{Hausner}, K.},
  \bibinfo{author}{{Stalford}, R.}, \bibinfo{author}{{Nasca}, R.},
  \bibinfo{author}{{Cacciapuoti}, L.}, \bibinfo{author}{{Much}, R.},
  \bibinfo{author}{{Feltham}, S.}, \bibinfo{author}{{Vudali}, T.},
  \bibinfo{author}{{L{\'e}ger}, B.}, \bibinfo{author}{{Picard}, F.},
  \bibinfo{author}{{Massonnet}, D.}, \bibinfo{author}{{Rochat}, P.},
  \bibinfo{author}{{Goujon}, D.}, \bibinfo{author}{{Sch{\"a}fer}, W.},
  \bibinfo{author}{{Laurent}, P.}, \bibinfo{author}{{Lemonde}, P.},
  \bibinfo{author}{{Clairon}, A.}, \bibinfo{author}{{Wolf}, P.},
  \bibinfo{author}{{Salomon}, C.}, \bibinfo{author}{{Proch{\'a}zka}, I.},
  \bibinfo{author}{{Schreiber}, U.}, \& \bibinfo{author}{{Montenbruck}, O.}
  (\bibinfo{year}{2011}).
\newblock \bibinfo{title}{{The ACES mission: System development and test
  status}}.
\newblock {\it \bibinfo{journal}{Acta Astronautica}\/},  {\it
  \bibinfo{volume}{69}\/}, \bibinfo{pages}{929--938}.
  \DOIprefix\doi{10.1016/j.actaastro.2011.07.002}.
\bibitem[{{Hirabayashi} et~al.(1998){Hirabayashi}, {Hirosawa}, {Kobayashi},
  {Murata}, {Edwards}, {Fomalont}, {Fujisawa}, {Ichikawa}, {Kii}, {Lovell},
  {Moellenbrock}, {Okayasu}, {Inoue}, {Kawaguchi}, {Kameno}, {Shibata},
  {Asaki}, {Bushimata}, {Enome}, {Horiuchi}, {Miyaji}, {Umemoto}, {Migenes},
  {Wajima}, {Nakajima}, {Morimoto}, {Ellis}, {Meier}, {Murphy}, {Preston},
  {Smith}, {Tingay}, {Traub}, {Wietfeldt}, {Benson}, {Claussen}, {Flatters},
  {Romney}, {Ulvestad}, {D'Addario}, {Langston}, {Minter}, {Carlson},
  {Dewdney}, {Jauncey}, {Reynolds}, {Taylor}, {McCulloch}, {Cannon}, {Gurvits},
  {Mioduszewski}, {Schilizzi} \& {Booth}}]{hirabayashi-1998-sci}
\bibinfo{author}{{Hirabayashi}, H.}, \bibinfo{author}{{Hirosawa}, H.},
  \bibinfo{author}{{Kobayashi}, H.}, \bibinfo{author}{{Murata}, Y.},
  \bibinfo{author}{{Edwards}, P.~G.}, \bibinfo{author}{{Fomalont}, E.~B.},
  \bibinfo{author}{{Fujisawa}, K.}, \bibinfo{author}{{Ichikawa}, T.},
  \bibinfo{author}{{Kii}, T.}, \bibinfo{author}{{Lovell}, J.~E.~J.},
  \bibinfo{author}{{Moellenbrock}, G.~A.}, \bibinfo{author}{{Okayasu}, R.},
  \bibinfo{author}{{Inoue}, M.}, \bibinfo{author}{{Kawaguchi}, N.},
  \bibinfo{author}{{Kameno}, S.}, \bibinfo{author}{{Shibata}, K.~M.},
  \bibinfo{author}{{Asaki}, Y.}, \bibinfo{author}{{Bushimata}, T.},
  \bibinfo{author}{{Enome}, S.}, \bibinfo{author}{{Horiuchi}, S.},
  \bibinfo{author}{{Miyaji}, T.}, \bibinfo{author}{{Umemoto}, T.},
  \bibinfo{author}{{Migenes}, V.}, \bibinfo{author}{{Wajima}, K.},
  \bibinfo{author}{{Nakajima}, J.}, \bibinfo{author}{{Morimoto}, M.},
  \bibinfo{author}{{Ellis}, J.}, \bibinfo{author}{{Meier}, D.~L.},
  \bibinfo{author}{{Murphy}, D.~W.}, \bibinfo{author}{{Preston}, R.~A.},
  \bibinfo{author}{{Smith}, J.~G.}, \bibinfo{author}{{Tingay}, S.~J.},
  \bibinfo{author}{{Traub}, D.~L.}, \bibinfo{author}{{Wietfeldt}, R.~D.},
  \bibinfo{author}{{Benson}, J.~M.}, \bibinfo{author}{{Claussen}, M.~J.},
  \bibinfo{author}{{Flatters}, C.}, \bibinfo{author}{{Romney}, J.~D.},
  \bibinfo{author}{{Ulvestad}, J.~S.}, \bibinfo{author}{{D'Addario}, L.~R.},
  \bibinfo{author}{{Langston}, G.~I.}, \bibinfo{author}{{Minter}, A.~H.},
  \bibinfo{author}{{Carlson}, B.~R.}, \bibinfo{author}{{Dewdney}, P.~E.},
  \bibinfo{author}{{Jauncey}, D.~L.}, \bibinfo{author}{{Reynolds}, J.~E.},
  \bibinfo{author}{{Taylor}, A.~R.}, \bibinfo{author}{{McCulloch}, P.~M.},
  \bibinfo{author}{{Cannon}, W.~H.}, \bibinfo{author}{{Gurvits}, L.~I.},
  \bibinfo{author}{{Mioduszewski}, A.~J.}, \bibinfo{author}{{Schilizzi},
  R.~T.}, \& \bibinfo{author}{{Booth}, R.~S.} (\bibinfo{year}{1998}).
\newblock \bibinfo{title}{{Overview and Initial Results of the Very Long
  Baseline Interferometry Space Observatory Programme}}.
\newblock {\it \bibinfo{journal}{Science}\/},  {\it \bibinfo{volume}{281}\/},
  \bibinfo{pages}{1825}. \DOIprefix\doi{10.1126/science.281.5384.1825}.
\bibitem[{{Hong} et~al.(2014){Hong}, {Shen}, {An} \& {Liu}}]{hong-2014-aa}
\bibinfo{author}{{Hong}, X.}, \bibinfo{author}{{Shen}, Z.},
  \bibinfo{author}{{An}, T.}, \& \bibinfo{author}{{Liu}, Q.}
  (\bibinfo{year}{2014}).
\newblock \bibinfo{title}{{The Chinese space Millimeter-wavelength VLBI array -
  A step toward imaging the most compact astronomical objects}}.
\newblock {\it \bibinfo{journal}{Acta Astronautica}\/},  {\it
  \bibinfo{volume}{102}\/}, \bibinfo{pages}{217--225}.
  \DOIprefix\doi{10.1016/j.actaastro.2014.05.026}.
  \href{http://arxiv.org/abs/1403.5188}{\tt arXiv:1403.5188}.
\bibitem[{{Johnson} et~al.(2016){Johnson}, {Kovalev}, {Gwinn}, {Gurvits},
  {Narayan}, {Macquart}, {Jauncey}, {Voitsik}, {Anderson}, {Sokolovsky} \&
  {Lisakov}}]{johnson-2016-apj}
\bibinfo{author}{{Johnson}, M.~D.}, \bibinfo{author}{{Kovalev}, Y.~Y.},
  \bibinfo{author}{{Gwinn}, C.~R.}, \bibinfo{author}{{Gurvits}, L.~I.},
  \bibinfo{author}{{Narayan}, R.}, \bibinfo{author}{{Macquart}, J.-P.},
  \bibinfo{author}{{Jauncey}, D.~L.}, \bibinfo{author}{{Voitsik}, P.~A.},
  \bibinfo{author}{{Anderson}, J.~M.}, \bibinfo{author}{{Sokolovsky}, K.~V.},
  \& \bibinfo{author}{{Lisakov}, M.~M.} (\bibinfo{year}{2016}).
\newblock \bibinfo{title}{{Extreme Brightness Temperatures and Refractive
  Substructure in 3C273 with RadioAstron}}.
\newblock {\it \bibinfo{journal}{Astrophys. J. Letters}\/},  {\it
  \bibinfo{volume}{820}\/}\bibinfo{issue}{(1)}, \bibinfo{pages}{L10}.
  \DOIprefix\doi{10.3847/2041-8205/820/1/L10}.
  \href{http://arxiv.org/abs/1601.05810}{\tt arXiv:1601.05810}.
\bibitem[{{Kardashev} et~al.(2013){Kardashev}, {Khartov}, {Abramov}, {Avdeev},
  {Alakoz}, {Aleksandrov}, {Ananthakrishnan}, {Andreyanov}, {Andrianov},
  {Antonov}, {Artyukhov}, {Arkhipov}, {Baan}, {Babakin}, {Babyshkin},
  {Bartel'}, {Belousov}, {Belyaev}, {Berulis}, {Burke}, {Biryukov}, {Bubnov},
  {Burgin}, {Busca}, {Bykadorov}, {Bychkova}, {Vasil'kov}, {Wellington},
  {Vinogradov}, {Wietfeldt}, {Voitsik}, {Gvamichava}, {Girin}, {Gurvits},
  {Dagkesamanskii}, {D'Addario}, {Giovannini}, {Jauncey}, {Dewdney}, {D'yakov},
  {Zharov}, {Zhuravlev}, {Zaslavskii}, {Zakhvatkin}, {Zinov'ev}, {Ilinen},
  {Ipatov}, {Kanevskii}, {Knorin}, {Casse}, {Kellermann}, {Kovalev}, {Kovalev},
  {Kovalenko}, {Kogan}, {Komaev}, {Konovalenko}, {Kopelyanskii}, {Korneev},
  {Kostenko}, {Kotik}, {Kreisman}, {Kukushkin}, {Kulishenko}, {Cooper},
  {Kut'kin}, {Cannon}, {Larionov}, {Lisakov}, {Litvinenko}, {Likhachev},
  {Likhacheva}, {Lobanov}, {Logvinenko}, {Langston}, {McCracken}, {Medvedev},
  {Melekhin}, {Menderov}, {Murphy}, {Mizyakina}, {Mozgovoi}, {Nikolaev},
  {Novikov}, {Novikov}, {Oreshko}, {Pavlenko}, {Pashchenko}, {Ponomarev},
  {Popov}, {Pravin-Kumar}, {Preston}, {Pyshnov}, {Rakhimov}, {Rozhkov},
  {Romney}, {Rocha}, {Rudakov}, {R{\"a}is{\"a}nen} et~al.}]{kardashev-2013-ar}
\bibinfo{author}{{Kardashev}, N.~S.}, \bibinfo{author}{{Khartov}, V.~V.},
  \bibinfo{author}{{Abramov}, V.~V.}, \bibinfo{author}{{Avdeev}, V.~Y.},
  \bibinfo{author}{{Alakoz}, A.~V.}, \bibinfo{author}{{Aleksandrov}, Y.~A.},
  \bibinfo{author}{{Ananthakrishnan}, S.}, \bibinfo{author}{{Andreyanov},
  V.~V.}, \bibinfo{author}{{Andrianov}, A.~S.}, \bibinfo{author}{{Antonov},
  N.~M.}, \bibinfo{author}{{Artyukhov}, M.~I.}, \bibinfo{author}{{Arkhipov},
  M.~Y.}, \bibinfo{author}{{Baan}, W.}, \bibinfo{author}{{Babakin}, N.~G.},
  \bibinfo{author}{{Babyshkin}, V.~E.}, \bibinfo{author}{{Bartel'}, N.},
  \bibinfo{author}{{Belousov}, K.~G.}, \bibinfo{author}{{Belyaev}, A.~A.},
  \bibinfo{author}{{Berulis}, J.~J.}, \bibinfo{author}{{Burke}, B.~F.},
  \bibinfo{author}{{Biryukov}, A.~V.}, \bibinfo{author}{{Bubnov}, A.~E.},
  \bibinfo{author}{{Burgin}, M.~S.}, \bibinfo{author}{{Busca}, G.},
  \bibinfo{author}{{Bykadorov}, A.~A.}, \bibinfo{author}{{Bychkova}, V.~S.},
  \bibinfo{author}{{Vasil'kov}, V.~I.}, \bibinfo{author}{{Wellington}, K.~J.},
  \bibinfo{author}{{Vinogradov}, I.~S.}, \bibinfo{author}{{Wietfeldt}, R.},
  \bibinfo{author}{{Voitsik}, P.~A.}, \bibinfo{author}{{Gvamichava}, A.~S.},
  \bibinfo{author}{{Girin}, I.~A.}, \bibinfo{author}{{Gurvits}, L.~I.},
  \bibinfo{author}{{Dagkesamanskii}, R.~D.}, \bibinfo{author}{{D'Addario}, L.},
  \bibinfo{author}{{Giovannini}, G.}, \bibinfo{author}{{Jauncey}, D.~L.},
  \bibinfo{author}{{Dewdney}, P.~E.}, \bibinfo{author}{{D'yakov}, A.~A.},
  \bibinfo{author}{{Zharov}, V.~E.}, \bibinfo{author}{{Zhuravlev}, V.~I.},
  \bibinfo{author}{{Zaslavskii}, G.~S.}, \bibinfo{author}{{Zakhvatkin}, M.~V.},
  \bibinfo{author}{{Zinov'ev}, A.~N.}, \bibinfo{author}{{Ilinen}, Y.},
  \bibinfo{author}{{Ipatov}, A.~V.}, \bibinfo{author}{{Kanevskii}, B.~Z.},
  \bibinfo{author}{{Knorin}, I.~A.}, \bibinfo{author}{{Casse}, J.~L.},
  \bibinfo{author}{{Kellermann}, K.~I.}, \bibinfo{author}{{Kovalev}, Y.~A.},
  \bibinfo{author}{{Kovalev}, Y.~Y.}, \bibinfo{author}{{Kovalenko}, A.~V.},
  \bibinfo{author}{{Kogan}, B.~L.}, \bibinfo{author}{{Komaev}, R.~V.},
  \bibinfo{author}{{Konovalenko}, A.~A.}, \bibinfo{author}{{Kopelyanskii},
  G.~D.}, \bibinfo{author}{{Korneev}, Y.~A.}, \bibinfo{author}{{Kostenko},
  V.~I.}, \bibinfo{author}{{Kotik}, A.~N.}, \bibinfo{author}{{Kreisman},
  B.~B.}, \bibinfo{author}{{Kukushkin}, A.~Y.}, \bibinfo{author}{{Kulishenko},
  V.~F.}, \bibinfo{author}{{Cooper}, D.~N.}, \bibinfo{author}{{Kut'kin},
  A.~M.}, \bibinfo{author}{{Cannon}, W.~H.}, \bibinfo{author}{{Larionov},
  M.~G.}, \bibinfo{author}{{Lisakov}, M.~M.}, \bibinfo{author}{{Litvinenko},
  L.~N.}, \bibinfo{author}{{Likhachev}, S.~F.}, \bibinfo{author}{{Likhacheva},
  L.~N.}, \bibinfo{author}{{Lobanov}, A.~P.}, \bibinfo{author}{{Logvinenko},
  S.~V.}, \bibinfo{author}{{Langston}, G.}, \bibinfo{author}{{McCracken}, K.},
  \bibinfo{author}{{Medvedev}, S.~Y.}, \bibinfo{author}{{Melekhin}, M.~V.},
  \bibinfo{author}{{Menderov}, A.~V.}, \bibinfo{author}{{Murphy}, D.~W.},
  \bibinfo{author}{{Mizyakina}, T.~A.}, \bibinfo{author}{{Mozgovoi}, Y.~V.},
  \bibinfo{author}{{Nikolaev}, N.~Y.}, \bibinfo{author}{{Novikov}, B.~S.},
  \bibinfo{author}{{Novikov}, I.~D.}, \bibinfo{author}{{Oreshko}, V.~V.},
  \bibinfo{author}{{Pavlenko}, Y.~K.}, \bibinfo{author}{{Pashchenko}, I.~N.},
  \bibinfo{author}{{Ponomarev}, Y.~N.}, \bibinfo{author}{{Popov}, M.~V.},
  \bibinfo{author}{{Pravin-Kumar}, A.}, \bibinfo{author}{{Preston}, R.~A.},
  \bibinfo{author}{{Pyshnov}, V.~N.}, \bibinfo{author}{{Rakhimov}, I.~A.},
  \bibinfo{author}{{Rozhkov}, V.~M.}, \bibinfo{author}{{Romney}, J.~D.},
  \bibinfo{author}{{Rocha}, P.}, \bibinfo{author}{{Rudakov}, V.~A.},
  \bibinfo{author}{{R{\"a}is{\"a}nen}, A.} et~al. (\bibinfo{year}{2013}).
\newblock \bibinfo{title}{{``RadioAstron'' - A telescope with a size of 300 000
  km: Main parameters and first observational results}}.
\newblock {\it \bibinfo{journal}{Astronomy Reports}\/},  {\it
  \bibinfo{volume}{57}\/}, \bibinfo{pages}{153--194}.
  \DOIprefix\doi{10.1134/S1063772913030025}.
  \href{http://arxiv.org/abs/1303.5013}{\tt arXiv:1303.5013}.
\bibitem[{{Kravchenko} et~al.(2020){Kravchenko}, {G{\'o}mez}, {Kovalev},
  {Lobanov}, {Savolainen}, {Bruni}, {Fuentes}, {Anderson}, {Jorstad},
  {Marscher}, {Tornikoski}, {L{\"a}hteenm{\"a}ki} \&
  {Lisakov}}]{kravchenko-2020-apj}
\bibinfo{author}{{Kravchenko}, E.~V.}, \bibinfo{author}{{G{\'o}mez}, J.~L.},
  \bibinfo{author}{{Kovalev}, Y.~Y.}, \bibinfo{author}{{Lobanov}, A.~P.},
  \bibinfo{author}{{Savolainen}, T.}, \bibinfo{author}{{Bruni}, G.},
  \bibinfo{author}{{Fuentes}, A.}, \bibinfo{author}{{Anderson}, J.~M.},
  \bibinfo{author}{{Jorstad}, S.~G.}, \bibinfo{author}{{Marscher}, A.~P.},
  \bibinfo{author}{{Tornikoski}, M.}, \bibinfo{author}{{L{\"a}hteenm{\"a}ki},
  A.}, \& \bibinfo{author}{{Lisakov}, M.~M.} (\bibinfo{year}{2020}).
\newblock \bibinfo{title}{{Probing the Innermost Regions of AGN Jets and Their
  Magnetic Fields with RadioAstron. III. Blazar S5 0716+71 at Microarcsecond
  Resolution}}.
\newblock {\it \bibinfo{journal}{Astrophysical Journal}\/},  {\it
  \bibinfo{volume}{893}\/}\bibinfo{issue}{(1)}, \bibinfo{pages}{68}.
  \DOIprefix\doi{10.3847/1538-4357/ab7dae}.
  \href{http://arxiv.org/abs/2003.08776}{\tt arXiv:2003.08776}.
\bibitem[{{Langston}(2012)}]{langston-2012-nrao-memo}
\bibinfo{author}{{Langston}, G.} (\bibinfo{year}{2012}).
\newblock {\it \bibinfo{title}{NRAO 43m Antenna Coordinates and Angular
  Limits}\/}.
\newblock \bibinfo{type}{Technical Report} \bibinfo{number}{EDIR Memo \#324}
  National Radio Astronomy Observatory \bibinfo{address}{Charlottesville,
  Virginia}.
\bibitem[{Litvinov \& Pilipenko(2021)}]{litvinov-2021-cqg}
\bibinfo{author}{Litvinov, D.}, \& \bibinfo{author}{Pilipenko, S.}
  (\bibinfo{year}{2021}).
\newblock \bibinfo{title}{Testing the {Einstein} equivalence principle with two
  earth-orbiting clocks}.
\newblock {\it \bibinfo{journal}{Classical and Quantum Gravity}\/},  {\it
  \bibinfo{volume}{38}\/}\bibinfo{issue}{(13)}, \bibinfo{pages}{135010}.
  \DOIprefix\doi{10.1088/1361-6382/abf895}.
  \href{http://arxiv.org/abs/2108.09723}{\tt arXiv:2108.09723}.
\bibitem[{{Litvinov} et~al.(2018){Litvinov}, {Rudenko}, {Alakoz}, {Bach},
  {Bartel}, {Belonenko}, {Belousov}, {Bietenholz}, {Biriukov}, {Carman},
  {Cim{\'o}}, {Courde}, {Dirkx}, {Duev}, {Filetkin}, {Granato}, {Gurvits},
  {Gusev}, {Haas}, {Herold}, {Kahlon}, {Kanevsky}, {Kauts}, {Kopelyansky},
  {Kovalenko}, {Kronschnabl}, {Kulagin}, {Kutkin}, {Lindqvist}, {Lovell},
  {Mariey}, {McCallum}, {Molera Calv{\'e}s}, {Moore}, {Moore}, {Neidhardt},
  {Pl{\"o}tz}, {Pogrebenko}, {Pollard}, {Porayko}, {Quick}, {Smirnov},
  {Sokolovsky}, {Stepanyants}, {Torre}, {de Vicente}, {Yang} \&
  {Zakhvatkin}}]{litvinov-2018-pla}
\bibinfo{author}{{Litvinov}, D.~A.}, \bibinfo{author}{{Rudenko}, V.~N.},
  \bibinfo{author}{{Alakoz}, A.~V.}, \bibinfo{author}{{Bach}, U.},
  \bibinfo{author}{{Bartel}, N.}, \bibinfo{author}{{Belonenko}, A.~V.},
  \bibinfo{author}{{Belousov}, K.~G.}, \bibinfo{author}{{Bietenholz}, M.},
  \bibinfo{author}{{Biriukov}, A.~V.}, \bibinfo{author}{{Carman}, R.},
  \bibinfo{author}{{Cim{\'o}}, G.}, \bibinfo{author}{{Courde}, C.},
  \bibinfo{author}{{Dirkx}, D.}, \bibinfo{author}{{Duev}, D.~A.},
  \bibinfo{author}{{Filetkin}, A.~I.}, \bibinfo{author}{{Granato}, G.},
  \bibinfo{author}{{Gurvits}, L.~I.}, \bibinfo{author}{{Gusev}, A.~V.},
  \bibinfo{author}{{Haas}, R.}, \bibinfo{author}{{Herold}, G.},
  \bibinfo{author}{{Kahlon}, A.}, \bibinfo{author}{{Kanevsky}, B.~Z.},
  \bibinfo{author}{{Kauts}, V.~L.}, \bibinfo{author}{{Kopelyansky}, G.~D.},
  \bibinfo{author}{{Kovalenko}, A.~V.}, \bibinfo{author}{{Kronschnabl}, G.},
  \bibinfo{author}{{Kulagin}, V.~V.}, \bibinfo{author}{{Kutkin}, A.~M.},
  \bibinfo{author}{{Lindqvist}, M.}, \bibinfo{author}{{Lovell}, J.~E.~J.},
  \bibinfo{author}{{Mariey}, H.}, \bibinfo{author}{{McCallum}, J.},
  \bibinfo{author}{{Molera Calv{\'e}s}, G.}, \bibinfo{author}{{Moore}, C.},
  \bibinfo{author}{{Moore}, K.}, \bibinfo{author}{{Neidhardt}, A.},
  \bibinfo{author}{{Pl{\"o}tz}, C.}, \bibinfo{author}{{Pogrebenko}, S.~V.},
  \bibinfo{author}{{Pollard}, A.}, \bibinfo{author}{{Porayko}, N.~K.},
  \bibinfo{author}{{Quick}, J.}, \bibinfo{author}{{Smirnov}, A.~I.},
  \bibinfo{author}{{Sokolovsky}, K.~V.}, \bibinfo{author}{{Stepanyants},
  V.~A.}, \bibinfo{author}{{Torre}, J.~M.}, \bibinfo{author}{{de Vicente}, P.},
  \bibinfo{author}{{Yang}, J.}, \& \bibinfo{author}{{Zakhvatkin}, M.~V.}
  (\bibinfo{year}{2018}).
\newblock \bibinfo{title}{{Probing the gravitational redshift with an
  Earth-orbiting satellite}}.
\newblock {\it \bibinfo{journal}{Physics Letters A}\/},  {\it
  \bibinfo{volume}{382}\/}\bibinfo{issue}{(33)}, \bibinfo{pages}{2192--2198}.
  \DOIprefix\doi{10.1016/j.physleta.2017.09.014}.
  \href{http://arxiv.org/abs/1710.10074}{\tt arXiv:1710.10074}.
\bibitem[{{Mezger} et~al.(1966){Mezger}, {Brown}, {Pauliny-Toth}, {Schraml} \&
  {Turlo}}]{mezger-1966-nrao-memo}
\bibinfo{author}{{Mezger}, P.~G.}, \bibinfo{author}{{Brown}, H.},
  \bibinfo{author}{{Pauliny-Toth}, I.}, \bibinfo{author}{{Schraml}, J.}, \&
  \bibinfo{author}{{Turlo}, Z.} (\bibinfo{year}{1966}).
\newblock {\it \bibinfo{title}{Radio Tests of the NRAO 140-foot Telescope in
  the Wavelength Range Between 11 and 0.95 cm}\/}.
\newblock \bibinfo{type}{Technical Report} \bibinfo{number}{NRAO Internal
  Report} National Radio Astronomy Observatory
  \bibinfo{address}{Charlottesville, Virginia}.
\bibitem[{{Moyer}(1971)}]{moyer-1971-techreport}
\bibinfo{author}{{Moyer}, T.~D.} (\bibinfo{year}{1971}).
\newblock {\it \bibinfo{title}{Mathematical formulation of the Double-Precision
  Orbit Determination Program (DPODP)}\/}.
\newblock \bibinfo{type}{Technical Report} \bibinfo{number}{32-1527} Jet
  Propulsion Laboratory, California Institute of Technology
  \bibinfo{address}{Pasadena, California}.
\bibitem[{{Moyer}(2005)}]{moyer-2005-book}
\bibinfo{author}{{Moyer}, T.~D.} (\bibinfo{year}{2005}).
\newblock {\it \bibinfo{title}{Formulation for observed and computed values of
  Deep Space Network data types for navigation}\/} volume~\bibinfo{volume}{3}
  of {\it \bibinfo{series}{Deep space communications and navigation}\/}.
\newblock \bibinfo{publisher}{Wiley-Interscience}.
\bibitem[{{Nunes} et~al.(2020){Nunes}, {Bartel}, {Bietenholz}, {Zakhvatkin},
  {Litvinov}, {Rudenko}, {Gurvits}, {Granato} \& {Dirkx}}]{nunes-2020-asr}
\bibinfo{author}{{Nunes}, N.~V.}, \bibinfo{author}{{Bartel}, N.},
  \bibinfo{author}{{Bietenholz}, M.~F.}, \bibinfo{author}{{Zakhvatkin}, M.~V.},
  \bibinfo{author}{{Litvinov}, D.~A.}, \bibinfo{author}{{Rudenko}, V.~N.},
  \bibinfo{author}{{Gurvits}, L.~I.}, \bibinfo{author}{{Granato}, G.}, \&
  \bibinfo{author}{{Dirkx}, D.} (\bibinfo{year}{2020}).
\newblock \bibinfo{title}{{The gravitational redshift monitored with
  RadioAstron from near Earth up to 350,000 km}}.
\newblock {\it \bibinfo{journal}{Advances in Space Research}\/},  {\it
  \bibinfo{volume}{65}\/}\bibinfo{issue}{(2)}, \bibinfo{pages}{790--797}.
  \DOIprefix\doi{10.1016/j.asr.2019.03.012}.
  \href{http://arxiv.org/abs/1904.01060}{\tt arXiv:1904.01060}.
\bibitem[{{Petrov}(2009)}]{petrov-2009-url}
\bibinfo{author}{{Petrov}, L.} (\bibinfo{year}{2009}).
\newblock \bibinfo{title}{{VLBI global solution asg2009d}}.
\newblock
  \bibinfo{howpublished}{\url{http://astrogeo.org/vlbi/solutions/2009d/}}.
\newblock \bibinfo{note}{{Accessed: 25/08/2021}}.
\bibitem[{{SKED antenna catalog}(2020)}]{sked-antenna-cat}
\bibinfo{author}{{SKED antenna catalog}} (\bibinfo{year}{2020}).
\newblock
  \bibinfo{howpublished}{\url{https://github.com/nvi-inc/sked_catalogs}}.
\newblock \bibinfo{note}{{Accessed: 25/08/2021}}.
\bibitem[{{Sovers} \& {Fanselow}(1987)}]{sovers-1987-techreport}
\bibinfo{author}{{Sovers}, O.~J.}, \& \bibinfo{author}{{Fanselow}, J.~L.}
  (\bibinfo{year}{1987}).
\newblock {\it \bibinfo{title}{Observation model and parameter partials for the
  JPL VLBI parameter estimation software MASTERFIT-1987}\/}.
\newblock \bibinfo{type}{Technical Report} \bibinfo{number}{88-18523, JPL
  Publication 83-39 rev. 3} Jet Propulsion Laboratory, California Institute of
  Technology \bibinfo{address}{Pasadena, California}.
\bibitem[{{Sovers} et~al.(1998){Sovers}, {Fanselow} \&
  {Jacobs}}]{sovers-1998-rmp}
\bibinfo{author}{{Sovers}, O.~J.}, \bibinfo{author}{{Fanselow}, J.~L.}, \&
  \bibinfo{author}{{Jacobs}, C.~S.} (\bibinfo{year}{1998}).
\newblock \bibinfo{title}{{Astrometry and geodesy with radio interferometry:
  experiments, models, results}}.
\newblock {\it \bibinfo{journal}{Reviews of Modern Physics}\/},  {\it
  \bibinfo{volume}{70}\/}, \bibinfo{pages}{1393--1454}.
  \DOIprefix\doi{10.1103/RevModPhys.70.1393}.
  \href{http://arxiv.org/abs/astro-ph/9712238}{\tt arXiv:astro-ph/9712238}.
\bibitem[{{Thompson} et~al.(2017){Thompson}, {Moran} \&
  {Swenson}}]{thompson-moran-swenson-2017-book}
\bibinfo{author}{{Thompson}, A.~R.}, \bibinfo{author}{{Moran}, J.~M.}, \&
  \bibinfo{author}{{Swenson}, G.~W., Jr.} (\bibinfo{year}{2017}).
\newblock {\it \bibinfo{title}{{Interferometry and Synthesis in Radio
  Astronomy, 3rd Edition}}\/}.
\newblock \DOIprefix\doi{10.1007/978-3-319-44431-4}.
\bibitem[{{Vessot} \& {Levine}(1979)}]{vessot-levine-1979-grg}
\bibinfo{author}{{Vessot}, R.~F.~C.}, \& \bibinfo{author}{{Levine}, M.~W.}
  (\bibinfo{year}{1979}).
\newblock \bibinfo{title}{{A test of the equivalence principle using a
  space-borne clock}}.
\newblock {\it \bibinfo{journal}{General Relativity and Gravitation}\/},  {\it
  \bibinfo{volume}{10}\/}, \bibinfo{pages}{181--204}.
  \DOIprefix\doi{10.1007/BF00759854}.
\bibitem[{{Vessot} et~al.(1980){Vessot}, {Levine}, {Mattison}, {Blomberg},
  {Hoffman}, {Nystrom}, {Farrel}, {Decher}, {Eby} \&
  {Baugher}}]{vessot-levine-1980-prl}
\bibinfo{author}{{Vessot}, R.~F.~C.}, \bibinfo{author}{{Levine}, M.~W.},
  \bibinfo{author}{{Mattison}, E.~M.}, \bibinfo{author}{{Blomberg}, E.~L.},
  \bibinfo{author}{{Hoffman}, T.~E.}, \bibinfo{author}{{Nystrom}, G.~U.},
  \bibinfo{author}{{Farrel}, B.~F.}, \bibinfo{author}{{Decher}, R.},
  \bibinfo{author}{{Eby}, P.~B.}, \& \bibinfo{author}{{Baugher}, C.~R.}
  (\bibinfo{year}{1980}).
\newblock \bibinfo{title}{{Test of relativistic gravitation with a space-borne
  hydrogen maser}}.
\newblock {\it \bibinfo{journal}{Physical Review Letters}\/},  {\it
  \bibinfo{volume}{45}\/}, \bibinfo{pages}{2081--2084}.
  \DOIprefix\doi{10.1103/PhysRevLett.45.2081}.
\bibitem[{Vremya-Ch(2006)}]{vremya-ch-vch-1010}
\bibinfo{author}{Vremya-Ch} (\bibinfo{year}{2006}).
\newblock \bibinfo{title}{{Active on-board hydrogen maser for Radioastron space
  mission VCH-1010}}.
\newblock
  \bibinfo{howpublished}{\url{https://www.vremya-ch.com/english/product/index6e49.html?Razdel=8&Id=39}}.
\newblock \bibinfo{note}{{Accessed: 25/08/2021}}.
\bibitem[{{Wade}(1970)}]{wade-1970-apj}
\bibinfo{author}{{Wade}, C.~M.} (\bibinfo{year}{1970}).
\newblock \bibinfo{title}{{Precise Positions of Radio Sources. I. Radio
  Measurements}}.
\newblock {\it \bibinfo{journal}{Astrophysical Journal}\/},  {\it
  \bibinfo{volume}{162}\/}, \bibinfo{pages}{381 -- 390}.
  \DOIprefix\doi{10.1086/150669}.
\bibitem[{Winternitz et~al.(2017)Winternitz, Bamford, Price, Carpenter, Long \&
  Farahmand}]{winternitz-2017-nav}
\bibinfo{author}{Winternitz, L.~B.}, \bibinfo{author}{Bamford, W.~A.},
  \bibinfo{author}{Price, S.~R.}, \bibinfo{author}{Carpenter, J.~R.},
  \bibinfo{author}{Long, A.~C.}, \& \bibinfo{author}{Farahmand, M.}
  (\bibinfo{year}{2017}).
\newblock \bibinfo{title}{Global positioning system navigation above 76,000 km
  for nasa's magnetospheric multiscale mission}.
\newblock {\it \bibinfo{journal}{NAVIGATION}\/},  {\it
  \bibinfo{volume}{64}\/}\bibinfo{issue}{(2)}, \bibinfo{pages}{289--300}.
  \DOIprefix\doi{https://doi.org/10.1002/navi.198}.
\bibitem[{{Zakhvatkin} et~al.(2020){Zakhvatkin}, {Andrianov}, {Avdeev},
  {Kostenko}, {Kovalev}, {Likhachev}, {Litovchenko}, {Litvinov}, {Rudnitskiy},
  {Shchurov}, {Sokolovsky}, {Stepanyants}, {Tuchin}, {Voitsik}, {Zaslavskiy},
  {Zharov} \& {Zuga}}]{zakhvatkin-2020-asr}
\bibinfo{author}{{Zakhvatkin}, M.~V.}, \bibinfo{author}{{Andrianov}, A.~S.},
  \bibinfo{author}{{Avdeev}, V.~Y.}, \bibinfo{author}{{Kostenko}, V.~I.},
  \bibinfo{author}{{Kovalev}, Y.~Y.}, \bibinfo{author}{{Likhachev}, S.~F.},
  \bibinfo{author}{{Litovchenko}, I.~D.}, \bibinfo{author}{{Litvinov}, D.~A.},
  \bibinfo{author}{{Rudnitskiy}, A.~G.}, \bibinfo{author}{{Shchurov}, M.~A.},
  \bibinfo{author}{{Sokolovsky}, K.~V.}, \bibinfo{author}{{Stepanyants},
  V.~A.}, \bibinfo{author}{{Tuchin}, A.~G.}, \bibinfo{author}{{Voitsik},
  P.~A.}, \bibinfo{author}{{Zaslavskiy}, G.~S.}, \bibinfo{author}{{Zharov},
  V.~E.}, \& \bibinfo{author}{{Zuga}, V.~A.} (\bibinfo{year}{2020}).
\newblock \bibinfo{title}{{RadioAstron orbit determination and evaluation of
  its results using correlation of space-VLBI observations}}.
\newblock {\it \bibinfo{journal}{Advances in Space Research}\/},  {\it
  \bibinfo{volume}{65}\/}\bibinfo{issue}{(2)}, \bibinfo{pages}{798--812}.
  \DOIprefix\doi{10.1016/j.asr.2019.05.007}.

\end{thebibliography}

\end{multicols}

\end{document}